\documentclass[usegraphicx,usenatbib]{mn2e}

\usepackage{lscape}
\usepackage{hyperref}

\newcommand{\subs}[1]{$_{\rm #1}$}
\newcommand{\sups}[1]{$^{\rm #1}$}

\newcommand{\degrees}{$^\circ$ }
\newcommand{\BE}{\begin{equation}}
\newcommand{\EE}{\end{equation}}
\newcommand{\Lsolar}{\mbox{\,$\rm L_{\odot}$}}        
\newcommand{\Msolar}{\mbox{\,$\rm M_{\odot}$} }	      
\newcommand{\Rsolar}{\mbox{\,$\rm R_{\odot}$}}	      
\newcommand{\Ca}{\mbox{Ca \small{II} }}
\newcommand{\Bl}{$B_{l}$ }
\newcommand{\Bla}{$|B_{l}|$ }
\newcommand{\kmsn}{km\ s$^{-1}$}
\newcommand{\kmss}{km\ s$^{-1}$ }
\newcommand{\vsinis}{$\!${\em v\,}sin{\em i} }
\newcommand{\vsini}{$\!${\em v\,}sin{\em i}}
\def\ga{\mathrel{\hbox{\rlap{\hbox{\lower4pt\hbox{$\sim$}}}\hbox{$>$}}}}
\def\la{\mathrel{\hbox{\rlap{\hbox{\lower4pt\hbox{$\sim$}}}\hbox{$<$}}}}

\title[A Bcool magnetic snapshot survey of solar-type stars]{A Bcool magnetic snapshot survey of solar-type stars}

\author[S. C. Marsden et al.]
{S.~C.~Marsden,$^{1}$ \thanks{Email: Stephen.Marsden@usq.edu.au, ppetit@irap.omp.eu, jeffers@astro.physik.uni-goettingen.de}  P.~Petit,$^{2,3\,\,\star}$ S.~V.~Jeffers,$^{4\,\,\star}$ J.~Morin,$^{4,5}$ R.~Fares,$^{6}$ A.~Reiners,$^{4}$ \and J.-D.~do~Nascimento~Jr.,$^{7,8}$  M.~Auri\`{e}re,$^{2,3}$ J.~Bouvier,$^{9}$ B.~D.~Carter,$^{1}$ C.~Catala,$^{10}$ \and B.~Dintrans,$^{2,3}$ J.-F.~Donati,$^{2,3}$ T.~Gastine,$^{11}$ M.~Jardine,$^{6}$ R.~Konstantinova-Antova,$^{12}$ \and J.~Lanoux,$^{2,3}$ F.~Ligni\`{e}res,$^{2,3}$ A.~Morgenthaler,$^{2,3}$ J.~C.~Ram\`{i}rez-V\`{e}lez,$^{13}$ \and S.~Th\'{e}ado,$^{2,3}$ V.~Van~Grootel,$^{14}$ and the Bcool Collaboration\\
   $^{1}$Computational Engineering and Science Research Centre, University of Southern Queensland, Toowoomba, 4350, Australia\\
   $^{2}$Universit\'{e} de Toulouse, UPS-OMP, Institut de Recherche en Astrophysique et Plan\'{e}tologie, Toulouse, France\\
   $^{3}$CNRS, Institut de Recherche en Astrophysique et Plan\'{e}tologie, 14 Avenue Edouard Belin, F-31400 Toulouse, France\\
   $^{4}$Institut f\"{u}r Astrophysik, Georg-August-Universit\"{a}t G\"{o}ttingen, Friedrich-Hund-Platz 1, 37077 G\"{o}ttingen, Germany\\
   $^{5}$LUPM-UMR 5299, CNRS \& Universit\'{e} Montpellier II, Place Eug\`{e}ne Bataillon, 34095 Montpellier Cedex 05, France\\
   $^{6}$SUPA, School of Physics \& Astronomy, University of St Andrews, North Haugh, St. Andrews KY16 9SS, UK\\
   $^{7}$Departamento de Fisica Te\'{o}rica e Experimental, Universidade Federal do Rio Grande do Norte, CEP: 59072-970 Natal, RN, Brazil\\ 
   $^{8}$Harvard-Smithsonian Center for Astrophysics, Cambridge, Massachusetts 02138, USA\\
   $^{9}$UJF-Grenoble 1 / CNRS-INSU, Institut de Plan\'{e}tologie et d'Astrophysique de Grenoble (IPAG) UMR 5274, Grenoble, F-38041\\
   $^{10}$Observatoire de Paris, LESIA, 5 place Jules Janssen, 92195 Meudon Cedex, France\\
   $^{11}$Max Planck Institut f\"{u}r Sonnensystemforschung, Max Planck Strasse 2, 37191 Katlenburg-Lindau, Germany\\
   $^{12}$Institute of Astronomy and NAO, Bulgarian Academy of Sciences, 72 Tsarigradsko shose, 1784 Sofia, Bulgaria\\
   $^{13}$Instituto de Astronomia, Universidad Nacional Autonoma de Mexico, 04510 Coyoacan, DF, Mexico\\
   $^{14}$Institut dAstrophysique et de G\'{e}ophysique, Universit\'{e} de Liege, All\'{e}e du 6 Aout 17, 4000 Liege, Belgium}

\date{Submitted version: 14th November 2013}

\pagerange{\pageref{firstpage}--\pageref{lastpage}} \pubyear{2013}

\begin{document}

\label{firstpage}

\maketitle

\begin{abstract}
Stellar magnetic field measurements obtained from spectropolarimetry offer key data for activity and dynamo studies, and we present the results of a major high-resolution spectropolarimetric Bcool project magnetic snapshot survey of 170 solar-type stars from observations with the Telescope Bernard Lyot and the Canada-France-Hawaii Telescope. For each target star a high signal-to-noise circularly polarised Stokes V profile has been obtained using Least-Squares Deconvolution, and used to detect surface magnetic fields and measure the corresponding mean surface longitudinal magnetic field ($B_{l}$).  Chromospheric activity indicators were also measured. 

Surface magnetic fields were detected for 67 stars, with 21 of these stars classified as mature solar-type stars, a result that increases by a factor of four the number of mature solar-type stars on which magnetic fields have been observed. In addition, a magnetic field was detected for 3 out of 18 of the subgiant stars surveyed. For the population of K-dwarfs the mean value of \Bla ($|B_{l}|_{mean}$) was also found to be higher (5.7 G) than $|B_{l}|_{mean}$ measured for the G-dwarfs (3.2 G) and the F-dwarfs (3.3 G). For the sample as a whole $|B_{l}|_{mean}$ increases with rotation rate and decreases with age, and the upper envelope for \Bla correlates well with the observed chromospheric emission.  Stars with a chromospheric S-index greater than about 0.2 show a high magnetic field detection rate and so offer optimal targets for future studies.

This survey constitutes the most extensive spectropolarimetric survey of cool stars undertaken to date, and suggests that it is feasible to pursue magnetic mapping of a wide range of moderately active solar-type stars to improve understanding of their surface fields and dynamos.
\end{abstract}


\section{Introduction} \label{Sec_int}

For the Sun and other slowly-rotating solar-type stars an interface-layer dynamo operating at the tachocline between the radiative and convective zones qualitatively explains the generation of toroidal and poloidal fields, the presence of a solar activity cycle, and the strong dependence of activity on stellar rotation \citep{ParkerEN:1993, CharbonneauP:2010}. Although there remain many aspects of solar magnetic field generation that cannot be comprehensively modelled, or fully understood. For solar-type stars that rotate much faster than the modern-day Sun, the dynamo appears to be generated by a fundamentally different type that operates throughout the whole convection zone, a so-called distributed dynamo whose existence is supported by the observational evidence accumulated throughout the last decade \citep[i.e.][]{DonatiJF:2003, PetitP:2004, MarsdenSC:2006}, and detailed numerical models \citep{BrownBP:2010}. To better understand dynamos across a range of solar-type stars and their evolution with the star, the observation of a variety of active stars offers an efficient way to probe how various physical stellar parameters (especially the depth of the convective zone or the rotation rate) can enhance or inhibit dynamo processes. 

Improved understanding of how the stellar dynamo operates in solar-type stars can be sought by probing the magnetic fields of cool stars across a range of stellar properties. However most techniques used to observe stellar magnetism either involve indirect proxies such as, Ca II H \& K, X-ray and radio measurements, or have involved a small number direct measurements obtained from Zeeman broadening \citep[see][for a review]{ReinersA:2012}. Despite these limitations, these surveys have shown a clear rotational dependence and saturation of the stellar dynamo, a relationship between stellar magnetic cycles and stellar rotation period, and a tight correlation between dynamo efficiency and Rossby number \citep[e.g.][]{SaarS:2002, BohmVitenseE:2007}.  

One of the most insightful ways to observe the dynamo in solar-type stars is through the use of spectropolarimetric observations. A significant amount of work has previously been achieved using spectropolarimetric observations of young, active, solar-type stars \citep[i.e.][]{DonatiJF:1997, DonatiJF:2003, JeffersSV:2011, MarsdenSC:2011a, MarsdenSC:2011b}, but it is only with recent advances in instrumentation, such as the development of the ESPaDOnS and NARVAL spectropolarimeters, that it is possible to directly measure solar strength magnetic fields on stars other than the Sun. This approach is complementary to other magnetic field proxies, since polarimetry is providing us with information on the strength and polarity of the large-scale magnetic field component.

The Bcool\footnote{\url{http://bcool.ast.obs-mip.fr/}, part of the MagIcS initiative \url{http://www.ast.obs-mip.fr/users/donati/magics/v1/}} project is an international collaboration of over 70 scientists with a common goal of understanding the magnetic activity of low-mass stars. The Bcool project approaches this question from both theoretical and observational viewpoints and has a number of threads of investigation, ranging from understanding how fully-convective stars generate magnetic fields in the absence of an interface layer \citep[i.e.][]{DonatiJF:2006a, MorinJ:2008} through to determining if the magnetic field in evolved stars has a fossil or dynamo origin \citep[i.e.][]{KonstantinovaAntovaR:2012}. In this paper we present the first large-scale spectropolarimetric survey of active and inactive solar-type stars to help our understanding of how the magnetic dynamo operates in solar-type main-sequence F, G and K stars as part of the Bcool cool star thread.

There are three main aspects to this Bcool project studies of solar-type stars: (1) a snapshot survey of a large number of solar-type stars, in order to determine the number of stars showing detectable magnetic fields and how the properties of these magnetic fields change with basic stellar parameters (this paper). (2) For a sample of interesting stars from the snapshot survey we have embarked on a program to obtain maps of their global magnetic topology (Petit et al. in prep.). (3) For a further, even smaller, stellar sample we have have undertaken long-term mapping, in order to look for stellar activity cycles \citep[i.e.][]{PetitP:2009, MorgenthalerA:2012}.   

The snapshot survey, which is the focus of this paper, has two main aims: (1) to detect magnetic fields on our target sample of stars so that we can select the most suitable targets for long-term mapping of their magnetic field topology (2) to determine if the large-scale magnetic field properties of solar-type stars vary with basic stellar parameters, such as age, temperature, rotation rate and \Ca H \& K emission. Additionally, this paper aims to determine if the detection rate of magnetic fields varies significantly with the signal-to-noise ratio of the observations (i.e. is there a minimum signal-to-noise ratio required to detect a magnetic field) and what is the minimum \Ca H \& K S-index for which we can get magnetic detections (i.e. is the S-index a good proxy for magnetic field detections).

In this paper we first describe the target selection and the spectropolarimetric observations in Sections~\ref{Sec_Targetselc} and~\ref{Sec_obs}. In Sections~\ref{Sec_Bl} and~\ref{Sec_activity} we detail the analysis undertaken on the data set and in Section~\ref{Sec_res} we discuss the results of the survey.

\section{The Bcool sample} \label{Sec_Targetselc}

The goal of the Bcool spectropolarimetric survey is to observe as many of the bright (V $\la$ 9.0) solar-type stars as possible to further our understanding of the magnetic activity of cool stars. In this first paper we present the spectropolarimetric snapshots of 170 solar-type stars that we have observed starting in 2006 until 2013 as part of the Bcool survey.  The initial targets were chosen from \citet{ValentiJA:2005} catalogues of ``Spectroscpic Properties of Cool Stars I" as they have accurate values for their stellar parameters.  Since \citet{ValentiJA:2005} mainly focused on G dwarfs, our target list shows a dearth of both F- and K-stars.  To address this observational bias, ten additional K stars were included from the \citet{WrightJT:2004} catalogue, and other sources as specified in Table~\ref{Bcool_params}, mostly based on their higher activity levels which introduced a bias towards younger K-stars. Additionally, stars that are already known to show magnetic cycles were included from \citet{BaliunasSL:1995} together with a sample of stars of around solar-mass of varying ages to represent the Sun across its evolutionary path (see references in Table~\ref{Bcool_params}). Most of the stars in our sample are mature solar-type stars, but there is a large spread in ages (ref Table~\ref{Bcool_params}). 

\setlength{\tabcolsep}{4pt}

\begin{landscape}
\begin{table}
\scriptsize
\begin{center}
\caption{The stellar parameters of the Bcool solar-type star sample. The 5th column gives the spectral type of the star according to SIMBAD. The 13th column gives the radius of the stellar convective zone, while the last two columns give the Log of the Rossby Number (calculated from equation 13 of \citet{WrightJT:2004} from the individual Log(R$^{\prime}_{\rm{HK}}$) measurements given in Table~\ref{Bcool_Activity}) and the list of references (see footnote at the end of the table) where the data has been taken from. \sups{SG}: Identifies the stars as a subgiant, see Figure~\ref{Fig_HR} and \sups{B} indicates a possible binary with two stars seen in the Stokes I LSD profile (the analysis has been done on the deeper of the two Stokes I profiles).  X indicates an undetermined parameter.}\label{Bcool_params}
\begin{tabular}{lccccccccccccccc}
\hline
HIP & SPOCS & HD & Other & Sp. & T\subs{eff} & Log(g) & Log(M/H) or & Log(Lum) & Age & Mass & Radius & Radius\subs{CZ} & \vsini & Log & Refs.\\
no. & no. & no. & names & Type & (K) & (cm s\sups{-2}) & Log(Fe/H)* & (\Lsolar)    & (Gyr) & (\Msolar) & (\Rsolar) & (\Rsolar) & (\kmsn) & (Rossby Number) & \\
\hline
400 & 1 & 225261 & - & G9V & 5265$^{+18}_{-18}$ & 4.54$^{+0.04}_{-0.02}$ & -0.31$^{+0.03}_{-0.03}$ & -0.341$^{+0.049}_{-0.049}$ & 12.28$^{+1.72}_{-7.08}$ & 0.794$^{+0.034}_{-0.018}$ & 0.80$^{+0.02}_{-0.03}$ & 0.240$^{+0.010}_{-0.016}$ & 0.0$^{+0.5}_{-0.0}$ & +0.303$^{+0.013}_{-0.009}$ & 1,2 \\[1mm]
544 & 4 & 166 & V439 And & K0V & 5577$^{+31}_{-31}$ & 4.57$^{+0.01}_{-0.02}$ & +0.12$^{+0.02}_{-0.02}$ & -0.215$^{+0.023}_{-0.023}$ & 0.00$^{+0.84}_{-0.00}$ & 0.977$^{+0.010}_{-0.014}$ & 0.88$^{+0.02}_{-0.02}$ & 0.254$^{+0.007}_{-0.004}$ & 4.1$^{+0.4}_{-0.4}$ & -0.395$^{+0.000}_{-0.000}$ & 1,2 \\[1mm]
682 & 6 & 377 & - & G2V & 5873$^{+44}_{-44}$ & 4.37$^{+0.04}_{-0.04}$ & +0.11$^{+0.03}_{-0.03}$ & 0.082$^{+0.078}_{-0.078}$ & 6.12$^{+1.28}_{-1.48}$ & 1.045$^{+0.028}_{-0.024}$ & 1.12$^{+0.05}_{-0.05}$ & 0.308$^{+0.020}_{-0.021}$ & 14.6$^{+0.5}_{-0.5}$ & -0.395$^{+0.000}_{-0.024}$ & 1,2 \\[1mm]
1499 & 13 & 1461 & - & G0V & 5765$^{+44}_{-44}$ & 4.37$^{+0.03}_{-0.03}$ & +0.16$^{+0.03}_{-0.03}$ & 0.078$^{+0.041}_{-0.041}$ & 7.12$^{+1.40}_{-1.56}$ & 1.026$^{+0.040}_{-0.030}$ & 1.11$^{+0.04}_{-0.04}$ & 0.323$^{+0.019}_{-0.020}$ & 1.6$^{+0.5}_{-0.5}$ & +0.336$^{+0.004}_{-0.004}$ & 1,2 \\[1mm]
1813 & 16 & 1832 & - & F8 & 5731$^{+44}_{-44}$ & 4.28$^{+0.04}_{-0.03}$ & -0.02$^{+0.03}_{-0.03}$ & 0.119$^{+0.074}_{-0.074}$ & 10.88$^{+1.36}_{-1.36}$ & 0.965$^{+0.020}_{-0.020}$ & 1.18$^{+0.06}_{-0.05}$ & 0.365$^{+0.021}_{-0.022}$ & 2.8$^{+0.5}_{-0.5}$ & +0.290$^{+0.005}_{-0.000}$ & 1,2 \\[1mm]
3093 & 26 & 3651 & 54 Psc & K0V & 5221$^{+25}_{-25}$ & 4.51$^{+0.02}_{-0.01}$ & +0.16$^{+0.02}_{-0.02}$ & -0.286$^{+0.018}_{-0.018}$ & X & 0.882$^{+0.026}_{-0.021}$ & 0.88$^{+0.03}_{-0.02}$ & 0.296$^{+0.006}_{-0.009}$ & 1.1$^{+0.3}_{-0.3}$ & +0.324$^{+0.004}_{-0.000}$ & 1,2 \\[1mm]
3203 & 31 & 3821 & - & - & 5828$^{+44}_{-44}$ & 4.51$^{+0.02}_{-0.03}$ & -0.07$^{+0.03}_{-0.03}$ & -0.088$^{+0.058}_{-0.058}$ & 0.00$^{+1.96}_{-0.00}$ & 1.011$^{+0.020}_{-0.024}$ & 0.95$^{+0.03}_{-0.03}$ & 0.242$^{+0.012}_{-0.007}$ & 4.3$^{+0.5}_{-0.5}$ & -0.112$^{+0.017}_{-0.017}$ & 1,2 \\[1mm]
3206 & 32 & 3765 & - & K2V & 5032$^{+44}_{-44}$ & 4.58$^{+0.03}_{-0.03}$ & +0.12$^{+0.03}_{-0.03}$ & -0.465$^{+0.035}_{-0.035}$ & 2.00$^{+9.20}_{-2.00}$ & 0.852$^{+0.020}_{-0.044}$ & 0.79$^{+0.02}_{-0.02}$ & 0.251$^{+0.016}_{-0.008}$ & 0.0$^{+0.5}_{-0.0}$ & +0.336$^{+0.019}_{-0.016}$ & 1,2 \\[1mm]
3765 & 40 & 4628 & - & K2.5V & 4994$^{+25}_{-25}$ & 4.59$^{+0.03}_{-0.02}$ & -0.19$^{+0.02}_{-0.02}$ & -0.532$^{+0.016}_{-0.016}$ & X & 0.756$^{+0.026}_{-0.013}$ & 0.76$^{+0.01}_{-0.03}$ & 0.238$^{+0.005}_{-0.011}$ & 2.0$^{+0.3}_{-0.3}$ & +0.294$^{+0.005}_{-0.009}$ & 1,2 \\[1mm]
3821 & 41 & 4614 & eta Cas & G3V & 5941$^{+31}_{-31}$ & 4.44$^{+0.03}_{-0.08}$ & -0.17$^{+0.02}_{-0.02}$ & 0.090$^{+0.012}_{-0.012}$ & 2.88$^{+5.00}_{-1.72}$ & 0.991$^{+0.084}_{-0.034}$ & 1.05$^{+0.06}_{-0.03}$ & 0.243$^{+0.040}_{-0.022}$ & 2.8$^{+0.4}_{-0.4}$ & +0.312$^{+0.000}_{-0.000}$ & 1,2 \\[1mm]
3979 & 43 & 4915 & - & G6V & 5650$^{+44}_{-44}$ & 4.53$^{+0.02}_{-0.03}$ & -0.18$^{+0.03}_{-0.03}$ & -0.164$^{+0.044}_{-0.044}$ & 0.76$^{+3.20}_{-0.76}$ & 0.939$^{+0.024}_{-0.032}$ & 0.88$^{+0.03}_{-0.02}$ & 0.235$^{+0.012}_{-0.009}$ & 1.8$^{+0.5}_{-0.5}$ & +0.222$^{+0.013}_{-0.007}$ & 1,2 \\[1mm]
4127 & 45 & 5065 & - & G0 & 5957$^{+44}_{-44}$ & 4.08$^{+0.04}_{-0.03}$ & -0.08$^{+0.03}_{-0.03}$ & 0.460$^{+0.074}_{-0.074}$ & 6.64$^{+0.52}_{-0.44}$ & 1.108$^{+0.020}_{-0.022}$ & 1.60$^{+0.07}_{-0.07}$ & 0.385$^{+0.030}_{-0.024}$ & 4.1$^{+0.5}_{-0.5}$ & +0.370$^{+0.000}_{-0.000}$ & 1,2 \\[1mm]
5315\sups{SG} & 53 & 6734 & 29 Cet & K0IV & 5067$^{+44}_{-44}$ & 3.76$^{+0.01}_{-0.01}$ & -0.28$^{+0.03}_{-0.03}$ & 0.753$^{+0.078}_{-0.078}$ & 7.3$^{+4.5}_{-3.3}$ & 2.25$^{+0.35}_{-0.35}$ & 3.10$^{+0.06}_{-0.09}$ & 1.515$^{+0.074}_{-0.004}$ & 2.2$^{+0.5}_{-0.5}$ & +0.399$^{+0.004}_{-0.004}$ & 1,2 \\[1mm]
5493\sups{SG} & - & 6920 & 4 And & F8V & 6028$^{+32}_{-32}$ & 3.78$^{+0.07}_{-0.07}$ & -0.01*$^{+0.04}_{-0.04}$ & 1.11$^{+X}_{-X}$ & X & 1.64$^{+X}_{-X}$ & 3.58$^{+0.19}_{-0.19}$ & X & 11.6$^{+0.8}_{-0.8}$ & +0.215$^{+0.000}_{-0.007}$ & 21,24 \\[1mm]
5985 & 64 & 7727 & 40 Cet & F8 & 6022$^{+44}_{-44}$ & 4.29$^{+0.03}_{-0.03}$ & +0.03$^{+0.03}_{-0.03}$ & 0.249$^{+0.047}_{-0.047}$ & 5.12$^{+0.92}_{-0.76}$ & 1.101$^{+0.020}_{-0.020}$ & 1.25$^{+0.06}_{-0.04}$ & 0.296$^{+0.023}_{-0.019}$ & 5.0$^{+0.5}_{-0.5}$ & +0.285$^{+0.005}_{-0.000}$ & 1,2 \\[1mm]
6405 & 68 & 8262 & - & G3V & 5761$^{+44}_{-44}$ & 4.45$^{+0.03}_{-0.04}$ & -0.16$^{+0.03}_{-0.03}$ & -0.035$^{+0.042}_{-0.042}$ & 5.88$^{+2.40}_{-2.60}$ & 0.953$^{+0.032}_{-0.034}$ & 0.98$^{+0.03}_{-0.03}$ & 0.264$^{+0.016}_{-0.020}$ & 1.6$^{+0.5}_{-0.5}$ & +0.303$^{+0.017}_{-0.014}$ & 1,2 \\[1mm]
7244 & 82 & 9472 & - & G0 & 5867$^{+44}_{-44}$ & 4.52$^{+0.01}_{-0.02}$ & +0.00$^{+0.03}_{-0.03}$ & -0.088$^{+0.063}_{-0.063}$ & 0.00$^{+0.64}_{-0.00}$ & 1.037$^{+0.014}_{-0.016}$ & 0.95$^{+0.03}_{-0.03}$ & 0.241$^{+0.006}_{-0.001}$ & 2.2$^{+0.5}_{-0.5}$ & -0.165$^{+0.035}_{-0.018}$ & 1,2 \\[1mm]
7276 & 83 & 9562 & - & G1V & 5939$^{+44}_{-44}$ & 4.04$^{+0.03}_{-0.03}$ & +0.19$^{+0.03}_{-0.03}$ & 0.556$^{+0.044}_{-0.044}$ & 5.04$^{+0.32}_{-0.28}$ & 1.242$^{+0.048}_{-0.038}$ & 1.80$^{+0.06}_{-0.06}$ & 0.452$^{+0.029}_{-0.030}$ & 4.2$^{+0.5}_{-0.5}$ & +0.390$^{+0.004}_{-0.000}$ & 1,2 \\[1mm]
7339 & 84 & 9407 & - & G6V & 5657$^{+44}_{-44}$ & 4.42$^{+0.03}_{-0.03}$ & +0.00$^{+0.03}_{-0.03}$ & -0.028$^{+0.027}_{-0.027}$ & 8.52$^{+2.12}_{-2.32}$ & 0.951$^{+0.026}_{-0.024}$ & 1.01$^{+0.04}_{-0.02}$ & 0.303$^{+0.016}_{-0.020}$ & 0.1$^{+0.5}_{-0.1}$ & +0.328$^{+0.004}_{-0.000}$ & 1,2 \\[1mm]
7513 & 85 & 9826 & ups And & F9V & 6213$^{+22}_{-22}$ & 4.16$^{+0.02}_{-0.04}$ & +0.12$^{+0.01}_{-0.01}$ & 0.522$^{+0.021}_{-0.021}$ & 3.12$^{+0.20}_{-0.24}$ & 1.310$^{+0.021}_{-0.014}$ & 1.64$^{+0.04}_{-0.05}$ & 0.315$^{+0.028}_{-0.073}$ & 9.6$^{+0.3}_{-0.3}$ & +0.316$^{+0.000}_{-0.004}$ & 1,2 \\[1mm]
7585 & 87 & 9986 & - & G5V & 5805$^{+44}_{-44}$ & 4.43$^{+0.03}_{-0.04}$ & +0.05$^{+0.03}_{-0.03}$ & 0.038$^{+0.044}_{-0.044}$ & 5.08$^{+1.88}_{-2.12}$ & 1.022$^{+0.032}_{-0.026}$ & 1.04$^{+0.04}_{-0.03}$ & 0.284$^{+0.019}_{-0.019}$ & 2.6$^{+0.5}_{-0.5}$ & +0.275$^{+0.000}_{-0.000}$ & 1,2 \\[1mm]
7734 & 90 & 10086 & - & G5IV & 5725$^{+44}_{-44}$ & 4.47$^{+0.04}_{-0.03}$ & +0.09$^{+0.03}_{-0.03}$ & -0.049$^{+0.035}_{-0.035}$ & 3.76$^{+2.60}_{-2.76}$ & 1.010$^{+0.038}_{-0.034}$ & 0.98$^{+0.03}_{-0.03}$ & 0.273$^{+0.018}_{-0.015}$ & 2.4$^{+0.5}_{-0.5}$ & +0.093$^{+0.000}_{-0.011}$ & 1,2 \\[1mm]
7918 & - & 10307 & - & G1.5V & 5834$^{+12}_{-12}$ & 4.30$^{+X}_{-X}$ & +0.012*$^{+X}_{-X}$ & 0.135$^{+0.009}_{-0.009}$ & 3.6$^{+4.2}_{-2.4}$ & 1.02$^{+X}_{-X}$ & X & X & 3$^{+X}_{-X}$ & +0.332$^{+0.000}_{-0.000}$ & 23,25 \\[1mm]
7981 & 95 & 10476 & 107 Psc & K1V & 5181$^{+25}_{-25}$ & 4.54$^{+0.03}_{-0.01}$ & -0.07$^{+0.02}_{-0.02}$ & -0.368$^{+0.016}_{-0.016}$ & X & 0.816$^{+0.022}_{-0.011}$ & 0.82$^{+0.02}_{-0.02}$ & 0.266$^{+0.006}_{-0.013}$ & 1.7$^{+0.3}_{-0.3}$ & +0.270$^{+0.005}_{-0.005}$ & 1,2 \\[1mm]
8159 & 97 & 10697 & 109 Psc & G5IV & 5680$^{+44}_{-44}$ & 4.03$^{+0.03}_{-0.03}$ & +0.10$^{+0.03}_{-0.03}$ & 0.448$^{+0.054}_{-0.054}$ & 7.84$^{+0.40}_{-0.48}$ & 1.112$^{+0.026}_{-0.020}$ & 1.73$^{+0.06}_{-0.07}$ & 0.568$^{+0.038}_{-0.032}$ & 2.5$^{+0.5}_{-0.5}$ & +0.366$^{+0.004}_{-0.004}$ & 1,2 \\[1mm]
8362 & 99 & 10780 & V987 Cas & K0V & 5327$^{+44}_{-44}$ & 4.52$^{+0.03}_{-0.03}$ & -0.06$^{+0.03}_{-0.03}$ & -0.291$^{+0.016}_{-0.016}$ & 10.12$^{+3.88}_{-5.20}$ & 0.846$^{+0.036}_{-0.020}$ & 0.85$^{+0.02}_{-0.02}$ & 0.267$^{+0.013}_{-0.015}$ & 1.3$^{+0.5}_{-0.5}$ & +0.124$^{+0.000}_{-0.000}$ & 1,2 \\[1mm]
8486 & - & 11131 & EZ Cet & G1Vk: & 5805$^{+12}_{-12}$ & 4.45$^{+X}_{-X}$ & -0.061*$^{+X}_{-X}$ & -0.046$^{+0.085}_{-0.085}$ & 4.9$^{+1.1}_{-3.1}$ & 1.00$^{+X}_{-X}$ & X & X & 4$^{+X}_{-X}$ & -0.147$^{+0.000}_{-0.000}$ & 23,25 \\[1mm]
9349 & - & 12264 & - & G5V & 5788$^{+12}_{-12}$ & 4.35$^{+X}_{-X}$ & +0.01*$^{+X}_{-X}$ & 0.007$^{+0.044}_{-0.044}$ & 5.4$^{+0.4}_{-3.4}$ & 1.00$^{+X}_{-X}$ & X & X & 3$^{+X}_{-X}$ & +0.114$^{+0.010}_{-0.021}$ & 23,25 \\[1mm]
9406\sups{SG} & 109 & 12328 & - & G5 & 4919$^{+44}_{-44}$ & 3.58$^{+0.04}_{-0.04}$ & -0.08$^{+0.03}_{-0.03}$ & 0.76$^{+0.10}_{-0.10}$ & 4.04$^{+0.68}_{-0.72}$ & 1.312$^{+0.068}_{-0.058}$ & 3.16$^{+0.13}_{-0.18}$ & 1.997$^{+X}_{-0.114}$ & 1.9$^{+0.5}_{-0.5}$ & +0.422$^{+0.000}_{-0.005}$ & 1,2 \\[1mm]
9829 & 112 & 12846 & - & G2V & 5626$^{+44}_{-44}$ & 4.42$^{+0.03}_{-0.03}$ & -0.20$^{+0.03}_{-0.03}$ & -0.084$^{+0.045}_{-0.045}$ & 11.28$^{+2.72}_{-2.72}$ & 0.877$^{+0.028}_{-0.024}$ & 0.97$^{+0.03}_{-0.03}$ & 0.290$^{+0.015}_{-0.020}$ & 2.2$^{+0.5}_{-0.5}$ & +0.307$^{+0.000}_{-0.004}$ & 1,2 \\[1mm]
9911 & 113 & 13043 & - & G2V & 5897$^{+44}_{-44}$ & 4.22$^{+0.04}_{-0.03}$ & +0.06$^{+0.03}_{-0.03}$ & 0.302$^{+0.064}_{-0.064}$ & 7.16$^{+0.76}_{-0.76}$ & 1.080$^{+0.022}_{-0.020}$ & 1.36$^{+0.05}_{-0.06}$ & 0.363$^{+0.024}_{-0.027}$ & 2.7$^{+0.5}_{-0.5}$ & +0.324$^{+0.000}_{-0.004}$ & 1,2 \\[1mm]
10339 & 119 & 13531A & V451 And & G0V & 5621$^{+44}_{-44}$ & 4.54$^{+0.02}_{-0.03}$ & -0.06$^{+0.03}_{-0.03}$ & -0.186$^{+0.048}_{-0.048}$ & 0.00$^{+3.52}_{-0.00}$ & 0.957$^{+0.018}_{-0.028}$ & 0.89$^{+0.02}_{-0.03}$ & 0.245$^{+0.012}_{-0.009}$ & 6.0$^{+0.5}_{-0.5}$ & -0.221$^{+0.019}_{-0.000}$ & 1,2 \\[1mm]
10505 & 121 & 13825 & - & G8IV & 5711$^{+44}_{-44}$ & 4.38$^{+0.04}_{-0.03}$ & +0.17$^{+0.03}_{-0.03}$ & 0.053$^{+0.052}_{-0.052}$ & 7.88$^{+1.56}_{-1.80}$ & 1.011$^{+0.046}_{-0.034}$ & 1.09$^{+0.04}_{-0.04}$ & 0.325$^{+0.022}_{-0.019}$ & 1.5$^{+0.5}_{-0.5}$ & +0.336$^{+0.004}_{-0.008}$ & 1,2 \\[1mm]
11548 & 127 & 15335 & 13 Tri & G0V & 5891$^{+44}_{-44}$ & 3.98$^{+0.03}_{-0.03}$ & -0.20$^{+0.03}_{-0.03}$ & 0.544$^{+0.053}_{-0.053}$ & 6.76$^{+0.36}_{-0.36}$ & 1.101$^{+0.022}_{-0.024}$ & 1.81$^{+0.06}_{-0.07}$ & 0.454$^{+0.029}_{-0.034}$ & 4.3$^{+0.5}_{-0.5}$ & +0.370$^{+0.008}_{-0.000}$ & 1,2 \\[1mm]
12048 & 128 & 16141 & 79 Cet & G5IV & 5794$^{+44}_{-44}$ & 4.19$^{+0.04}_{-0.04}$ & +0.09$^{+0.03}_{-0.03}$ & 0.30$^{+0.10}_{-0.10}$ & 8.68$^{+0.76}_{-0.76}$ & 1.052$^{+0.026}_{-0.022}$ & 1.39$^{+0.07}_{-0.07}$ & 0.416$^{+0.030}_{-0.028}$ & 1.9$^{+0.5}_{-0.5}$ & +0.362$^{+0.004}_{-0.004}$ & 1,2 \\[1mm]
12114 & 129 & 16160 & - & K3V & 4866$^{+31}_{-31}$ & 4.62$^{+X}_{-0.03}$ & +0.00$^{+0.02}_{-0.02}$ & -0.564$^{+0.018}_{-0.018}$ & 0.54$^{+5.84}_{-0.54}$ & 0.809$^{+0.013}_{-0.030}$ & 0.76$^{+0.01}_{-0.02}$ & 0.235$^{+0.011}_{-0.006}$ & 2.9$^{+0.4}_{-0.4}$ & +0.253$^{+0.000}_{-0.006}$ & 1,2 \\[1mm]
\hline
\end{tabular}
\end{center}
\end{table}
\end{landscape}	

\addtocounter {table} {-1}

\begin{landscape}
\begin{table}
\scriptsize
\begin{center}
\caption{continued.}
\begin{tabular}{lccccccccccccccc}
\hline
HIP & SPOCS & HD & Other & Sp. & T\subs{eff} & Log(g) & Log(M/H) or & Log(Lum) & Age & Mass & Radius & Radius\subs{CZ} & \vsini & Log & Refs.\\
no. & no. & no. & names & Type & (K) & (cm s\sups{-2}) & Log(Fe/H)* & (\Lsolar)    & (Gyr) & (\Msolar) & (\Rsolar) & (\Rsolar) & (\kmsn) & (Rossby Number) & \\
\hline
13702 & - & 18256 & rho Ari & F6V & 6380$^{+X}_{-X}$ & 4.17$^{+X}_{-X}$ & -0.23*$^{+X}_{-X}$ & 0.82$^{+X}_{-X}$ & X & 1.56$^{+X}_{-X}$ & 1.70$^{+X}_{-X}$ & X & X & +0.241$^{+0.000}_{-0.000}$ & 14,20 \\[1mm]
14150 & 148 & 18803 & 51 Ari & G8V & 5638$^{+44}_{-44}$ & 4.44$^{+0.04}_{-0.03}$ & +0.09$^{+0.03}_{-0.03}$ & -0.061$^{+0.039}_{-0.039}$ & 7.60$^{+2.48}_{-2.72}$ & 0.962$^{+0.034}_{-0.028}$ & 0.99$^{+0.04}_{-0.03}$ & 0.297$^{+0.019}_{-0.018}$ & 0.8$^{+0.5}_{-0.5}$ & +0.264$^{+0.000}_{-0.005}$ & 1,2 \\[1mm]
15457 & 161 & 20630 & kap\sups{1} Cet & G5Vv & 5742$^{+31}_{-31}$ & 4.52$^{+X}_{-0.03}$ & +0.10$^{+0.02}_{-0.02}$ & -0.084$^{+0.017}_{-0.017}$ & 0.00$^{+2.76}_{-0.00}$ & 1.034$^{+0.014}_{-0.030}$ & 0.95$^{+0.03}_{-0.02}$ & 0.260$^{+0.013}_{-0.008}$ & 5.2$^{+0.4}_{-0.4}$ & -0.129$^{+0.000}_{-0.017}$ & 1,2 \\[1mm]
15776\sups{SG} & 165 & 21019 & - & G2V & 5529$^{+44}_{-44}$ & 3.98$^{+0.02}_{-0.02}$ & -0.36$^{+0.03}_{-0.03}$ & 0.606$^{+0.062}_{-0.062}$ & 7.16$^{+0.56}_{-0.52}$ & 1.654$^{+0.026}_{-0.068}$ & 2.23$^{+0.05}_{-0.10}$ & 0.724$^{+0.050}_{-0.061}$ & 1.7$^{+0.5}_{-0.5}$ & +0.359$^{+0.004}_{-0.000}$ & 1,2 \\[1mm]
16537 & 171 & 22049 & eps Eri & K2Vk: & 5146$^{+31}_{-31}$ & 4.61$^{+X}_{-0.02}$ & +0.00$^{+0.02}_{-0.02}$ & -0.486$^{+0.011}_{-0.011}$ & 0.00$^{+0.60}_{-0.00}$ & 0.856$^{+0.006}_{-0.008}$ & 0.77$^{+0.02}_{-0.01}$ & 0.235$^{+0.005}_{-0.006}$ & 2.4$^{+0.4}_{-0.4}$ & -0.262$^{+0.000}_{-0.000}$ & 1,2 \\[1mm]
16641\sups{SG} & 172 & 22072 & - & K1IVe & 5027$^{+44}_{-44}$ & 3.57$^{+0.03}_{-0.04}$ & -0.19$^{+0.03}_{-0.03}$ & 0.799$^{+0.080}_{-0.080}$ & 2.72$^{+0.52}_{-0.44}$ & 1.396$^{+0.100}_{-0.044}$ & 3.28$^{+0.17}_{-0.14}$ & 1.913$^{+X}_{-0.112}$ & 2.4$^{+0.5}_{-0.5}$ & +0.412$^{+0.000}_{-0.005}$ & 1,2 \\[1mm]
17027\sups{SG} & 175 & 22713 & 21 Eri & K1V & 5065$^{+44}_{-44}$ & 3.68$^{+0.02}_{-0.03}$ & -0.02$^{+0.03}_{-0.03}$ & 0.699$^{+0.055}_{-0.055}$ & 3.04$^{+0.40}_{-0.12}$ & 1.426$^{+0.002}_{-0.050}$ & 2.91$^{+0.12}_{-0.10}$ & 1.640$^{+0.090}_{-0.130}$ & 2.4$^{+0.5}_{-0.5}$ & +0.408$^{+0.000}_{-0.000}$ & 1,2 \\[1mm]
17147 & 178 & 22879 & - & F9V & 5688$^{+44}_{-44}$ & 4.4$^{+X}_{-X}$ & -0.76$^{+0.03}_{-0.03}$ & 0.052$^{+0.043}_{-0.043}$ & X & 0.79$^{+0.02}_{-0.02}$ & 1.092$^{+0.027}_{-0.027}$ & X & 1.3$^{+0.5}_{-0.5}$ & +0.316$^{+0.000}_{-0.000}$ & 1 \\[1mm]
17183\sups{SG} & 179 & 22918 & - & G5 & 4939$^{+44}_{-44}$ & 3.73$^{+0.03}_{-0.04}$ & +0.01$^{+0.03}_{-0.03}$ & 0.536$^{+0.082}_{-0.082}$ & 6.88$^{+0.84}_{-1.24}$ & 1.161$^{+0.040}_{-0.046}$ & 2.48$^{+0.11}_{-0.11}$ & 1.463$^{+0.105}_{-0.083}$ & 2.4$^{+0.5}_{-0.5}$ & +0.412$^{+0.005}_{-0.005}$ & 1,2 \\[1mm]
17378\sups{SG} & 180 & 23249 & del Eri & K1III-IV & 5095$^{+44}_{-44}$ & 4.01$^{+0.01}_{-0.01}$ & +0.03$^{+0.03}_{-0.03}$ & 0.496$^{+0.018}_{-0.018}$ & 6.28$^{+0.08}_{-0.40}$ & 1.193$^{+0.634}_{-0.018}$ & 2.27$^{+0.03}_{-0.03}$ & 1.189$^{+0.020}_{-0.006}$ & 2.6$^{+0.5}_{-0.5}$ & +0.399$^{+0.000}_{-0.000}$ & 1,2 \\[1mm]
18106 & 187 & 24213 & - & G0 & 6044$^{+44}_{-44}$ & 4.20$^{+0.03}_{-0.03}$ & +0.05$^{+0.03}_{-0.03}$ & 0.394$^{+0.071}_{-0.071}$ & 4.28$^{+0.36}_{-0.36}$ & 1.144$^{+0.072}_{-0.016}$ & 1.45$^{+0.06}_{-0.06}$ & 0.325$^{+0.016}_{-0.014}$ & 3.7$^{+0.5}_{-0.5}$ & +0.362$^{+0.008}_{-0.004}$ & 1,2 \\[1mm]
18267 & 189 & 24496 & - & G7V & 5572$^{+44}_{-44}$ & 4.52$^{+0.03}_{-0.04}$ & -0.01$^{+0.03}_{-0.03}$ & -0.152$^{+0.043}_{-0.043}$ & 3.16$^{+3.88}_{-3.16}$ & 0.956$^{+0.030}_{-0.036}$ & 0.91$^{+0.03}_{-0.03}$ & 0.257$^{+0.018}_{-0.011}$ & 0.0$^{+0.5}_{-0.0}$ & +0.290$^{+0.009}_{-0.010}$ & 1,2 \\[1mm]
18606\sups{SG} & 194 & 25069 & - & G9V & 4994$^{+44}_{-44}$ & 3.48$^{+0.04}_{-0.03}$ & +0.10$^{+0.03}_{-0.03}$ & 0.944$^{+0.066}_{-0.066}$ & 1.72$^{+0.24}_{-0.12}$ & 1.701$^{+0.040}_{-0.044}$ & 3.98$^{+0.16}_{-0.15}$ & 1.999$^{+X}_{-0.016}$ & 3.3$^{+0.5}_{-0.5}$ & +0.449$^{+0.006}_{-0.000}$ & 1,2 \\[1mm]
19076 & 199 & 25680 & 39 Tau & G5V & 5874$^{+31}_{-31}$ & 4.50$^{+0.02}_{-0.03}$ & +0.04$^{+0.02}_{-0.02}$ & 0.007$^{+0.030}_{-0.030}$ & 0.00$^{+2.16}_{-0.00}$ & 1.071$^{+0.016}_{-0.025}$ & 1.01$^{+0.02}_{-0.03}$ & 0.252$^{+0.012}_{-0.009}$ & 3.8$^{+0.4}_{-0.4}$ & -0.112$^{+0.000}_{-0.000}$ & 1,2 \\[1mm]
19849 & 207 & 26965 & 40 Eri & K0.5V & 5151$^{+22}_{-22}$ & 4.55$^{+0.03}_{-0.02}$ & -0.08$^{+0.01}_{-0.01}$ & -0.380$^{+0.013}_{-0.013}$ & X & 0.808$^{+0.025}_{-0.008}$ & 0.82$^{+0.02}_{-0.03}$ & 0.263$^{+0.007}_{-0.013}$ & 0.5$^{+0.3}_{-0.3}$ & +0.264$^{+0.000}_{-0.000}$ & 1,2 \\[1mm]
19925 & - & 27063 & - & G0 & 5767$^{+12}_{-12}$ & 4.53$^{+X}_{-X}$ & +0.071*$^{+X}_{-X}$ & -0.096$^{+0.034}_{-0.034}$ & 3.0$^{+0.7}_{-2.3}$ & 1.02$^{+X}_{-X}$ & X & X & 2$^{+X}_{-X}$ & +0.201$^{+0.040}_{-0.040}$ & 23,25 \\[1mm]
20800 & 213 & 28005 & - & G0 & 5819$^{+44}_{-44}$ & 4.33$^{+0.03}_{-0.03}$ & +0.29$^{+0.03}_{-0.03}$ & 0.181$^{+0.052}_{-0.052}$ & 5.48$^{+1.52}_{-0.60}$ & 1.103$^{+0.092}_{-0.044}$ & 1.22$^{+0.05}_{-0.05}$ & 0.353$^{+0.020}_{-0.019}$ & 2.4$^{+0.5}_{-0.5}$ & +0.355$^{+0.004}_{-0.004}$ & 1,2 \\[1mm]
22319\sups{SG} & 227 & 30508 & - & G5 & 5206$^{+44}_{-44}$ & 3.68$^{+0.03}_{-0.03}$ & -0.21$^{+0.03}_{-0.03}$ & 0.726$^{+0.082}_{-0.082}$ & 3.52$^{+0.40}_{-0.44}$ & 1.326$^{+0.050}_{-0.044}$ & 2.77$^{+0.12}_{-0.11}$ & 1.341$^{+0.074}_{-0.078}$ & 2.6$^{+0.5}_{-0.5}$ & +0.332$^{+0.000}_{-0.004}$ & 1,2 \\[1mm]
22336 & 228 & 30562 & -	& F8V & 5937$^{+44}_{-44}$ & 4.16$^{+0.02}_{-0.02}$ & +0.19$^{+0.03}_{-0.03}$ & 0.450$^{+0.048}_{-0.048}$ & 4.00$^{+1.76}_{-0.28}$ & 1.277$^{+0.046}_{-0.130}$ & 1.58$^{+0.05}_{-0.06}$ & 0.390$^{+0.004}_{-0.024}$ & 4.3$^{+0.5}_{-0.5}$ & +0.366$^{+0.004}_{-0.000}$ & 1,2 \\[1mm]
22449 & 230 & 30652 & 1 Ori & F6V & 6424$^{+31}_{-31}$ & 4.26$^{+0.10}_{-0.02}$ & +0.00$^{+0.02}_{-0.02}$ & 0.433$^{+0.018}_{-0.018}$ & 2.84$^{+0.36}_{-2.44}$ & 1.236$^{+0.079}_{-0.016}$ & 1.32$^{+0.12}_{-0.03}$ & 0.223$^{+0.008}_{-0.005}$ & 16.8$^{+0.4}_{-0.4}$ & +0.161$^{+0.009}_{-0.000}$ & 1,2 \\[1mm]
22633\sups{SG} & 234 & 30825 & - & G5 & 5176$^{+44}_{-44}$ & 3.63$^{+0.04}_{-0.03}$ & -0.21$^{+0.03}_{-0.03}$ & 0.82$^{+0.12}_{-0.12}$ & 2.32$^{+0.52}_{-0.20}$ & 1.443$^{+0.082}_{-0.072}$ & 3.16$^{+0.15}_{-0.25}$ & 1.566$^{+0.107}_{-0.125}$ & 2.5$^{+0.5}_{-0.5}$ & +0.399$^{+0.000}_{-0.000}$ & 1,2 \\[1mm]
23311 & 242 & 32147 & - & K3V & 4827$^{+31}_{-31}$ & 4.61$^{+0.02}_{-0.03}$ & +0.30$^{+0.02}_{-0.02}$ & -0.551$^{+0.017}_{-0.017}$ & 0.00$^{+5.45}_{-0.00}$ & 0.838$^{+0.034}_{-0.033}$ & 0.78$^{+0.02}_{-0.03}$ & 0.256$^{+0.009}_{-0.009}$ & 1.7$^{+0.4}_{-0.4}$ & +0.253$^{+0.000}_{-0.000}$ & 1,2 \\[1mm]
24813 & 256 & 34411 & 15 Aur & G1.5IV-V & 5911$^{+22}_{-22}$ & 4.29$^{+0.03}_{-0.04}$ & +0.09$^{+0.01}_{-0.01}$ & 0.245$^{+0.025}_{-0.025}$ & 6.48$^{+1.32}_{-1.92}$ & 1.081$^{+0.054}_{-0.029}$ & 1.28$^{+0.04}_{-0.04}$ & 0.335$^{+0.027}_{-0.026}$ & 2.0$^{+0.3}_{-0.3}$ & +0.347$^{+0.534}_{-0.119}$ & 1,2 \\[1mm]
25278 & - & 35296 & V1119 Tau & F8V & 6167$^{+40}_{-40}$ & 4.26$^{+0.08}_{-0.08}$ & +0.05*$^{+0.04}_{-0.04}$ & 0.26$^{+X}_{-X}$ & X & 1.22$^{+X}_{-X}$ & 1.20$^{+X}_{-X}$ & X & X & -0.241$^{+0.000}_{-0.000}$ & 14,21 \\[1mm]
25486 & 261 & 35850 & HR 1817 & F8V(n)k: & 6496$^{+44}_{-44}$ & 4.37$^{+X}_{-0.01}$ & -0.07$^{+0.03}_{-0.03}$ & 0.231$^{+0.046}_{-0.046}$ & 0.4$^{+0.6}_{-0.2}$ & 1.268$^{+0.012}_{-0.014}$ & 1.22$^{+0.04}_{-0.02}$ & 0.190$^{+0.007}_{-0.005}$ & 54.7$^{+0.5}_{-0.5}$ & -0.856$^{+0.034}_{-0.000}$ & 1,2 \\[1mm]
27913 & 288 & 39587 & 54 Ori & G0VCH & 5882$^{+31}_{-31}$ & 4.44$^{+0.03}_{-0.03}$ & +0.00$^{+0.02}_{-0.02}$ & 0.041$^{+0.021}_{-0.021}$ & 4.32$^{+1.88}_{-2.04}$ & 1.028$^{+0.030}_{-0.028}$ & 1.05$^{+0.02}_{-0.03}$ & 0.272$^{+0.018}_{-0.018}$ & 9.8$^{+0.4}_{-0.4}$ & -0.221$^{+0.000}_{-0.000}$ & 1,2 \\[1mm]
29568 & - & 43162 & V352 Cma & G5V & 5633$^{+35}_{-35}$ & 4.48$^{+0.07}_{-0.07}$ & -0.01*$^{+0.04}_{-0.04}$ & X & X & 1.0$^{+X}_{-X}$ & X & X & 9.6$^{+X}_{-X}$ & -0.262$^{+0.060}_{-0.042}$ & 12,16 \\[1mm]
30476 & 309 & 45289 & - & G2V & 5737$^{+25}_{-25}$ & 4.29$^{+0.03}_{-0.03}$ & +0.02$^{+0.02}_{-0.02}$ & 0.154$^{+0.030}_{-0.030}$ & 9.88$^{+1.36}_{-1.36}$ & 0.993$^{+0.023}_{-0.020}$ & 1.22$^{+0.04}_{-0.04}$ & 0.367$^{+0.020}_{-0.022}$ & 1.6$^{+0.3}_{-0.3}$ & +0.351$^{+0.019}_{-0.019}$ & 1,2 \\[1mm]
31965 & - & 47309 & - & G0 & 5770$^{+12}_{-12}$ & 4.31$^{+X}_{-X}$ & +0.05*$^{+X}_{-X}$ & 0.136$^{+0.035}_{-0.035}$ & 7.4$^{+1.3}_{-4.4}$ & 1.00$^{+X}_{-X}$ & X & X & 3$^{+X}_{-X}$ & +0.351$^{+0.019}_{-0.019}$ & 23,25 \\[1mm]
32673 & - & 49178 & - & G0 & 5724$^{+12}_{-12}$ & 4.57$^{+X}_{-X}$ & +0.06*$^{+X}_{-X}$ & -0.030$^{+0.045}_{-0.045}$ & 3.2$^{+0.4}_{-1.8}$ & 1.00$^{+X}_{-X}$ & X & X & 2$^{+X}_{-X}$ & +0.264$^{+0.016}_{-0.017}$ & 23,25 \\[1mm]
32851 & - & 49933 & - & F2V & 6522$^{+38}_{-38}$ & 4.00$^{+0.06}_{-0.06}$ & -0.49$^{+0.02}_{-0.02}$ & 0.59$^{+X}_{-X}$ & X & 1.21$^{+X}_{-X}$ & X & X & X & +0.093$^{+0.000}_{-0.000}$ & 21 \\[1mm]
33277 & 330 & 50692 & 37 Gem & G0V & 5891$^{+44}_{-44}$ & 4.36$^{+0.03}_{-0.03}$ & -0.13$^{+0.03}_{-0.03}$ & 0.104$^{+0.032}_{-0.032}$ & 6.76$^{+1.44}_{-1.52}$ & 0.992$^{+0.026}_{-0.026}$ & 1.10$^{+0.04}_{-0.03}$ & 0.281$^{+0.020}_{-0.018}$ & 2.7$^{+0.5}_{-0.5}$ & +0.299$^{+0.004}_{-0.009}$ & 1,2 \\[1mm]
35185 & - & 56202 & - & G5 & 5793$^{+12}_{-12}$ & 4.19$^{+X}_{-X}$ & +0.00*$^{+X}_{-X}$ & -0.006$^{+0.052}_{-0.052}$ & 5.8$^{+0.2}_{-3.7}$ & 1.00$^{+X}_{-X}$ & X & X & 7$^{+X}_{-X}$ & -0.202$^{+0.019}_{-0.039}$ & 23,25 \\[1mm]
35265 & 346 & 56124 & - & G0 & 5848$^{+44}_{-44}$ & 4.45$^{+0.04}_{-0.04}$ & -0.02$^{+0.03}_{-0.03}$ & 0.027$^{+0.059}_{-0.059}$ & 3.36$^{+2.16}_{-2.36}$ & 1.029$^{+0.032}_{-0.028}$ & 1.01$^{+0.05}_{-0.03}$ & 0.260$^{+0.019}_{-0.015}$ & 1.5$^{+0.5}_{-0.5}$ & +0.280$^{+0.005}_{-0.000}$ & 1,2 \\[1mm]
36704 & 350 & 59747 & DX Lyn & G5V & 5177$^{+44}_{-44}$ & 4.61$^{+0.01}_{-0.03}$ & -0.03$^{+0.03}_{-0.03}$ & -0.501$^{+0.052}_{-0.052}$ & 0.00$^{+0.92}_{-0.00}$ & 0.846$^{+0.014}_{-0.018}$ & 0.78$^{+0.02}_{-0.02}$ & 0.232$^{+0.008}_{-0.006}$ & 2.6$^{+0.5}_{-0.5}$ & -0.372$^{+0.000}_{-0.000}$ & 1,2 \\[1mm]
38018 & - & 61994 & - & G6V & 5476$^{+42}_{-42}$ & X & -0.25*$^{+X}_{-X}$ & X & X & X & 1.157$^{+0.079}_{-0.079}$ & X & X & +0.103$^{+0.010}_{-0.000}$ & 10 \\[1mm]
38228 & 357 & 63433 & V377 Gem & G5IV & 5742$^{+44}_{-44}$ & 4.53$^{+0.02}_{-0.01}$ & +0.02$^{+0.03}_{-0.03}$ & -0.153$^{+0.040}_{-0.040}$ & 0.00$^{+0.84}_{-0.00}$ & 1.001$^{+0.012}_{-0.014}$ & 0.91$^{+0.02}_{-0.02}$ & 0.245$^{+0.008}_{-0.004}$ & 7.0$^{+0.5}_{-0.5}$ & -0.262$^{+0.000}_{-0.021}$ & 1,2 \\[1mm]
38647 & - & 64324 & - & G0 & 5714$^{+12}_{-12}$ & 4.43$^{+X}_{-X}$ & +0.01*$^{+X}_{-X}$ & -0.092$^{+0.033}_{-0.033}$ & 2.7$^{+2.7}_{-0.4}$ & 0.99$^{+X}_{-X}$ & X & X & 3$^{+X}_{-X}$ & +0.021$^{+0.013}_{-0.013}$ & 23,25 \\[1mm]
38747 & - & 64942 & - & G5V & 5804$^{+12}_{-12}$ & 4.42$^{+X}_{-X}$ & +0.07*$^{+X}_{-X}$ & -0.061$^{+0.047}_{-0.047}$ & 2.6$^{+2.3}_{-1.6}$ & 1.04$^{+X}_{-X}$ & X & X & 6$^{+X}_{-X}$ & -0.304$^{+0.063}_{-0.044}$ & 23,25 \\[1mm]
\hline
\end{tabular}
\end{center}
\end{table}
\end{landscape}	

\addtocounter {table} {-1}

\begin{landscape}
\begin{table}
\scriptsize
\begin{center}
\caption{continued.}
\begin{tabular}{lccccccccccccccc}
\hline
HIP & SPOCS & HD & Other & Sp & T\subs{eff} & Log(g) & Log(M/H) or & Log(Lum) & Age & Mass & Radius & Radius\subs{CZ} & \vsini & Log & Refs.\\
no. & no. & no. & names & Type & (K) & (cm s\sups{-2}) & Log(Fe/H)* & (\Lsolar)    & (Gyr) & (\Msolar) & (\Rsolar) & (\Rsolar) & (\kmsn) & (Rossby Number) & \\
\hline
41484 & 386 & 71148 & - & G5V & 5818$^{+44}_{-44}$ & 4.37$^{+0.03}_{-0.03}$ & -0.01$^{+0.03}_{-0.03}$ & 0.074$^{+0.038}_{-0.038}$ & 7.00$^{+1.56}_{-1.64}$ & 1.001$^{+0.022}_{-0.022}$ & 1.09$^{+0.04}_{-0.04}$ & 0.301$^{+0.017}_{-0.020}$ & 2.6$^{+0.5}_{-0.5}$ & +0.290$^{+0.009}_{-0.010}$ & 1,2 \\[1mm]
41526 & - & 71227 & - & G0V & 5801$^{+12}_{-12}$ & 4.27$^{+X}_{-X}$ & -0.02*$^{+X}_{-X}$ & 0.030$^{+0.039}_{-0.039}$ & 6.0$^{+0.5}_{-3.2}$ & 0.99$^{+X}_{-X}$ & X & X & 3$^{+X}_{-X}$ & +0.324$^{+0.008}_{-0.008}$ & 23,25 \\[1mm]
41844 & 387 & 71881 & - & G1V & 5822$^{+44}_{-44}$ & 4.28$^{+0.04}_{-0.03}$ & -0.02$^{+0.03}_{-0.03}$ & 0.170$^{+0.077}_{-0.077}$ & 8.72$^{+1.16}_{-1.16}$ & 1.005$^{+0.020}_{-0.020}$ & 1.21$^{+0.05}_{-0.05}$ & 0.344$^{+0.024}_{-0.022}$ & 2.1$^{+0.5}_{-0.5}$ & +0.355$^{+0.040}_{-0.031}$ & 1,2 \\[1mm]
42333 & 397 & 73350 & V401 Hya & G5V & 5802$^{+44}_{-44}$ & 4.49$^{+0.02}_{-0.04}$ & +0.04$^{+0.03}_{-0.03}$ & -0.024$^{+0.050}_{-0.050}$ & 1.88$^{+2.56}_{-1.88}$ & 1.038$^{+0.024}_{-0.034}$ & 0.98$^{+0.04}_{-0.02}$ & 0.259$^{+0.016}_{-0.011}$ & 4.0$^{+0.5}_{-0.5}$ & +0.058$^{+0.000}_{-0.000}$ & 1,2 \\[1mm]
42403 & 398 & 73344 & - & F8 & 6173$^{+44}_{-44}$ & 4.40$^{+0.02}_{-0.04}$ & +0.10$^{+0.03}_{-0.03}$ & 0.239$^{+0.067}_{-0.067}$ & 1.04$^{+1.24}_{-1.04}$ & 1.189$^{+0.026}_{-0.026}$ & 1.17$^{+0.04}_{-0.04}$ & 0.231$^{+0.015}_{-0.012}$ & 6.3$^{+0.5}_{-0.5}$ & +0.103$^{+0.010}_{-0.000}$ & 1,2 \\[1mm]
42438 & - & 72905 & 3 Uma & G1.5Vb & 5873$^{+34}_{-34}$ & 4.44$^{+0.05}_{-0.05}$ & -0.04*$^{+0.03}_{-0.03}$ & X & 2.1$^{+1.9}_{-1.9}$ & 1.00$^{+0.030}_{-0.030}$ & X & X & 10.0$^{+0.6}_{-0.6}$ & -0.419$^{+0.000}_{-0.000}$ & 8,11 \\[1mm]
43410 & 406 & 75332 & - & F7Vn & 6258$^{+44}_{-44}$ & 4.34$^{+0.03}_{-0.03}$ & +0.05$^{+0.03}_{-0.03}$ & 0.330$^{+0.050}_{-0.050}$ & 1.88$^{+0.72}_{-0.92}$ & 1.211$^{+0.026}_{-0.020}$ & 1.24$^{+0.05}_{-0.03}$ & 0.212$^{+0.018}_{-0.014}$ & 9.0$^{+0.5}_{-0.5}$ & -0.096$^{+0.000}_{-0.017}$ & 1,2 \\[1mm]
43557 & - & 75767 & - & G0 & 5805$^{+12}_{-12}$ & 4.42$^{+X}_{-X}$ & -0.06*$^{+X}_{-X}$ & 0.056$^{+0.025}_{-0.025}$ & 6.3$^{+1.6}_{-3.1}$ & 0.97$^{+X}_{-X}$ & X & X & 4$^{+X}_{-X}$ & +0.161$^{+0.017}_{-0.009}$ & 23,25 \\[1mm]
43726 & 410 & 76151 & - & G3V & 5790$^{+31}_{-31}$ & 4.49$^{+0.02}_{-0.03}$ & +0.07$^{+0.02}_{-0.02}$ & -0.012$^{+0.031}_{-0.031}$ & 1.32$^{+2.52}_{-1.32}$ & 1.056$^{+0.023}_{-0.036}$ & 1.00$^{+0.02}_{-0.03}$ & 0.262$^{+0.014}_{-0.012}$ & 1.2$^{+0.4}_{-0.4}$ & +0.169$^{+0.008}_{-0.000}$ & 1,2 \\[1mm]
44897 & 416 & 78366 & - & F9V & 6014$^{+31}_{-31}$ & 4.46$^{+0.02}_{-0.02}$ & +0.03$^{+0.02}_{-0.02}$ & 0.096$^{+0.035}_{-0.035}$ & 0.00$^{+0.68}_{-0.00}$ & 1.133$^{+0.014}_{-0.016}$ & 1.06$^{+0.03}_{-0.02}$ & 0.243$^{+0.009}_{-0.006}$ & 3.9$^{+0.4}_{-0.4}$ & -0.049$^{+0.000}_{-0.000}$ & 1,2 \\[1mm]
44997 & - & 78660 & - & G5 & 5696$^{+12}_{-12}$ & 4.54$^{+X}_{-X}$ & +0.044*$^{+X}_{-X}$ & -0.009$^{+0.048}_{-0.048}$ & 3.1$^{+3.2}_{-0.4}$ & 0.97$^{+X}_{-X}$ & X & X & 2$^{+X}_{-X}$ & +0.312$^{+0.008}_{-0.009}$ & 23,25 \\[1mm]
46066 & - & 80533 & - & G0 & 5709$^{+50}_{-50}$ & 4.49$^{+0.07}_{-0.07}$ & -0.07*$^{+0.04}_{-0.04}$ & X & X & X & X & X & X & +0.222$^{+0.031}_{-0.036}$ & 13 \\[1mm]
46580 & 424 & 82106 & - & K3V & 4868$^{+44}_{-44}$ & 4.63$^{+0.01}_{-0.02}$ & -0.04$^{+0.03}_{-0.03}$ & -0.635$^{+0.028}_{-0.028}$ & 0.00$^{+0.60}_{-0.00}$ & 0.786$^{+0.008}_{-0.012}$ & 0.73$^{+0.02}_{-0.02}$ & 0.223$^{+0.005}_{-0.007}$ & 3.1$^{+0.5}_{-0.5}$ & -0.129$^{+0.000}_{-0.017}$ & 1,2 \\[1mm]
46903 & - & 82460 & - & G0 & 5746$^{+12}_{-12}$ & 4.40$^{+X}_{-X}$ & -0.03*$^{+X}_{-X}$ & -0.095$^{+0.046}_{-0.046}$ & 3.4$^{+0.3}_{-2.5}$ & 1.02$^{+X}_{-X}$ & X & X & X & +0.247$^{+0.017}_{-0.012}$ & 23,25 \\[1mm]
49081 & 436 & 86728 & 20 Lmi & G3Va & 5700$^{+31}_{-31}$ & 4.40$^{+0.05}_{-0.04}$ & +0.11$^{+0.02}_{-0.02}$ & -0.005$^{+0.095}_{-0.095}$ & 9.52$^{+2.36}_{-3.12}$ & 0.960$^{+0.040}_{-0.027}$ & 1.06$^{+0.04}_{-0.04}$ & 0.326$^{+0.020}_{-0.024}$ & 3.4$^{+0.4}_{-0.4}$ & +0.340$^{+0.004}_{-0.000}$ & 1,2 \\[1mm]
49350 & 437 & 87359 & - & G5 & 5676$^{+44}_{-44}$ & 4.46$^{+0.04}_{-0.04}$ & +0.04$^{+0.03}_{-0.03}$ & -0.057$^{+0.060}_{-0.060}$ & 5.44$^{+2.76}_{-3.16}$ & 0.979$^{+0.034}_{-0.030}$ & 0.97$^{+0.04}_{-0.03}$ & 0.277$^{+0.020}_{-0.019}$ & 1.9$^{+0.5}_{-0.5}$ & +0.259$^{+0.011}_{-0.011}$ & 1,2 \\[1mm]
49580 & - & 87680 & - & G2V & 5782$^{+12}_{-12}$ & 4.41$^{+X}_{-X}$ & +0.02*$^{+X}_{-X}$ & -0.077$^{+0.034}_{-0.034}$ & 3.8$^{+0.6}_{-2.5}$ & 1.04$^{+X}_{-X}$ & X & X & 1$^{+X}_{-X}$ & +0.270$^{+0.015}_{-0.016}$ & 23,25 \\[1mm]
49728 & - & 88084 & - & G2V & 5744$^{+12}_{-12}$ & 4.40$^{+X}_{-X}$ & -0.07*$^{+X}_{-X}$ & 0.009$^{+0.025}_{-0.025}$ & 4.0$^{+0.4}_{-1.4}$ & 0.94$^{+X}_{-X}$ & X & X & 3$^{+X}_{-X}$ & +0.328$^{+0.015}_{-0.016}$ & 23,25 \\[1mm]
49756 & 441 & 88072 & - & G0 & 5751$^{+44}_{-44}$ & 4.38$^{+0.04}_{-0.04}$ & -0.01$^{+0.03}_{-0.03}$ & 0.048$^{+0.073}_{-0.073}$ & 8.28$^{+1.68}_{-1.96}$ & 0.975$^{+0.024}_{-0.020}$ & 1.07$^{+0.05}_{-0.05}$ & 0.308$^{+0.021}_{-0.022}$ & 1.5$^{+0.5}_{-0.5}$ & +0.299$^{+0.009}_{-0.009}$ & 1,2 \\[1mm]
49908 & - & 88230 & - & K8V & 4085$^{+14}_{-14}$ & 4.61$^{+0.05}_{-0.05}$ & -0.29*$^{+0.05}_{-0.05}$ & -0.989$^{+0.004}_{-0.004}$ & X & 0.60$^{+X}_{-X}$ & 0.6398$^{+0.0046}_{-0.0046}$ & X & 1.9$^{+0.7}_{-0.7}$ & +0.169$^{+0.000}_{-0.000}$ & 3,4,5,6 \\[1mm]
50316 & 446 & 88986 & 24 Lmi & G0V & 5838$^{+44}_{-44}$ & 4.16$^{+0.03}_{-0.03}$ & +0.03$^{+0.03}_{-0.03}$ & 0.348$^{+0.053}_{-0.053}$ & 8.00$^{+0.60}_{-0.60}$ & 1.070$^{+0.020}_{-0.016}$ & 1.46$^{+0.06}_{-0.06}$ & 0.411$^{+0.028}_{-0.025}$ & 2.6$^{+0.5}_{-0.5}$ & +0.340$^{+0.011}_{-0.012}$ & 1,2 \\[1mm]
50505 & 449 & 89269 & - & G5 & 5586$^{+44}_{-44}$ & 4.42$^{+0.03}_{-0.03}$ & -0.18$^{+0.03}_{-0.03}$ & -0.091$^{+0.036}_{-0.036}$ & 12.08$^{+1.92}_{-2.76}$ & 0.874$^{+0.026}_{-0.020}$ & 0.97$^{+0.03}_{-0.03}$ & 0.295$^{+0.012}_{-0.019}$ & 0.8$^{+0.5}_{-0.5}$ & +0.275$^{+0.000}_{-0.005}$ & 1,2 \\[1mm]
53721 & 472 & 95128 & 47 Uma & G1V & 5882$^{+16}_{-16}$ & 4.31$^{+0.03}_{-0.04}$ & +0.02$^{+0.01}_{-0.01}$ & 0.206$^{+0.021}_{-0.021}$ & 6.48$^{+1.44}_{-1.04}$ & 1.063$^{+0.022}_{-0.029}$ & 1.24$^{+0.04}_{-0.04}$ & 0.325$^{+0.028}_{-0.026}$ & 2.8$^{+0.2}_{-0.2}$ & +0.343$^{+0.004}_{-0.004}$ & 1,2 \\[1mm]
54952 & - & 97584 & - & K5 & 4670$^{+X}_{-X}$ & 4.05$^{+X}_{-X}$ & -0.04*$^{+0.06}_{-0.06}$ & X & X & X & X & X & X & -0.034$^{+0.000}_{-0.015}$ & 26 \\[1mm]
55459 & 488 & 98618 & - & G5V & 5812$^{+44}_{-44}$ & 4.42$^{+0.04}_{-0.04}$ & +0.00$^{+0.03}_{-0.03}$ & 0.042$^{+0.067}_{-0.067}$ & 5.64$^{+1.92}_{-2.20}$ & 1.009$^{+0.028}_{-0.024}$ & 1.04$^{+0.05}_{-0.04}$ & 0.281$^{+0.022}_{-0.019}$ & 2.1$^{+0.5}_{-0.5}$ & +0.359$^{+0.012}_{-0.008}$ & 1,2 \\[1mm]
56242 & 494 & 100180 & 88 Leo & G0V & 5989$^{+44}_{-44}$ & 4.39$^{+0.03}_{-0.04}$ & -0.02$^{+0.03}_{-0.03}$ & 0.132$^{+0.052}_{-0.052}$ & 3.80$^{+1.40}_{-1.56}$ & 1.069$^{+0.026}_{-0.024}$ & 1.10$^{+0.04}_{-0.04}$ & 0.261$^{+0.018}_{-0.018}$ & 3.3$^{+0.5}_{-0.5}$ & +0.290$^{+0.000}_{-0.000}$ & 1,2 \\[1mm]
56948 & - & 101364 & - & G5 & 5781$^{+12}_{-12}$ & 4.41$^{+X}_{-X}$ & +0.02*$^{+X}_{-X}$ & 0.099$^{+0.046}_{-0.046}$ & X & X & X & X & 1$^{+X}_{-X}$ & +0.259$^{+0.053}_{-0.058}$ & 7,8,25 \\[1mm]
56997 & 500 & 101501 & 61 Uma & G8V & 5488$^{+22}_{-22}$ & 4.48$^{+0.10}_{-0.02}$ & -0.03$^{+0.01}_{-0.01}$ & -0.217$^{+0.017}_{-0.017}$ & X & 0.850$^{+0.124}_{-0.012}$ & 0.90$^{+0.02}_{-0.04}$ & 0.287$^{+0.006}_{-0.042}$ & 2.4$^{+0.3}_{-0.3}$ & -0.202$^{+0.000}_{-0.000}$ & 1,2 \\[1mm]
57939 & 511 & 103095 & - & G8Vp & 4950$^{+44}_{-44}$ & 4.63$^{+0.03}_{-0.01}$ & -1.16$^{+0.03}_{-0.03}$ & -0.628$^{+0.017}_{-0.017}$ & X & 0.661$^{+0.028}_{-0.006}$ & 0.66$^{+0.02}_{-0.02}$ & 0.170$^{+0.006}_{-0.006}$ & 0.5$^{+0.5}_{-0.5}$ & +0.264$^{+0.011}_{-0.005}$ & 1,2 \\[1mm]
58708\sups{SG} & 516 & 104556 & - & G8Vw & 5051$^{+44}_{-44}$ & 3.76$^{+0.01}_{-0.01}$ & -0.27$^{+0.03}_{-0.03}$ & 0.828$^{+0.094}_{-0.094}$ & 2.72$^{+0.64}_{-0.24}$ & 2.55$^{+0.420}_{-0.420}$ & 3.12$^{+0.05}_{-0.05}$ & 1.572$^{+0.119}_{-0.031}$ & 2.6$^{+0.5}_{-0.5}$ & +0.427$^{+0.005}_{-0.005}$ & 1,2 \\[1mm]
60098 & 529 & 107213 & 9 Com & F8Vs & 6274$^{+44}_{-44}$ & 4.03$^{+0.03}_{-0.05}$ & +0.24$^{+0.03}_{-0.03}$ & 0.737$^{+0.077}_{-0.077}$ & 2.04$^{+0.12}_{-0.16}$ & 1.508$^{+0.068}_{-0.048}$ & 2.00$^{+0.09}_{-0.09}$ & 0.320$^{+0.028}_{-0.010}$ & 8.4$^{+0.5}_{-0.5}$ & +0.399$^{+0.000}_{-0.004}$ & 1,2 \\[1mm]
60353 & 530 & 107705 & 17 Vir & F8V & 6130$^{+44}_{-44}$ & 4.34$^{+0.03}_{-0.04}$ & +0.08$^{+0.03}_{-0.03}$ & 0.272$^{+0.079}_{-0.079}$ & 3.28$^{+0.68}_{-0.92}$ & 1.163$^{+0.026}_{-0.022}$ & 1.23$^{+0.06}_{-0.05}$ & 0.257$^{+0.019}_{-0.019}$ & 6.3$^{+0.5}_{-0.5}$ & +0.303$^{+0.000}_{-0.000}$ & 1,2 \\[1mm]
61901 & - & 110315 & - & K4.5V & 4488$^{+11}_{-11}$ & 4.63$^{+0.01}_{-0.02}$ & -0.15*$^{+0.05}_{-0.05}$ & X & 0.00$^{+0.72}_{-0.00}$ & 0.730$^{+0.010}_{-0.014}$ & 0.69$^{+0.02}_{-0.02}$ & 0.215$^{+0.006}_{-0.006}$ & X & +0.299$^{+0.004}_{-0.005}$ & 2,27 \\[1mm]
62523 & 542 & 111395 & LW Com & G5V & 5654$^{+44}_{-44}$ & 4.51$^{+0.03}_{-0.03}$ & +0.06$^{+0.03}_{-0.03}$ & -0.112$^{+0.035}_{-0.035}$ & 1.40$^{+3.24}_{-1.40}$ & 1.004$^{+0.020}_{-0.036}$ & 0.93$^{+0.03}_{-0.02}$ & 0.259$^{+0.015}_{-0.010}$ & 2.8$^{+0.5}_{-0.5}$ & +0.008$^{+0.013}_{-0.000}$ & 1,2 \\[1mm]
64797 & - & 115404A & - & K1V & 4932$^{+X}_{-X}$ & 4.63$^{+X}_{-X}$ & -0.18*$^{+0.05}_{-0.05}$ & -0.114$^{+X}_{-X}$ & X & 0.860$^{+X}_{-X}$ & 0.77$^{+X}_{-X}$ & X & X & -0.165$^{+0.000}_{-0.000}$ & 14,28\\[1mm]
65347 & - & 116514 & - & G4V & 6374$^{+X}_{-X}$ & X & +0.03*$^{+X}_{-X}$ & 0.304$^{+0.086}_{-0.086}$ & X & X & X & X & X & -0.262$^{+0.040}_{-0.021}$ & 7,9,15 \\[1mm]
66147 & 571 & 117936 & - & K3V & 4834$^{+44}_{-44}$ & 4.62$^{+0.01}_{-0.02}$ & +0.07$^{+0.03}_{-0.03}$ & -0.612$^{+0.040}_{-0.040}$ & 0.00$^{+0.80}_{-0.00}$ & 0.805$^{+0.010}_{-0.010}$ & 0.74$^{+0.02}_{-0.02}$ & 0.233$^{+0.006}_{-0.005}$ & 1.4$^{+0.5}_{-0.5}$ & +0.008$^{+0.000}_{-0.014}$ & 1,2 \\[1mm]
66774\sups{SG} & - & 119347 & - & K0IV & 5343$^{+X}_{-X}$ & X & X & 0.450$^{+0.065}_{-0.065}$ & X & 0.95$^{+X}_{-X}$ & X & X & X & +0.454$^{+0.032}_{-0.022}$ & 7,9 \\[1mm]
67275 & 577 & 120126 & Tau Boo & F6IV & 6387$^{+25}_{-25}$ & 4.27$^{+0.04}_{-0.03}$ & +0.25$^{+0.02}_{-0.02}$ & 0.481$^{+0.024}_{-0.024}$ & 1.64$^{+0.44}_{-0.52}$ & 1.341$^{+0.054}_{-0.039}$ & 1.46$^{+0.05}_{-0.05}$ & 0.230$^{+0.010}_{-0.005}$ & 15.0$^{+0.3}_{-0.3}$ & +0.259$^{+0.000}_{-0.000}$ & 1,2 \\[1mm]
\hline
\end{tabular}
\end{center}
\end{table}
\end{landscape}

\addtocounter {table} {-1}

\begin{landscape}
\begin{table}
\scriptsize
\begin{center}
\caption{continued.}
\begin{tabular}{lccccccccccccccc}
\hline
HIP & SPOCS & HD & Other & Sp. & T\subs{eff} & Log(g) & Log(M/H) or & Log(Lum) & Age & Mass & Radius & Radius\subs{CZ} & \vsini & Log & Refs.\\
no. & no. & no. & names & Type & (K) & (cm s\sups{-2}) & Log(Fe/H)* & (\Lsolar)    & (Gyr) & (\Msolar) & (\Rsolar) & (\Rsolar) & (\kmsn) & (Rossby Number) & \\
\hline
67422 & 579 & 120476A & - & K5V & 4729$^{+44}_{-44}$ & 4.38$^{+0.01}_{-0.01}$ & -0.35$^{+0.03}_{-0.03}$ & -0.482$^{+0.038}_{-0.038}$ & X & 0.870$^{+0.010}_{-0.018}$ & 1.01$^{+0.03}_{-0.03}$ & 0.393$^{+0.005}_{-0.007}$ & 0.3$^{+0.5}_{-0.3}$ & +0.021$^{+0.102}_{-0.101}$ & 1,2 \\[1mm]
68184 & 585 & 122064 & - & K3V & 4757$^{+44}_{-44}$ & 4.56$^{+0.02}_{-0.01}$ & +0.10$^{+0.03}_{-0.03}$ & -0.527$^{+0.015}_{-0.015}$ & X & 0.800$^{+0.026}_{-0.012}$ & 0.78$^{+0.02}_{-0.02}$ & 0.263$^{+0.003}_{-0.009}$ & 1.3$^{+0.5}_{-0.5}$ & +0.332$^{+0.000}_{-0.004}$ & 1,2 \\[1mm]
71181 & 604 & 128165 & - & K3V & 4809$^{+44}_{-44}$ & 4.60$^{+0.02}_{-0.03}$ & -0.09$^{+0.03}_{-0.03}$ & -0.591$^{+0.021}_{-0.021}$ & 1.96$^{+7.96}_{-1.96}$ & 0.780$^{+0.016}_{-0.032}$ & 0.74$^{+0.02}_{-0.02}$ & 0.232$^{+0.009}_{-0.009}$ & 1.8$^{+0.5}_{-0.5}$ & +0.169$^{+0.000}_{-0.000}$ & 1,2 \\[1mm]
71631 & 608 & 129333 & EK Dra & G1.5V & 5845$^{+44}_{-44}$ & 4.50$^{+0.02}_{-0.02}$ & +0.03$^{+0.03}_{-0.03}$ & -0.057$^{+0.043}_{-0.043}$ & 0.00$^{+1.44}_{-0.00}$ & 1.044$^{+0.014}_{-0.020}$ & 0.97$^{+0.02}_{-0.03}$ & 0.249$^{+0.011}_{-0.006}$ & 16.8$^{+0.5}_{-0.5}$ & -1.111$^{+0.000}_{-0.000}$ & 1,2 \\[1mm]
72848 & 625 & 131511 & DE Boo & K2V & 5335$^{+31}_{-31}$ & 4.58$^{+X}_{-0.03}$ & +0.11$^{+0.02}_{-0.02}$ & -0.314$^{+0.021}_{-0.021}$ & 0.00$^{+2.08}_{-0.00}$ & 0.926$^{+0.017}_{-0.032}$ & 0.84$^{+0.02}_{-0.02}$ & 0.253$^{+0.012}_{-0.007}$ & 4.5$^{+0.4}_{-0.4}$ & -0.304$^{+0.000}_{-0.000}$ & 1,2 \\[1mm]
74432 & 641 & 135101 & - & G5V & 5705$^{+44}_{-44}$ & 4.30$^{+0.04}_{-0.04}$ & +0.09$^{+0.03}_{-0.03}$ & 0.16$^{+0.11}_{-0.11}$ & 10.16$^{+1.32}_{-1.36}$ & 0.992$^{+0.026}_{-0.024}$ & 1.19$^{+0.06}_{-0.06}$ & 0.370$^{+0.025}_{-0.024}$ & 1.8$^{+0.5}_{-0.5}$ & +0.362$^{+0.004}_{-0.008}$ & 1,2 \\[1mm]
76114 & 660 & 138573 & - & G5IV-V & 5722$^{+44}_{-44}$ & 4.39$^{+0.04}_{-0.04}$ & -0.05$^{+0.03}_{-0.03}$ & 0.025$^{+0.054}_{-0.054}$ & 8.44$^{+1.88}_{-2.08}$ & 0.957$^{+0.026}_{-0.022}$ & 1.05$^{+0.04}_{-0.05}$ & 0.302$^{+0.020}_{-0.021}$ & 1.0$^{+0.5}_{-0.5}$ & +0.320$^{+0.012}_{-0.013}$ & 1,2 \\[1mm]
79578 & 695 & 145825 & - & G3V & 5803$^{+25}_{-25}$ & 4.49$^{+0.03}_{-0.04}$ & +0.03$^{+0.02}_{-0.02}$ & -0.014$^{+0.043}_{-0.043}$ & 1.92$^{+2.76}_{-1.92}$ & 1.042$^{+0.027}_{-0.042}$ & 1.00$^{+0.03}_{-0.03}$ & 0.260$^{+0.017}_{-0.012}$ & 1.4$^{+0.3}_{-0.3}$ & +0.186$^{+0.008}_{-0.008}$ & 1,2 \\[1mm]
79672 & 698 & 146233 & 18 Sco & G2Va & 5791$^{+44}_{-44}$ & 4.42$^{+0.03}_{-0.03}$ & +0.03$^{+0.03}_{-0.03}$ & 0.025$^{+0.027}_{-0.027}$ & 5.84$^{+1.88}_{-1.96}$ & 1.005$^{+0.028}_{-0.024}$ & 1.04$^{+0.03}_{-0.03}$ & 0.289$^{+0.017}_{-0.018}$ & 2.6$^{+0.5}_{-0.5}$ & +0.299$^{+0.004}_{-0.000}$ & 1,2 \\[1mm]
81300 & 711 & 149661 & V2133 Oph & K2V & 5277$^{+31}_{-31}$ & 4.59$^{+X}_{-0.03}$ & +0.05$^{+0.02}_{-0.02}$ & -0.359$^{+0.019}_{-0.019}$ & 0.00$^{+4.16}_{-0.00}$ & 0.892$^{+0.028}_{-0.051}$ & 0.82$^{+0.02}_{-0.02}$ & 0.250$^{+0.012}_{-0.008}$ & 2.2$^{+0.4}_{-0.4}$ & -0.020$^{+0.014}_{-0.000}$ & 1,2 \\[1mm]
82588 & 729 & 152391 & V2292 Oph & G8.5Vk: & 5479$^{+44}_{-44}$ & 4.56$^{+0.02}_{-0.03}$ & -0.05$^{+0.03}_{-0.03}$ & -0.255$^{+0.031}_{-0.031}$ & 0.72$^{+3.92}_{-0.72}$ & 0.927$^{+0.016}_{-0.034}$ & 0.85$^{+0.03}_{-0.02}$ & 0.245$^{+0.012}_{-0.008}$ & 3.8$^{+0.5}_{-0.5}$ & -0.202$^{+0.106}_{-0.102}$ & 1,2 \\[1mm]
86193 & 759 & 159909 & - & G5 & 5723$^{+44}_{-44}$ & 4.30$^{+0.03}_{-0.04}$ & +0.06$^{+0.03}_{-0.03}$ & 0.149$^{+0.064}_{-0.064}$ & 9.64$^{+1.36}_{-1.32}$ & 0.994$^{+0.024}_{-0.022}$ & 1.19$^{+0.05}_{-0.05}$ & 0.359$^{+0.023}_{-0.020}$ & 1.9$^{+0.5}_{-0.5}$ & +0.285$^{+0.014}_{-0.010}$ & 1,2 \\[1mm]
86400 & - & 160346 & - & K3V & 4937$^{+86}_{-86}$ & 4.44$^{+X}_{-X}$ & -0.02*$^{+X}_{-X}$ & -0.46$^{+X}_{-X}$ & X & 0.86$^{+X}_{-X}$ & 0.77$^{+X}_{-X}$ & X & X & +0.208$^{+0.007}_{-0.000}$ & 14,17 \\[1mm]
86974 & 764 & 161797 & 86 Her & G5IV & 5597$^{+22}_{-22}$ & 4.04$^{+0.03}_{-0.04}$ & +0.24$^{+0.01}_{-0.01}$ & 0.415$^{+0.014}_{-0.014}$ & 8.08$^{+0.32}_{-0.40}$ & 1.091$^{+0.109}_{-0.004}$ & 1.73$^{+0.06}_{-0.04}$ & 0.607$^{+0.048}_{-0.025}$ & 2.7$^{+0.3}_{-0.3}$ & +0.403$^{+0.014}_{-0.013}$ & 1,2 \\[1mm]
88194 & 777 & 164595 & - & G2V & 5745$^{+44}_{-44}$ & 4.41$^{+0.03}_{-0.04}$ & -0.06$^{+0.03}_{-0.03}$ & 0.026$^{+0.043}_{-0.043}$ & 7.12$^{+1.96}_{-2.08}$ & 0.971$^{+0.028}_{-0.024}$ & 1.04$^{+0.04}_{-0.04}$ & 0.289$^{+0.019}_{-0.019}$ & 0.0$^{+0.5}_{-0.0}$ & +0.320$^{+0.016}_{-0.013}$ & 1,2 \\[1mm]
88945 & 783 & 166435 & - & G1IV & 5843$^{+44}_{-44}$ & 4.47$^{+0.03}_{-0.03}$ & +0.01$^{+0.03}_{-0.03}$ & -0.008$^{+0.036}_{-0.036}$ & 2.04$^{+2.32}_{-2.04}$ & 1.039$^{+0.026}_{-0.032}$ & 0.99$^{+0.04}_{-0.03}$ & 0.256$^{+0.016}_{-0.012}$ & 7.9$^{+0.5}_{-0.5}$ & -0.419$^{+0.000}_{-0.000}$ & 1,2 \\[1mm]
88972 & 784 & 166620 & - & K2V & 5000$^{+18}_{-18}$ & 4.56$^{+0.01}_{-0.01}$ & -0.05$^{+0.01}_{-0.01}$ & -0.454$^{+0.015}_{-0.015}$ & X & 0.791$^{+0.014}_{-0.008}$ & 0.79$^{+0.02}_{-0.01}$ & 0.259$^{+0.004}_{-0.007}$ & 2.1$^{+0.2}_{-0.2}$ & +0.324$^{+0.000}_{-0.004}$ & 1,2 \\[1mm]
90729\sups{SG} & 803 & 170829 & - & G8IV & 5398$^{+44}_{-44}$ & 3.91$^{+0.02}_{-0.03}$ & +0.08$^{+0.03}_{-0.03}$ & 0.465$^{+0.049}_{-0.049}$ & 7.60$^{+0.44}_{-0.68}$ & 1.133$^{+0.024}_{-0.024}$ & 1.98$^{+0.07}_{-0.07}$ & 0.836$^{+0.039}_{-0.042}$ & 2.7$^{+0.5}_{-0.5}$ & +0.386$^{+0.008}_{-0.008}$ & 1,2 \\[1mm]
91043 & - & 171488 & V889 Her & G2V & 5830$^{+30}_{-30}$ & 4.5$^{+X}_{-X}$ & -0.5*$^{+X}_{-X}$ & 0.12$^{+0.03}_{-0.03}$ & 0.03$^{+0.02}_{-0.00}$ & 1.06$^{+0.02}_{-0.02}$ & 1.09$^{+0.05}_{-0.05}$ & X & 39.0$^{+0.5}_{-0.5}$ & -0.961$^{+0.036}_{-0.000}$ & 22 \\[1mm]
92984 & 821 & 175726 & - & G5 & 5998$^{+44}_{-44}$ & 4.43$^{+0.03}_{-0.04}$ & -0.10$^{+0.03}_{-0.03}$ & 0.096$^{+0.046}_{-0.046}$ & 2.72$^{+1.56}_{-1.84}$ & 1.058$^{+0.030}_{-0.030}$ & 1.06$^{+0.04}_{-0.04}$ & 0.239$^{+0.018}_{-0.014}$ & 12.3$^{+0.5}_{-0.5}$ & -0.202$^{+0.000}_{-0.000}$ & 1,2 \\[1mm]
95253 & - & 182101 & - & F6V & 6344$^{+X}_{-X}$ & 4.22$^{+X}_{-X}$ & -0.29$^{+X}_{-X}$ & 0.489$^{+0.017}_{-0.017}$ & X & 1.30$^{+X}_{-X}$ & X & X & 13$^{+X}_{-X}$ & +0.124$^{+0.010}_{-0.000}$ & 7,20 \\[1mm]
95962 & 844 & 183658 & - & G0 & 5798$^{+44}_{-44}$ & 4.38$^{+0.04}_{-0.04}$ & +0.04$^{+0.03}_{-0.03}$ & 0.088$^{+0.069}_{-0.069}$ & 7.00$^{+1.52}_{-1.76}$ & 1.010$^{+0.024}_{-0.020}$ & 1.09$^{+0.05}_{-0.04}$ & 0.305$^{+0.021}_{-0.022}$ & 1.2$^{+0.5}_{-0.5}$ & +0.336$^{+0.011}_{-0.012}$ & 1,2 \\[1mm]
96085 & 845 & 183870 & - & K2V & 5067$^{+44}_{-44}$ & 4.60$^{+0.02}_{-0.02}$ & -0.05$^{+0.03}_{-0.03}$ & -0.501$^{+0.034}_{-0.034}$ & 0.00$^{+1.36}_{-0.00}$ & 0.831$^{+0.012}_{-0.018}$ & 0.77$^{+0.02}_{-0.02}$ & 0.231$^{+0.009}_{-0.006}$ & 3.0$^{+0.5}_{-0.5}$ & -0.165$^{+0.018}_{-0.000}$ &1,2 \\[1mm]
96100 & 846 & 185144 & sig Dra & G9V & 5246$^{+18}_{-18}$ & 4.54$^{+0.04}_{-0.01}$ & -0.16$^{+0.01}_{-0.01}$ & -0.374$^{+0.011}_{-0.011}$ & 9.2$^{+3.6}_{-4.7}$ & 0.803$^{+0.045}_{-0.014}$ & 0.81$^{+0.02}_{-0.03}$ & 0.255$^{+0.007}_{-0.021}$ & 1.4$^{+0.2}_{-0.2}$ & +0.253$^{+0.006}_{-0.000}$ & 1,2 \\[1mm]
96895 & 854 & 186408 & 16 Cyg A & G1.5Vb & 5781$^{+17}_{-17}$ & 4.28$^{+0.04}_{-0.05}$ & +0.08$^{+0.01}_{-0.01}$ & 0.200$^{+0.024}_{-0.024}$ & 8.36$^{+2.92}_{-1.92}$ & 1.022$^{+0.047}_{-0.035}$ & 1.26$^{+0.05}_{-0.04}$ & 0.365$^{+0.029}_{-0.028}$ & 2.8$^{+0.2}_{-0.2}$ & +0.332$^{+0.000}_{-0.004}$ & 1,2 \\[1mm]
96901 & 855 & 186427 & 16 Cyg B & G3V & 5674$^{+17}_{-17}$ & 4.30$^{+0.04}_{-0.02}$ & +0.02$^{+0.01}_{-0.01}$ & 0.095$^{+0.024}_{-0.024}$ & 11.80$^{+2.20}_{-2.00}$ & 0.956$^{+0.026}_{-0.025}$ & 1.17$^{+0.04}_{-0.03}$ & 0.371$^{+0.023}_{-0.024}$ & 2.2$^{+0.2}_{-0.2}$ & +0.340$^{+0.000}_{-0.000}$ & 1,2 \\[1mm]
98921 & 873 & 190771 & - & G5IV & 5834$^{+44}_{-44}$ & 4.47$^{+0.03}_{-0.03}$ & +0.14$^{+0.03}_{-0.03}$ & 0.001$^{+0.023}_{-0.023}$ & 1.76$^{+2.32}_{-1.76}$ & 1.065$^{+0.030}_{-0.040}$ & 1.01$^{+0.03}_{-0.03}$ & 0.268$^{+0.016}_{-0.012}$ & 4.3$^{+0.5}_{-0.5}$ & -0.165$^{+0.000}_{-0.000}$ & 1,2 \\[1mm]
100511 & - & 194012 & - & F8V & 6223$^{+54}_{-54}$ & X & -0.19*$^{+X}_{-X}$ & 0.257$^{+0.009}_{-0.009}$ & 2.8$^{+1.3}_{-2.2}$ & 1.20$^{+X}_{-X}$ & 1.20$^{+0.03}_{-0.03}$ & X & 6$^{+X}_{-X}$ & +0.178$^{+0.000}_{-0.008}$ & 7,8,10 \\[1mm]
100970 & 894 & 195019 & - & G3IV-V & 5788$^{+44}_{-44}$ & 4.18$^{+0.03}_{-0.04}$ & +0.00$^{+0.03}_{-0.03}$ & 0.286$^{+0.069}_{-0.069}$ & 9.32$^{+0.76}_{-0.72}$ & 1.025$^{+0.020}_{-0.018}$ & 1.38$^{+0.06}_{-0.05}$ & 0.413$^{+0.026}_{-0.028}$ & 2.5$^{+0.5}_{-0.5}$ & +0.343$^{+0.011}_{-0.012}$ & 1,2 \\[1mm]
101875 & 899 & 196850 & - & G0 & 5790$^{+44}_{-44}$ & 4.34$^{+0.03}_{-0.03}$ & -0.11$^{+0.03}_{-0.03}$ & 0.086$^{+0.035}_{-0.035}$ & 8.96$^{+1.40}_{-1.52}$ & 0.964$^{+0.022}_{-0.022}$ & 1.11$^{+0.04}_{-0.04}$ & 0.311$^{+0.020}_{-0.018}$ & 2.0$^{+0.5}_{-0.5}$ & +0.312$^{+0.004}_{-0.004}$ & 1,2 \\[1mm]
104214 & - & 201091 & 61 Cyg A & K5V & 4525$^{+140}_{-140}$ & 4.70$^{+X}_{-0.02}$ & -0.37$^{+0.19}_{-0.19}$ & X & 0.00$^{+0.44}_{-0.00}$ & 0.660$^{+0.010}_{-0.002}$ & 0.62$^{+0.01}_{-0.02}$ & 0.181$^{+0.006}_{-0.001}$ & 1.1$^{+X}_{-X}$ & +0.034$^{+0.000}_{-0.000}$ & 2,18 \\[1mm]
107350 & 942 & 206860 & HN Peg & G0V & 5974$^{+25}_{-25}$ & 4.48$^{+0.01}_{-0.03}$ & -0.01$^{+0.02}_{-0.02}$ & 0.062$^{+0.033}_{-0.033}$ & 0.00$^{+0.88}_{-0.00}$ & 1.103$^{+0.012}_{-0.016}$ & 1.04$^{+0.02}_{-0.03}$ & 0.239$^{+0.010}_{-0.007}$ & 10.6$^{+0.3}_{-0.3}$ & -0.262$^{+0.000}_{-0.000}$ & 1,2 \\[1mm]
108473 & 948 & 208776 & - & G0V & 5977$^{+44}_{-44}$ & 4.11$^{+0.03}_{-0.04}$ & -0.07$^{+0.03}_{-0.03}$ & 0.387$^{+0.089}_{-0.089}$ & 6.92$^{+0.52}_{-0.52}$ & 1.092$^{+0.024}_{-0.024}$ & 1.55$^{+0.08}_{-0.07}$ & 0.385$^{+0.026}_{-0.029}$ & 5.2$^{+0.5}_{-0.5}$ & +0.374$^{+0.008}_{-0.008}$ & 1,2 \\[1mm]
108506\sups{SG} & 949 & 208801 & - & K2V & 4915$^{+25}_{-25}$ & 3.79$^{+0.01}_{-0.01}$ & +0.08$^{+0.02}_{-0.02}$ & 0.637$^{+0.059}_{-0.059}$ & 4.24$^{+0.80}_{-0.40}$ & 1.813$^{+0.046}_{-0.053}$ & 2.93$^{+0.05}_{-0.06}$ & 1.736$^{+0.022}_{-0.022}$ & 2.6$^{+0.3}_{-0.3}$ & +0.427$^{+0.005}_{-0.000}$ & 1,2 \\[1mm]
109378 & 960 & 210277 & - & G0 & 5555$^{+44}_{-44}$ & 4.39$^{+0.03}_{-0.03}$ & +0.20$^{+0.03}_{-0.03}$ & -0.020$^{+0.035}_{-0.035}$ & 10.64$^{+2.04}_{-2.20}$ & 0.986$^{+0.038}_{-0.052}$ & 1.06$^{+0.03}_{-0.04}$ & 0.343$^{+0.019}_{-0.022}$ & 1.8$^{+0.5}_{-0.5}$ & +0.351$^{+0.000}_{-0.004}$ & 1,2 \\[1mm]
109439\sups{SG} & 962 & 210460 & - & G0V & 5658$^{+44}_{-44}$ & 3.79$^{+0.04}_{-0.02}$ & -0.22$^{+0.03}_{-0.03}$ & 0.954$^{+0.074}_{-0.074}$ & 2.64$^{+0.28}_{-0.28}$ & 1.49$^{+0.23}_{-0.20}$ & 2.95$^{+0.11}_{-0.02}$ & 0.819$^{+0.001}_{-0.044}$ & 3.7$^{+0.5}_{-0.5}$ & +0.161$^{+0.000}_{-0.009}$ & 1,2 \\[1mm]
109572 & 964 & 210855 & - & F8V & 6261$^{+31}_{-31}$ & 3.84$^{+0.07}_{-0.03}$ & +0.12$^{+0.02}_{-0.02}$ & 0.939$^{+0.037}_{-0.037}$ & 2.44$^{+0.12}_{-0.12}$ & 1.510$^{+0.034}_{-0.020}$ & 2.52$^{+0.08}_{-0.09}$ & 0.000$^{+0.003}_{-0.000}$ & 11.6$^{+0.4}_{-0.4}$ & +0.340$^{+0.004}_{-0.000}$ & 1,2 \\[1mm]
109647\sups{SG,B} & - & 210763 & - & F7V & 6388$^{+91}_{-91}$ & 3.86$^{+0.03}_{-0.02}$ & X & X & 1.52$^{+0.08}_{-0.12}$ & 1.738$^{+0.042}_{-0.042}$ & 2.63$^{+0.10}_{-0.09}$ & 0.000$^{+0.003}_{-0.000}$ & X & +0.247$^{+0.000}_{-0.000}$ & 29 \\[1mm]
111274 & 978 & 213575 & - & G2V & 5668$^{+44}_{-44}$ & 4.16$^{+0.03}_{-0.03}$ & -0.05$^{+0.03}_{-0.03}$ & 0.269$^{+0.062}_{-0.062}$ & 11.40$^{+0.84}_{-0.76}$ & 0.979$^{+0.016}_{-0.018}$ & 1.38$^{+0.06}_{-0.06}$ & 0.449$^{+0.024}_{-0.032}$ & 2.0$^{+0.5}_{-0.5}$ & +0.355$^{+0.004}_{-0.008}$ & 1,2 \\[1mm]
\hline
\end{tabular}
\end{center}
\end{table}
\end{landscape}

\addtocounter {table} {-1}

\begin{landscape}
\begin{table}
\scriptsize
\begin{center}
\caption{continued.}
\begin{tabular}{lccccccccccccccc}
\hline
HIP & SPOCS & HD & Other & Sp. & T\subs{eff} & Log(g) & Log(M/H) or & Log(Lum) & Age & Mass & Radius & Radius\subs{CZ} & \vsini & Log & Refs.\\
no. & no. & no. & names & Type & (K) & (cm s\sups{-2}) & Log(Fe/H)* & (\Lsolar)    & (Gyr) & (\Msolar) & (\Rsolar) & (\Rsolar) & (\kmsn) & (Rossby Number) & \\
\hline
113357 & 990 & 217014 & 51 Peg & G2.5IVa & 5787$^{+25}_{-25}$ & 4.36$^{+0.04}_{-0.03}$ & +0.15$^{+0.02}_{-0.02}$ & 0.117$^{+0.025}_{-0.025}$ & 6.76$^{+1.64}_{-1.48}$ & 1.054$^{+0.039}_{-0.036}$ & 1.15$^{+0.04}_{-0.04}$ & 0.327$^{+0.058}_{-0.024}$ & 2.6$^{+0.3}_{-0.3}$ & +0.347$^{+0.008}_{-0.004}$ & 1,2 \\[1mm]
113421 & 992 & 217107 & - & G8IV & 5704$^{+44}_{-44}$ & 4.42$^{+0.04}_{-0.03}$ & +0.27$^{+0.03}_{-0.03}$ & 0.050$^{+0.031}_{-0.031}$ & 5.84$^{+1.92}_{-2.44}$ & 1.108$^{+0.034}_{-0.052}$ & 1.08$^{+0.04}_{-0.03}$ & 0.316$^{+0.022}_{-0.018}$ & 0.0$^{+0.5}_{-0.0}$ & +0.359$^{+0.008}_{-0.008}$ & 1,2 \\[1mm]
113829 & 995 & 217813 & MT Peg & G1V & 5885$^{+44}_{-44}$ & 4.47$^{+0.02}_{-0.04}$ & +0.02$^{+0.03}_{-0.03}$ & 0.031$^{+0.044}_{-0.044}$ & 1.20$^{+2.16}_{-1.20}$ & 1.066$^{+0.020}_{-0.030}$ & 1.01$^{+0.04}_{-0.02}$ & 0.252$^{+0.014}_{-0.011}$ & 3.2$^{+0.5}_{-0.5}$ & -0.221$^{+0.019}_{-0.000}$ & 1,2 \\[1mm]
113896 & 996 & 217877 & - & F8V & 5953$^{+44}_{-44}$ & 4.28$^{+0.03}_{-0.04}$ & -0.10$^{+0.03}_{-0.03}$ & 0.224$^{+0.058}_{-0.058}$ & 6.88$^{+0.96}_{-0.96}$ & 1.036$^{+0.020}_{-0.024}$ & 1.24$^{+0.05}_{-0.05}$ & 0.303$^{+0.024}_{-0.020}$ & 2.8$^{+0.5}_{-0.5}$ & +0.324$^{+0.004}_{-0.000}$ & 1,2 \\[1mm]
113994\sups{SG} & 1000 & 218101 & - & G8IV & 5301$^{+44}_{-44}$ & 3.96$^{+0.02}_{-0.01}$ & +0.08$^{+0.03}_{-0.03}$ & 0.603$^{+0.076}_{-0.076}$ & 4.68$^{+0.72}_{-0.32}$ & 2.02$^{+0.31}_{-0.31}$ & 2.48$^{+0.02}_{-0.09}$ & 0.971$^{+0.007}_{-0.007}$ & 0.9$^{+0.5}_{-0.5}$ & +0.340$^{+0.000}_{-0.000}$ & 1,2 \\[1mm]
114378 & 1005 & 218687 & - & G0V & 5905$^{+44}_{-44}$ & 4.34$^{+0.03}_{-0.03}$ & -0.10$^{+0.03}_{-0.03}$ & 0.131$^{+0.044}_{-0.044}$ & 6.72$^{+1.28}_{-1.36}$ & 1.009$^{+0.022}_{-0.024}$ & 1.13$^{+0.04}_{-0.04}$ & 0.287$^{+0.021}_{-0.018}$ & 10.3$^{+0.5}_{-0.5}$ & -0.326$^{+0.000}_{-0.000}$ & 1,2 \\[1mm]
114456 & 1007 & 218868 & - & K0 & 5540$^{+44}_{-44}$ & 4.45$^{+0.04}_{-0.04}$ & +0.19$^{+0.03}_{-0.03}$ & -0.111$^{+0.036}_{-0.036}$ & 9.16$^{+3.00}_{-3.40}$ & 0.939$^{+0.058}_{-0.036}$ & 0.97$^{+0.04}_{-0.03}$ & 0.308$^{+0.019}_{-0.020}$ & 2.1$^{+0.5}_{-0.5}$ & +0.303$^{+0.000}_{-0.004}$ & 1,2 \\[1mm]
114622 & 1008 & 219134 & - & K3V & 4835$^{+17}_{-17}$ & 4.59$^{+X}_{-0.02}$ & +0.09$^{+0.01}_{-0.01}$ & -0.556$^{+0.013}_{-0.013}$ & 12.46$^{+1.54}_{-11.96}$ & 0.794$^{+0.037}_{-0.022}$ & 0.77$^{+0.02}_{-0.02}$ & 0.254$^{+0.006}_{-0.013}$ & 1.8$^{+0.2}_{-0.2}$ & +0.285$^{+0.005}_{-0.000}$ & 1,2 \\[1mm]
115951 & 1019 & 221146 & 12 Psc & G1V & 5903$^{+44}_{-44}$ & 4.22$^{+0.04}_{-0.04}$ & +0.06$^{+0.03}_{-0.03}$ & 0.358$^{+0.080}_{-0.080}$ & 6.76$^{+0.68}_{-0.72}$ & 1.098$^{+0.024}_{-0.022}$ & 1.39$^{+0.06}_{-0.06}$ & 0.360$^{+0.028}_{-0.032}$ & 2.9$^{+0.5}_{-0.5}$ & +0.351$^{+0.004}_{-0.000}$ & 1,2 \\[1mm]
116085 & 1020 & 221354 & - & K2V & 5133$^{+44}_{-44}$ & 4.49$^{+0.02}_{-0.01}$ & -0.01$^{+0.03}_{-0.03}$ & -0.262$^{+0.024}_{-0.024}$ & X & 0.839$^{+0.010}_{-0.004}$ & 0.88$^{+0.01}_{-0.02}$ & 0.286$^{+0.005}_{-0.003}$ & 2.1$^{+0.5}_{-0.5}$ & +0.332$^{+0.000}_{-0.000}$ & 1,2 \\[1mm]
116106 & 1021 & 221356 & - & F8V & 5976$^{+44}_{-44}$ & 4.33$^{+0.04}_{-0.03}$ & -0.16$^{+0.03}_{-0.03}$ & 0.160$^{+0.047}_{-0.047}$ & 6.28$^{+1.16}_{-1.24}$ & 1.015$^{+0.026}_{-0.026}$ & 1.15$^{+0.04}_{-0.04}$ & 0.274$^{+0.018}_{-0.020}$ & 3.7$^{+0.5}_{-0.5}$ & +0.316$^{+0.004}_{-0.004}$ & 1,2 \\[1mm]
116421 & 1023 & 221830 & - & F9V & 5740$^{+44}_{-44}$ & 4.22$^{+0.03}_{-0.03}$ & -0.20$^{+0.03}_{-0.03}$ & 0.213$^{+0.048}_{-0.048}$ & 11.52$^{+1.04}_{-1.08}$ & 0.947$^{+0.020}_{-0.024}$ & 1.27$^{+0.05}_{-0.05}$ & 0.363$^{+0.032}_{-0.019}$ & 1.6$^{+0.5}_{-0.5}$ & +0.340$^{+0.004}_{-0.000}$ & 1,2 \\[1mm]
116613 & 1025 & 222143 & V454 And & G3/4V & 5869$^{+44}_{-44}$ & 4.47$^{+0.03}_{-0.03}$ & +0.12$^{+0.03}_{-0.03}$ & 0.015$^{+0.038}_{-0.038}$ & 1.16$^{+2.32}_{-1.16}$ & 1.075$^{+0.024}_{-0.034}$ & 1.01$^{+0.04}_{-0.02}$ & 0.263$^{+0.015}_{-0.011}$ & 3.0$^{+0.5}_{-0.5}$ & -0.112$^{+0.000}_{-0.017}$ & 1,2 \\[1mm]
- & 621 & 131156A & ksi Boo A & G8V & 5570$^{+31}_{-31}$ & 4.57$^{+0.02}_{-0.01}$ & -0.07$^{+0.02}_{-0.02}$ & -0.284$^{+0.049}_{-0.049}$ & 0.00$^{+0.76}_{-0.00}$ & 0.931$^{+0.015}_{-0.020}$ & 0.84$^{+0.03}_{-0.02}$ & 0.237$^{+0.006}_{-0.006}$ & 4.6$^{+0.4}_{-0.4}$ & -0.522$^{+0.000}_{-0.000}$ & 1,2 \\[1mm]
- & - & 131156B & ksi Boo B & K4V & 4350$^{+150}_{-150}$ & 4.40$^{+0.01}_{-0.02}$ & +0.012$^{+0.00}_{-0.00}$ & -1.01$^{+0.08}_{-0.08}$ & X & 0.992$^{+0.010}_{-0.040}$ & 1.07$^{+0.02}_{-0.02}$ & 0.408$^{+0.005}_{-0.006}$ & X & -0.221$^{+0.019}_{-0.020}$ & 2,19 \\[1mm]
- & 830 & 179958 & - & G4V & 5760$^{+44}_{-44}$ & 4.35$^{+0.03}_{-0.03}$ & +0.05$^{+0.03}_{-0.03}$ & 0.081$^{+0.042}_{-0.042}$ & 8.36$^{+1.40}_{-1.44}$ & 0.994$^{+0.022}_{-0.020}$ & 1.11$^{+0.04}_{-0.03}$ & 0.323$^{+0.021}_{-0.017}$ & 2.2$^{+0.5}_{-0.5}$ & +0.324$^{+0.016}_{-0.017}$ & 1,2 \\[1mm]
- & 0 & - & Moon & G2V & 5770$^{+18}_{-18}$ & 4.44$^{+X}_{-X}$ & +0.00$^{+0.01}_{-0.01}$ & 0.000$^{+0.000}_{-0.000}$ & 4.32$^{+1.68}_{-1.72}$ & 1.01$^{+0.02}_{-0.02}$ & 1.000$^{+0.000}_{-0.000}$ & X & 1.7$^{+0.2}_{-0.2}$ & +0.307$^{+0.017}_{-0.013}$ & 1 \\[1mm]
\hline
\end{tabular}
\end{center}
References: 1: \citet{ValentiJA:2005}, 2: \citet{TakedaG:2007}, 3: \citet{BoyajianTS:2012}, 4: \citet{WuY:2011}, 5: \citet{HerreroE:2012}, 6: \citet{DelfosseX:1998}, 7: do Nascimento Jr (private communication), 8: \citet{HolmbergJ:2009}, 9: \citet{Bailer-JonesCAL:2011}, 10: \citet{MasanaE:2006}, 11: \citet{GonzalezG:2010}, 12: \citet{SantosNC:2004}, 13: \citet*{RamirezI:2009}, 14: \citet{WrightNJ:2011}, 15: \citet{YossKM:1997}, 16: \citet{Martinez-ArnaizR:2010}, 17: \citet{SoubiranC:2008}, 18: \citet{AfferL:2005}, 19: \citet{FernandesJ:1998}, 20: \citet{BalachandranS:1990}, 21: \citet{TakedaY:2005}, 22: \citet{StrassmeierKG:2003}, 23: \citet{TakedaY:2007}, 24: \citet{FuhrmannK:1998}, 25: \citet{doNascimentoJD:2010}, 26: \citet{LuckRE:2005}, 27: \citet{MisheninaTV:2012}, 28: \citet*{RamirezI:2007} and 29: \citet{CasagrandeL:2011}.
\end{table}
\end{landscape}

The stellar parameters for each star are summarised in Table~\ref{Bcool_params}.  For most stars, the stellar parameters of the targets were taken from either \citet{ValentiJA:2005} or \citet{TakedaG:2007}. For the remaining $\sim$20\% of our sample the stellar parameters are taken from other papers, as indicated in Table~\ref{Bcool_params}. The Rossby number in Table~\ref{Bcool_params} was calculated for each star using the formulations of \citet[][Equation 13]{WrightJT:2004} which relates the chromospheric activity, Log(R$^{\prime}_{\rm{HK}}$), of the star to its Rossby number. The Log(R$^{\prime}_{\rm{HK}}$) values used for this calculation are the individual measurements from this paper, see Section~\ref{Sec_CaHK} and Table~\ref{Bcool_Activity}.

\begin{figure}
  \centering
  \includegraphics[trim =40 25 0 30, clip, angle=-90, width=\columnwidth]{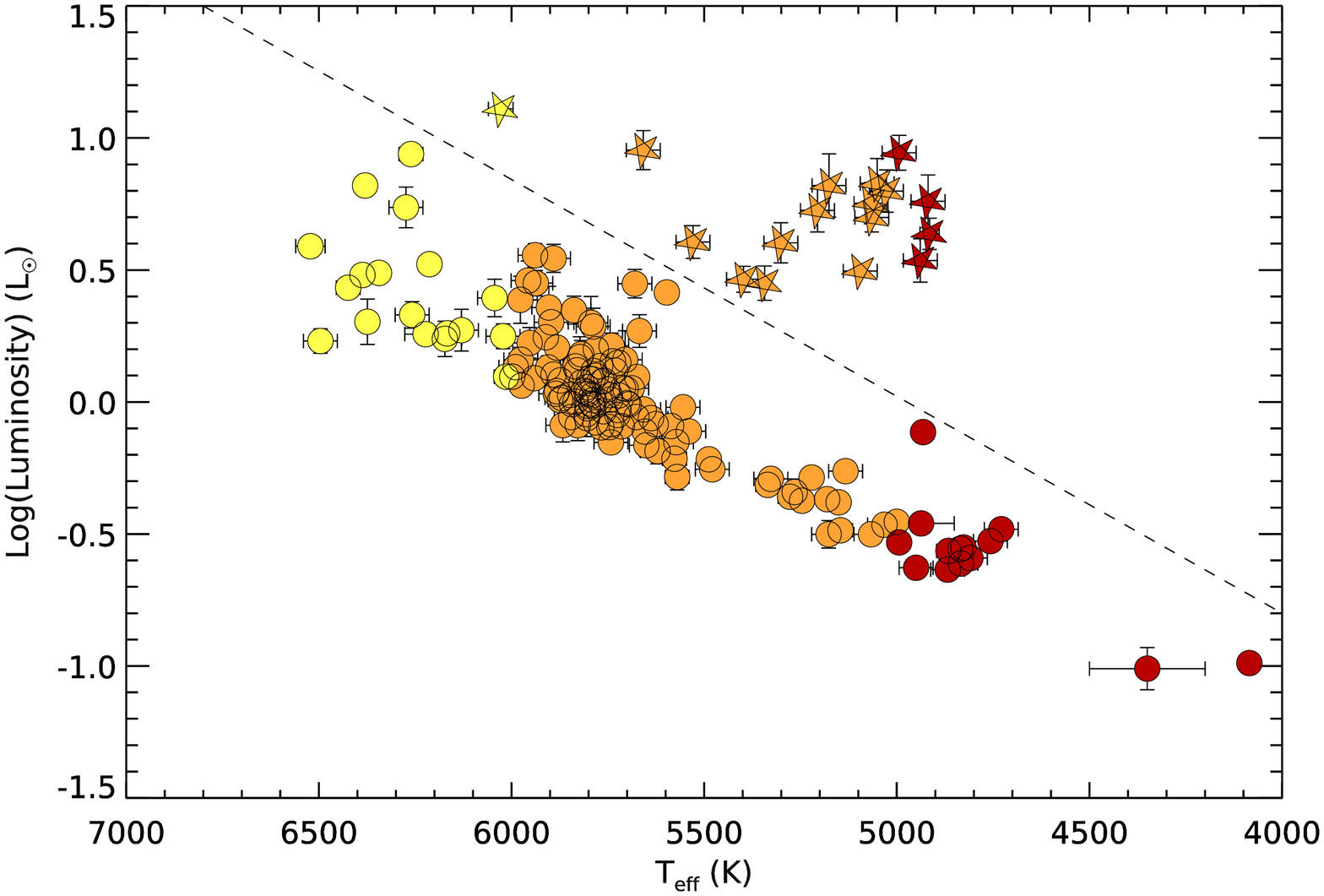}
  \includegraphics[trim =30 25 10 3, clip, angle=-90, width=\columnwidth]{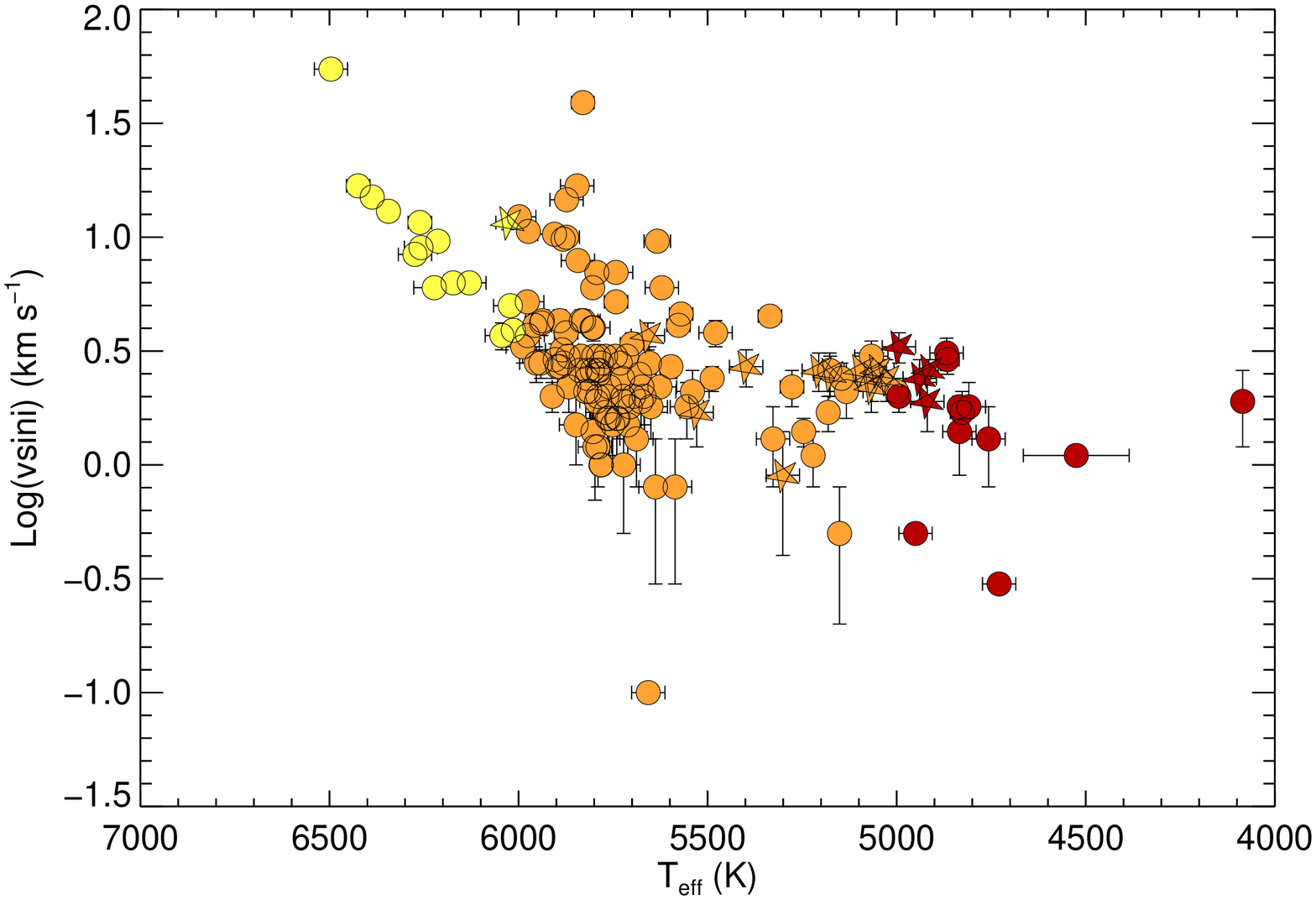}
  \caption{Plot of the HR diagram for our stars (upper panel) and a plot of Log(\vsini) against effective temperature (lower panel). Stars in red have T\subs{eff} $<$ 5000 K, those in orange have 5000 K $\le$ T\subs{eff} $\le$ 6000 K and those in yellow have T\subs{eff} $>$ 6000 K. Filled circles represent dwarf stars and five-pointed stars represent sub-giants, with the dividing line between the two shown as a dashed line (upper panel).} 
  \label{Fig_HR}
\end{figure}

Using the stellar parameters given in Table~\ref{Bcool_params} and directly extracted from our input catalogues, we placed the stars contained in the Bcool sample on a Hertzsprung-Russell (HR) diagram, which is shown in Figure~\ref{Fig_HR} (upper panel). The Bcool sample predominantly contains dwarf stars, but a number (18) of subgiant stars have also been observed as part of the program. The HR diagram was used to classify our stars as either dwarfs or subgiants using the dashed line shown in Figure~\ref{Fig_HR}. All stars in our sample have a T\subs{eff} measure, however there are eight stars for which we do not have any luminosity information, and thus are not included in Figure~\ref{Fig_HR}. For these eight stars six have Log(g) values of greater than 4.0 cm s\sups{-2} and are thus classified as dwarfs, while a seventh (HIP 109647) has a Log(g) value below 4.0 cm s\sups{-2} and has thus been classified as a subgiant. The last target (HIP 38018) has neither Log(g) nor luminosity information, but we have assumed that it is a dwarf based on its radius (1.157$^{+0.079}_{-0.079}$ \Rsolar, \citealt{MasanaE:2006}). Using the simple radius and temperature relationship to luminosity, this radius, along with a T\subs{eff} = 5476 K, gives a luminosity of Log(Lum) = +0.033, making HIP 38018 a dwarf according to Figure~\ref{Fig_HR}.  A plot of the \vsinis values of the stars against their effective temperatures (Figure~\ref{Fig_HR} (lower panel)) shows that the mean rotation rate of the stars contained in the Bcool sample decreases with a decreasing effective temperature. The mean \vsinis value of the F-stars in the sample is significantly higher (12.0 \kmsn) than that of the G-stars (3.4 \kmsn) and the K-stars (1.8 \kmsn).

\section{Observations} \label{Sec_obs}

In this Section we describe the instrument and the observational procedure for securing the data and how the exposures are combined to produce Stokes I, Stokes V, and N (null) profile from the polarimetric sequence.

\subsection{Polarimetric Observations}
\subsubsection{Instrument Description}

The observations were secured with the T\'{e}lescope Bernard Lyot (Observatoire du Pic du Midi, France) using the NARVAL \'{e}chelle spectropolarimeter or with the Canada-France-Hawaii telescope (Mauna Kea, Hawaii) using the ESPaDOnS \'{e}chelle spectropolarimeter with both instruments in polarimetric mode. NARVAL is a twin copy of ESPaDOnS and both have almost identical capabilities. The two spectropolarimeters provide full optical wavelength coverage (370 - 1000 nm) in a single exposure (with only a few nanometers missing in the near infrared). The resolving power of both instruments is approximately 65,000, with a pixel size of 1.8 \kmsn, and both instruments comprise bench mounted spectrographs that are fibre-fed from a Cassegrain mounted polarimetric module. The polarimetric module consists of a series of three Fresnel rhombs that perform polarimetry over the entire spectral range. These are followed by a Wollaston prism which splits the incident light into two beams, each containing the opposite polarisation state, which are then fed into the high-resolution spectrograph, so that both polarisation states are recorded simultaneously. Thus both Stokes I (unpolarised) and Stokes V (circularly polarised) spectra can be recovered from each observation.  More information on the operation of NARVAL and ESPaDOnS can be found in \citet{AuriereM:2003} and \citet{DonatiJF:2006b}, respectively.

\subsubsection{Observational Procedure}\label{Sec_polar}

Each Stokes V observation is obtained from a sequence of four subexposures. For each of these subexposures the half-wave rhombs are rotated to different angles (see \citealt{PetitP:2003}) so that the polarisation state transmitted in each fibre is alternated, to help remove any instrumental effects. This results in the 1\sups{st} and 4\sups{th} and the 2\sups{nd} and 3\sups{rd} subexposures having the same polarisation states being transmitted along the fibres. Consequently, each series of 4 exposures results in 8 spectra (4 left-hand circularly polarised spectra and 4 right-hand circularly polarised spectra). 

The intensity (Stokes I) spectrum is obtained by adding the 8 spectra together, while the polarised (Stokes V) spectra is obtained by constructively adding the left-hand circularly polarised spectra and the right-hand circularly polarised spectra, see Equation~\ref{Eqn_StokesV}. 
\BE
\frac{V}{I} = \frac{R_{v}-1}{R_{v}+1}\\
{\rm where}\\
R^{4}_{v} = \frac{i_{1\perp}/i_{1\parallel}}{i_{2\perp}/i_{2\parallel}}\frac{i_{4\perp}/i_{4\parallel}}{i_{3\perp}/i_{3\parallel}}
\label{Eqn_StokesV}
\EE
where $i_{k\perp}$ and $i_{k\parallel}$ are the two spectra obtained in subexposure $k$.

Additionally, a ``Null" (N) spectrum is also computed to test for ``spurious" magnetic signatures. The Null spectrum is created by destructively adding the left- and right-hand circularly polarised light and thus should cancel out any real magnetic signals, see Equation~\ref{Eqn_Null}. 
\BE
\frac{N}{I} = \frac{R_{n}-1}{R_{n}+1}\\
{\rm where}\\
R^{4}_{n} = \frac{i_{1\perp}/i_{1\parallel}}{i_{4\perp}/i_{4\parallel}}\frac{i_{2\perp}/i_{2\parallel}}{i_{3\perp}/i_{3\parallel}}
\label{Eqn_Null}
\EE
These formulas are from \citet{DonatiJF:1997} where more detail is given on the computation of V, I and N profiles from spectropolarimetric data.

\subsection{Observations and Data Processing}

The data for the Bcool spectropolarimetric survey have been obtained over a total of 25 observing runs at TBL and CFHT starting in late-2006 and extending up to until mid-2013, with the vast majority of data being secured at the TBL.  The journal of observations including the number of observations for each star obtained at each observing epoch is shown in Table~\ref{Bcool_obs}. Some of stars (35/170 = 21\%) have been observed at multiple epochs to look for temporal evolution in their magnetic properties, however the majority of the targets have only been observed once. The data were reduced using the \mbox{\scriptsize{LIBRE-ESPRIT}} data reduction package which is an automatic reduction software package that has been specifically designed for the reduction of spectropolarimetric observations. \mbox{\scriptsize{LIBRE-ESPRIT}} is based on \mbox{\scriptsize{ESPRIT}} which is described in \citet{DonatiJF:1997}.
For each stellar observation both the Stokes I and V reduced spectra were computed with a sampling of 1.8 \kmsn.

\subsection{Least-Squares Deconvolution}

Because the S/N in the reduced spectra is not high enough to detect Zeeman signatures in individual lines, it was necessary to apply the Least-Squares Deconvolution \citep[LSD,][]{DonatiJF:1997} technique to extract the polarimetric signal from the data. The LSD technique is a multi-line technique that extracts the Stokes I and V information from each spectral line and builds one high signal-to-noise LSD profile.  It is described in more detail in \citet{DonatiJF:1997} and \citet{KochukhovO:2010}. LSD spectral profiles were computed for both the polarised and unpolarised spectra to produce both Stokes V and I LSD profiles for each observation with a resolved element of 1.8 \kmsn.  To create LSD profiles, a line list or mask, is generated from stellar atmosphere and spectral synthesis models for a set stellar parameters similar to that of the observed star: T\subs{eff}, Log(g) and Log(M/H), with the range and step size of each of the three parameters given in Table~\ref{Mask_params}.  The model spectral lines are then selected based on central wavelength, central depth and Land\'{e} factor.  

\begin{table}
\begin{center}
\caption{Table of the stellar parameters covered by the LSD line masks.}
\label{Mask_params}
\begin{tabular}{lcc}
\hline
Parameter & Range & Step size\\
\hline
T\subs{eff} (K) & 4000 --- 6500 & 250 \\
Log(g) (cm s\sups{-2}) & 3.5 --- 4.5 & 0.5 \\
Log(M/H) & -0.2 --- +0.2 & 0.2 \\
\hline
\end{tabular}
\end{center}
\end{table}

For the Bcool snapshot survey we created a grid of $\sim$100 model masks using the Vienna Atomic Line Database (VALD\footnote{http://vald.astro.unive.ac.at/$\sim$vald/php/vald.php}, \citealt{KupkaFG:2000}).  The LSD line masks were created from VALD using all lines of depth greater than or equal to 10\% of the continuum. This is shallower than the the minimum line depth that LSD traditionally uses \citep[i.e.][]{DonatiJF:1997} but several tests showed that the difference is compatible within the error bars.  Strong lines such as the \Ca H \& K, and H$\alpha$ lines were excluded from the line lists.  The number of lines used in the calculation of the LSD profile ranges from 5200 to 14700 depending on the stellar parameters and the masks cover a wavelength range of 350 nm to 1100 nm. The number of lines used per LSD profile for each observation is tabulated in Table~\ref{Bcool_LSD}.

When creating the LSD profiles the weighting of the spectral lines were adjusted so that the mean weight of both the Stokes V and Stokes I profile was close to unity. The formula used to determining the mean weights of the LSD profiles is given in Equation~\ref{Mean_weight_eqn}:
\BE
{\rm Mean\ weight} = \frac{\displaystyle\sum\limits_{j=1}^n S^{2}_{j} w^{2}_{j}}{\displaystyle\sum\limits_{j=1}^n S^{2}_{j} w_{j}}, \label{Mean_weight_eqn}
\EE
where $n$ is the number of lines used to create the LSD profile, $S_{j}$ is the signal-to-noise ratio and $w_{j}$ is the line weight of the $j$\sups{th} line given by Equation~\ref{Weight_pol_eqn} (for Stokes V data) or Equation~\ref{Weight_int_eqn} (for Stokes I data):
\BE
w_{j} = \frac{d_{j} \lambda_{j} g_{j}}{d_{0} \lambda_{0} g_{0}}, {\rm for\ Stokes\ V\ data\ or}, \label{Weight_pol_eqn}
\EE
\BE
w_{j} = \frac{d_{j}}{d_{0}}, {\rm for\ Stokes\ I\ data}, \label{Weight_int_eqn}
\EE
where $d$ is the line central depth, $\lambda$ is the line central wavelength and $g$ is the line's Land\'{e} factor. The values of $d_{0}$, $\lambda_{0}$ and $g_{0}$ are the normalisation parameters and were varied for every 250 K step in effective temperature and are given in Table~\ref{Weight_params}.

\begin{landscape}
\begin{table}
\scriptsize
\begin{center}
\caption{Results from the analysis of the stellar LSD profiles of the Bcool solar-type stars sample. The 2\sups{nd} and 3\sups{rd} columns gives the observation date and observation number on that date. Columns 6 and 7 give the signal-to-noise ratio of the Stokes V LSD profile and the number of lines used to create the profile, while columns 8 and 9 notes if there was a Definite (D), Marginal (M) or No (N) detection of a magnetic field in the profile along with the False Alarm Probability for the detection. Columns 10 and 11 give the velocity range over which Equation~\ref{Blong_eqn} was calculated, while columns 12 and 13 give the measured \Bl and $N_{l}$ (Equation~\ref{Blong_eqn} calculated using the Null profile) measures. The final four columns give the total number of observations we have for each star and the number of definite, marginal and non detections we have from these observations. N\sups{D}: Indicates that the definite detection was unreliable due to noise and so the a non detection was chosen instead. \sups{SG}: Indicates a subgiant, see Figure~\ref{Fig_HR}. \sups{VF}: Indicates that the previous radial velocity is from from \citet{ValentiJA:2005} rather than \citet{NideverDL:2002}. \sups{NS}: Indicates a non-radial velocity standard star (i.e. $\sigma$\subs{rms} $\ge$ 100 m s\sups{-1}) according to \citet{NideverDL:2002}, all other stars with \citet{NideverDL:2002} radial velocities are considered to be radial velocity standard stars with $\sigma$\subs{rms} $<$ 100 m s\sups{-1}.} \label{Bcool_LSD} 
\begin{tabular}{lcccccccccccccccc}
\hline
HIP & obs. & obs. & rad. vel. (\kmsn) & rad. vel. (\kmsn) & SNR\subs{LSD} & \# lines & Det? & FAP & v\subs{min} & v\subs{max} & \Bl & N$_{l}$ & \# obs. & \# def. & \# mar. & \# non. \\
no. & date & \# & (This Work) &  \citep{NideverDL:2002} &  & used & & & (\kmsn) &  (\kmsn) & (Gauss) & (Gauss) & tot. & det. & det. & det. \\
\hline
400 & 10dec10 & 01 & +7.7 & +7.511 & 13540 & 10384 & N & 3.428 $\times$ 10\sups{-01} & -2 & +16 & +2.1 $\pm$ 1.0 & -0.3 $\pm$ 1.0 & 1 & 0 & 0 & 1 \\
544 & 27sep10 & 01 & -6.4 & -6.537 & 45559 & 12137 & D & 0.000 $\times$ 10\sups{+00} & -16 & +4 & +2.7 $\pm$ 0.3 & -0.2 $\pm$ 0.3 & 1 & 1 & 0 & 0 \\
682 & 13dec10 & 01 & +1.5 & +1.184 &  18853 & 11135 & D & 7.377 $\times$ 10\sups{-06} & -16 & +18 & +4.4 $\pm$ 1.8 & -0.6 $\pm$ 1.8 & 1 & 1 & 0 & 0 \\
1499 & 19oct10 & 01 & -10.0 & -10.166 & 23795 & 11140 & N & 6.059 $\times$ 10\sups{-01} & -18 & -2 & -0.6 $\pm$ 0.5 & -0.6 $\pm$ 0.5 & 1 & 0 & 0 & 1 \\
1813 & 03jan11 & 01 & -30.4 & -30.550 & 22433 & 9712 & N & 4.001 $\times$ 10\sups{-02} & -40 & -20 & +2.4 $\pm$ 0.7 & +0.0 $\pm$ 0.7 & 1 & 0 & 0 & 1 \\
3093 & 10aug10 & 01 & -32.7 & -32.961 & 44842 & 13129 & D & 0.000 $\times$ 10\sups{+00} & -40 & -25 & -3.2 $\pm$ 0.2 & -0.1 $\pm$ 0.2 & 28 & 22 & 0 & 6 \\
3203 & 15dec10 & 01 & +12.4 & +13.2\sups{VF} & 7788 & 9752 & N & 1.375 $\times$ 10\sups{-01} & +2 & +23 & -7.2 $\pm$ 2.3 & +3.2 $\pm$ 2.3 & 1 & 0 & 0 & 1 \\
3206 & 13jan11 & 01 & -63.0 & -63.202 & 10004 & 14006 & N & 7.278 $\times$ 10\sups{-01} & -70 & -56 & -3.1 $\pm$ 1.1 & +1.2 $\pm$ 1.1 & 1 & 0 & 0 & 1 \\
3765 & 26aug07 & 01 & -9.9 & -10.230 & 11142 & 11238 & M & 1.046 $\times$ 10\sups{-04} & -16 & -4 & +3.6 $\pm$ 0.8 & -0.2 $\pm$ 0.8 & 2 & 0 & 1 & 1 \\
3821 & 27sep10 & 01 & +8.7 & +8.314 & 47965 & 7608 & N & 6.348 $\times$ 10\sups{-01} & +0 & +16 & -0.5 $\pm$ 0.2 & -0.4 $\pm$ 0.3 & 1 & 0 & 0 & 1 \\
3979 & 12dec10 & 01 & -3.5 & -3.742 & 11337 & 8492 & N & 4.683 $\times$ 10\sups{-01} & -16 & +9 & -2.2 $\pm$ 1.7 & +0.0 $\pm$ 1.7 & 1 & 0 & 0 & 1 \\
4127 & 29sep10 & 01 & -73.6 & -73.844 & 25057 & 9228 & N & 8.028 $\times$ 10\sups{-01} & -90 & -54 & +3.5 $\pm$ 1.4 & +2.0 $\pm$ 1.4 & 1 & 0 & 0 & 1 \\
5315\sups{SG} & 12nov10 & 01 & -94.4 & -94.511 & 26681 & 12062 & N & 6.359 $\times$ 10\sups{-01} & -108 & -79 & +1.2 $\pm$ 0.9 & +0.4 $\pm$ 0.9 & 1 & 0 & 0 & 1 \\
5493\sups{SG} & 18aug07 & 01 & -13.8 & - & 39830 & 8757 & M & 1.057 $\times$ 10\sups{-03} & -32 & +4 & +1.8 $\pm$ 0.8 & -0.9 $\pm$ 0.8 & 4 & 0 & 2 & 2 \\
5985 & 04dec10 & 01 & +6.2 & +5.898 & 17616 & 8796 & N & 4.086 $\times$ 10\sups{-01} & -9 & +22 & -1.4 $\pm$ 1.5 & +2.0 $\pm$ 1.5 & 1 & 0 & 0 & 1 \\
6405 & 15dec10 & 01 & +5.9 & +5.636 & 7249 & 8495 & N & 9.500 $\times$ 10\sups{-01} & +0 & +13 & +0.8 $\pm$ 1.2 & +1.1 $\pm$ 1.2 & 1 & 0 & 0 & 1 \\
7244 & 25sep09 & 01 & +11.7 & +8.9\sups{VF} & 31411 & 9987 & D & 0.000 $\times$ 10\sups{+00} & +4 & +20 & -3.5 $\pm$ 0.4 & -0.6 $\pm$ 0.4 & 1 & 1 & 0 & 0 \\
7276 & 19dec10 & 01 & -14.8 & -14.989 & 45843 & 10664 & N & 9.650 $\times$ 10\sups{-01} & -32 & +4 & -0.8 $\pm$ 0.7 & +0.6 $\pm$ 0.7 & 1 & 0 & 0 & 1 \\
7339 & 27sep10 & 01 & -33.1 & -33.291 & 37272 & 9692 & N & 8.986 $\times$ 10\sups{-02} & -49 & -18 & +2.0 $\pm$ 0.7 & -0.7 $\pm$ 0.7 & 1 & 0 & 0 & 1 \\
7513 & 14nov06 & 01 & -28.3 & -28.674 & 36749 & 9872 & N & 7.340 $\times$ 10\sups{-01} & -49 & -9 & +2.5 $\pm$ 1.1 & -0.3 $\pm$ 1.1 & 3 & 0 & 0 & 3 \\
7585 & 11feb08 & 01 & -20.9 & -21.047 & 27989 & 9723 & D & 2.222 $\times$ 10\sups{-06} & -31 & -11 & +2.5 $\pm$ 0.5 & +0.4 $\pm$ 0.5 & 44 & 3 & 2 & 39 \\
7734 & 16nov10 & 01 & +2.3 & +2.143 & 31146 & 9730 & N & 7.294 $\times$ 10\sups{-02} & -4 & +9 & -0.6 $\pm$ 0.3 & -0.1 $\pm$ 0.3 & 1 & 0 & 0 & 1 \\
7918 & 29nov06 & 01 & +4.9 & - & 45461 & 9984 & N & 9.834 $\times$ 10\sups{-01} & -14 & +23 & -0.7 $\pm$ 0.8 & -0.1 $\pm$ 0.8 & 2 & 0 & 0 & 2 \\
7981 & 29jan07 & 01 & -33.4 & -33.647 & 34210 & 11714 & D & 0.000 $\times$ 10\sups{+00} & -41 & -25 & -3.3 $\pm$ 0.3 & +0.8 $\pm$ 0.3 & 39 & 17 & 2 & 20 \\
8159 & 27jan11 & 01 & -45.9 & -46.022 & 31246 & 10271 & N & 9.024 $\times$ 10\sups{-01} & -54 & -38 & -0.4 $\pm$ 0.3 & +0.1 $\pm$ 0.3 & 1 & 0 & 0 & 1 \\
8362 & 16nov10 & 01 & +3.0 & +2.764 & 54108 & 11735 & M & 2.995 $\times$ 10\sups{-03} & -5 & +11 & -0.9 $\pm$ 0.2 & +0.0 $\pm$ 0.2 & 1 & 0 & 1 & 0 \\
8486 & 05sep09 & 01 & +3.4 & - & 31597 & 9978 & D & 2.377 $\times$ 10\sups{-05} & -4 & +11 & +1.9 $\pm$ 0.3 & -0.1 $\pm$ 0.3 & 1 & 1 & 0 & 0 \\
9349 & 28sep09 & 01 & +19.7 & - & 24257 & 9964 & D & 0.000 $\times$ 10\sups{+00} & +13 & +27 & -6.1 $\pm$ 0.4 & +0.3 $\pm$ 0.4 & 1 & 1 & 0 & 0 \\
9406\sups{SG} & 14oct10 & 01 & +43.6 & +43.6\sups{VF} & 38438 & 14601 & N & 9.748 $\times$ 10\sups{-01} & +34 & +52 & -0.3 $\pm$ 0.3 & -0.1 $\pm$ 0.3 & 1 & 0 & 0 & 1 \\
9829 & 28sep10 & 01 & -4.5 & -4.662 & 23385 & 8477 & N & 9,292 $\times$ 10\sups{-01} & -22 & +13 & +2.3 $\pm$ 1.3 & -2.7 $\pm$ 1.3 & 1 & 0 & 0 & 1 \\
9911 & 12nov10 & 01 & -39.0 & -39.333 & 25080 & 9278 & N & 9.121 $\times$ 10\sups{-01} & -58 & -20 & -0.8 $\pm$ 1.3 & -0.8 $\pm$ 1.3 & 1 & 0 & 0 & 1 \\
10339 & 03sep09 & 01 & +7.0 & +7.203 & 30908 & 11086 & D & 0.000 $\times$ 10\sups{+00} & -4 & +18 & +10.9 $\pm$ 0.5 & +0.0 $\pm$ 0.5 & 1 & 1 & 0 & 0 \\
10505 & 08oct10 & 01 & -2.1 & -2.237 & 14301 & 11131 & N & 9.920 $\times$ 10\sups{-01} & -9 & +4 & -0.8 $\pm$ 0.6 & -0.5 $\pm$ 0.6 & 1 & 0 & 0 & 1 \\
11548 & 26sep10 & 01 & +41.5 & +41.202 & 35126 & 8120 & N & 9.041 $\times$ 10\sups{-01} & +31 & +52 & -0.3 $\pm$ 0.5 & +0.0 $\pm$ 0.5 & 1 & 0 & 0 & 1 \\
12048 & 18dec10 & 01 & -50.7 & -50.971 & 22279 & 10271 & N & 4.743 $\times$ 10\sups{-02} & -58 & -45 & +1.0 $\pm$ 0.4 & +0.4 $\pm$ 0.4 & 1 & 0 & 0 & 1 \\
12114 & 26jan07 & 01 & +26.0 & +25.766 & 20575 & 13344 & N & 1.686 $\times$ 10\sups{-01} & +16 & +36 & -1.2 $\pm$ 0.7 & -0.7 $\pm$ 0.7 & 2 & 0 & 0 & 2 \\
13702 & 21jan07 & 01 & +14.2 & - & 22099 & 6119 & N & 1.292 $\times$ 10\sups{-02} & -5 & +34 & +4.5 $\pm$ 1.6 & +0.6 $\pm$ 1.6 & 1 & 0 & 0 & 1 \\
14150 & 27sep10 & 01 & +10.1 & +9.878 & 34646 & 9724 & N & 6.583 $\times$ 10\sups{-02} & +4 & +16 & +0.9 $\pm$ 0.2 & +0.0 $\pm$ 0.2 & 1 & 0 & 0 & 1 \\
15457 & 01oct12 & 01 & +19.2 & +19.021 & 22917 & 9702 & D & 0.000 $\times$ 10\sups{+00} & +9 & +29 & +7.7 $\pm$ 0.6 & +0.1 $\pm$ 0.6 & 16 & 14 & 0 & 2 \\
15776\sups{SG} & 04oct10 & 01 & +41.8 & +41.630 & 25463 & 10092 & N & 6.145 $\times$ 10\sups{-01} & +25 & +58 & -1.4 $\pm$ 1.2 & +0.4 $\pm$ 1.2 & 1 & 0 & 0 & 1 \\
16537 & 01feb07 & 01 & +16.5 & +16.332 & 39693 & 11754 & D & 0.000 $\times$ 10\sups{+00} & +11 & +23 & -10.9 $\pm$ 0.2 & -0.2 $\pm$ 0.2 & 58 & 52 & 3 & 3 \\
16641\sups{SG} & 28sep10 & 01 & +11.2 & +11.050 & 42863 & 13202 & N & 7.448 $\times$ 10\sups{-01} & -5 & +27 & -0.6 $\pm$ 0.6 & +0.1 $\pm$ 0.6 & 1 & 0 & 0 & 1 \\
17027\sups{SG} & 14oct10 & 01 & +40.8 & +46.9\sups{VF} & 60896 & 14604 & N & 7.665 $\times$ 10\sups{-01} & +22 & +59 & +0.7 $\pm$ 0.5 & -0.4 $\pm$ 0.5 & 1 & 0 & 0 & 1 \\
17147 & 12nov10 & 01 & +120.6 & +120.356 & 26800 & 8391 & N & 6.535 $\times$ 10\sups{-01} & +115 & +126 & -0.6 $\pm$ 0.5 & +0.3 $\pm$ 0.5 & 1 & 0 & 0 & 1 \\
17183\sups{SG} & 19jan11 & 01 & +43.2 & +49.2\sups{VF} & 22155 & 14611 & N & 6.341 $\times$ 10\sups{-01} & +27 & +59 & -0.9 $\pm$ 1.2 & +1.1 $\pm$ 1.2 & 1 & 0 & 0 & 1 \\
\hline
\end{tabular}
\end{center}
\end{table}
\end{landscape}

\addtocounter {table} {-1}

\begin{landscape}
\begin{table}
\scriptsize
\begin{center}
\caption{continued.} 
\begin{tabular}{lcccccccccccccccc}
\hline
HIP & obs. & obs. & rad. vel. (\kmsn) & rad. vel. (\kmsn) & SNR\subs{LSD} & \# lines & Det? & FAP & v\subs{min} & v\subs{max} & \Bl & N$_{l}$ & \# obs. & \# def. & \# mar. & \# non. \\
no. & date & \# & (This Work) &  \citep{NideverDL:2002} & & used & & & (\kmsn) &  (\kmsn) & (Gauss) & (Gauss) & tot. & det. & det. & det. \\
\hline
17378\sups{SG} & 04oct10 & 01 & -6.1 & -6.295 & 69175 & 13606 & N & 9.311 $\times$ 10\sups{-01} & -16 & +4 & -0.2 $\pm$ 0.2 & +0.3 $\pm$ 0.2 & 1 & 0 & 0 & 1 \\
18106 & 25aug08 & 01 & -39.4 & -39.608 & 36400 & 9264 & N\sups{D} & 7.964 $\times$ 10\sups{-01} & -59 & -20 & +1.5 $\pm$ 1.0 & +0.9 $\pm$ 1.0 & 7 & 1 & 0 & 6 \\
18267 & 08oct10 & 01 & +19.1 & +18.938 & 14448 & 10784 & M & 8.639 $\times$ 10\sups{-03} & +13 & +25 & -2.6 $\pm$ 0.6 & +0.1 $\pm$ 0.6 & 1 & 0 & 1 & 0 \\
18606\sups{SG} & 19jan11 & 01 & +38.7 & +38.484 & 30323 & 14620 & N & 9.579 $\times$ 10\sups{-01} & +23 & +54 & +1.0 $\pm$ 0.8 & +0.1 $\pm$ 0.8 & 1 & 0 & 0 & 1 \\
19076 & 26sep10 & 01 & +24.3 & +24.034 & 45274 & 9714 & D & 0.000 $\times$ 10\sups{+00} & +18 & +31 & -5.6 $\pm$ 0.2 & +0.1 $\pm$ 0.2 & 1 & 1 & 0 & 0 \\
19849 & 27sep10 & 01 & -42.1 & -42.331 & 57421 & 11679 & D & 7.915 $\times$ 10\sups{-12} & -50 & -34 & +1.3 $\pm$ 0.2 & +0.2 $\pm$ 0.2 & 1 & 1 & 0 & 0 \\
19925 & 02oct09 & 01 & -9.5 & - & 19595 & 9977 & N & 3.599 $\times$ 10\sups{-02} & -16 & -4 & +1.7 $\pm$ 0.4 & -0.6 $\pm$ 0.4 & 1 & 0 & 0 & 1 \\
20800 & 05oct10 & 01 & +35.0 & +34.768 & 25196 & 11103 & N & 7.149 $\times$ 10\sups{-01} & +16 & +54 & +0.9 $\pm$ 1.3 & +1.5 $\pm$ 1.3 & 1 & 0 & 0 & 1 \\
22319\sups{SG} & 16nov10 & 01 & +39.2 & +40.9\sups{VF} & 26590 & 12012 & N & 7.220 $\times$ 10\sups{-02} & +29 & +50 & -0.6 $\pm$ 0.6 & +0.5 $\pm$ 0.6 & 1 & 0 & 0 & 1 \\
22336 & 22jan11 & 01 & +77.4 & +77.189 & 34140 & 10614 & N & 6.960 $\times$ 10\sups{-01} & +59 & +90 & -0.5 $\pm$ 0.7 & +0.1 $\pm$ 0.7 & 1 & 0 & 0 & 1 \\
22449 & 04oct12 & 01 & +24.5 & +24.9\sups{VF} & 41710 & 7174 & N & 2.049 $\times$ 10\sups{-01} & -4 & +52 & -4.3 $\pm$ 1.5 & -0.3 $\pm$ 1.5 & 1 & 0 & 0 & 1 \\
22633\sups{SG} & 05feb11 & 01 & -12.6 & -10.7\sups{VF} & 32600 & 12043 & N & 5.709 $\times$ 10\sups{-01} & -18 & -7 & +0.4 $\pm$ 0.2 & +0.2 $\pm$ 0.2 & 1 & 0 & 0 & 1 \\
23311 & 11jan13 & 01 & +21.7 & +21.552 & 49149 & 14742 & N & 9.881 $\times$ 10\sups{-01} & +7 & +36 & -0.6 $\pm$ 0.4 & +0.5 $\pm$ 0.4 & 1 & 0 & 0 & 1 \\
24813 & 16feb08 & 01 & +66.7 & +66.511 & 40737 & 8756 & N & 9.485 $\times$ 10\sups{-01} & +58 & +76 & -0.3 $\pm$ 0.3 & +0.3 $\pm$ 0.3 & 1 & 0 & 0 & 1 \\
25278 & 09feb08 & 01 & +38.0 & - & 31949 & 7904 & D & 3.331 $\times$ 10\sups{-16} & +11 & +65 & -8.4 $\pm$ 1.8 & +0.8 $\pm$ 1.8 & 31 & 27 & 3 & 1 \\
25486 & 27dec07 & 08 & +24.2 & +17.0\sups{VF} & 12760 & 7150 & D & 0.000 $\times$ 10\sups{+00} & -50 & +90 & -288.9 $\pm$ 17.2 & +39.6 $\pm$ 17.1 & 60 & 54 & 4 & 2 \\
27913 & 26jan07 & 01 & -12.8 & -12.171 $\pm$ 1.294\sups{NS} & 25572 & 8788 & D & 0.000 $\times$ 10\sups{+00} & -23 & -2 & -6.9 $\pm$ 0.7 & +1.0 $\pm$ 0.7 & 44 & 42 & 1 & 1 \\
29568 & 28sep09 & 01 & +22.3 & - & 40301 & 9955 & D & 0.000 $\times$ 10\sups{+00} & +14 & +31 & -5.1 $\pm$ 0.3 & -0.1 $\pm$ 0.3 & 1 & 1 & 0 & 0 \\
30476 & 20mar08 & 01 & +56.6 & +57.4\sups{VF} & 26501 & 9993 & N & 8.615 $\times$ 10\sups{-02} & +50 & +63 & +0.7 $\pm$ 0.3 & +0.4 $\pm$ 0.3 & 1 & 0 & 0 & 1 \\
31965 & 27sep09 & 01 & +28.1 & - & 30435 & 9948 & N & 2.050 $\times$ 10\sups{-01} & +18 & +38 & +1.2 $\pm$ 0.5 & -0.2 $\pm$ 0.5 & 1 & 0 & 0 & 1 \\
32673 & 01oct09 & 01 & +48.1 & - & 24738 & 9932 & N & 6.959 $\times$ 10\sups{-01} & +31 & +65 & +1.7 $\pm$ 1.2 & -0.2 $\pm$ 1.2 & 1 & 0 & 0 & 1 \\
32851 & 16jan12 & 01 & -12.0 & - & 30460 & 6087 & N & 4.931 $\times$ 10\sups{-01} & -23 & -2 & +1.1 $\pm$ 0.8 & -0.8 $\pm$ 0.8 & 1 & 0 & 0 & 1 \\
33277 & 06feb07 & 01 & -14.8 & -15.023 & 13898 & 7610 & N & 1.883 $\times$ 10\sups{-02} & -23 & -7 & +3.2 $\pm$ 0.8 & +0.6 $\pm$ 0.8 & 3 & 0 & 0 & 3 \\
35185 & 01oct09 & 01 & +22.2 & - & 19252 & 9957 & D & 0.000 $\times$ 10\sups{+00} & +13 & +31 & -4.5 $\pm$ 0.7 & -1.0 $\pm$ 0.7 & 1 & 1 & 0 & 0 \\
35265 & 19jan08 & 01 & +22.8 & +22.526 & 18360 & 9729 & D & 4.240 $\times$ 10\sups{-08} & +13 & +32 & +4.9 $\pm$ 0.8 & +0.6 $\pm$ 0.8 & 36 & 21 & 4 & 11 \\
36704 & 14nov06 & 01 & -15.4 & -15.744 & 14418 & 12025 & D & 0.000 $\times$ 10\sups{+00} & -22 & -9 & -5.4 $\pm$ 0.6 & +0.0 $\pm$ 0.6 & 1 & 1 & 0 & 0 \\
38018 & 30jan09 & 01 & -30.0 & - & 17039 & 9393 & N & 1.3666 $\times$ 10\sups{-01} & -36 & -23 & +1.6 $\pm$ 0.5 & +0.1 $\pm$ 0.5 & 1 & 0 & 0 & 1 \\
38228 & 02jan10 & 01 & -15.7 & -15.888 & 33388 & 9962 & D & 0.000 $\times$ 10\sups{+00} & -27 & -5 & +2.5 $\pm$ 0.5 & -0.2 $\pm$ 0.5 & 1 & 1 & 0 & 0 \\
38647 & 29sep09 & 01 & +17.2 & - & 28148 & 9967 & D & 0.000 $\times$ 10\sups{+00} & +11 & +23 & +4.4 $\pm$ 0.3 & -0.1 $\pm$ 0.3 & 1 & 1 & 0 & 0 \\
38747 & 04jan10 & 01 & -7.9 & - & 18112 & 9974 & D & 0.000 $\times$ 10\sups{+00} & -31 & +14 & +3.4 $\pm$ 2.4 & +2.0 $\pm$ 2.4 & 1 & 1 & 0 & 0 \\
41484 & 15mar08 & 01 & -32.2 & -32.342 & 15598 & 10006 & N & 7.235 $\times$ 10\sups{-01} & -38 & -25 & +1.5 $\pm$ 0.6 & +5.8 $\pm$ 0.6 & 4 & 0 & 0 & 4 \\
41526 & 03jan10 & 01 & -0.8 & - & 26398 & 9967 & N & 8.609 $\times$ 10\sups{-02} & -9 & +7 & -1.6 $\pm$ 0.5 & +0.5 $\pm$ 0.5 & 1 & 0 & 0 & 1 \\
41844 & 23jan10 & 01 & +13.7 & +13.629 & 31730 & 9979 & M & 3.466 $\times$ 10\sups{-04} & +5 & +22 & -2.0 $\pm$ 0.4 & -0.3 $\pm$ 0.4 & 1 & 0 & 1 & 0 \\
42333 & 22jan12 & 01 & +35.6 & +35.370 & 21666 & 9670 & D & 0.000 $\times$ 10\sups{+00} & +29 & +43 & +7.7 $\pm$ 0.5 & -0.9 $\pm$ 0.5 & 20 & 12 & 2 & 6 \\
42403 & 16feb08 & 01 & +6.4 & +6.097 & 6541 & 7918 & N & 3.148 $\times$ 10\sups{-01} & -2 & +14 & -4.2 $\pm$ 1.8 & +1.9 $\pm$ 1.8 & 1 & 0 & 0 & 1 \\
42438 & 26jan07 & 01 & -12.7 & - & 18685 & 9731 & D & 0.000 $\times$ 10\sups{+00} & -25 & +0 & -11.0 $\pm$ 1.2 & +0.8 $\pm$ 1.2 & 12 & 12 & 0 & 0 \\
43410 & 23nov11 & 01 & +4.5 & +4.256 & 15946 & 7876 & D & 1.111 $\times$ 10\sups{-08} & -18 & +27 & -8.1 $\pm$ 2.6 & -0.6 $\pm$ 2.6 & 31 & 13 & 4 & 14 \\
43557 & 01jan10 & 01 & +10.6 & - & 32356 & 9972 & M & 6.680 $\times$ 10\sups{-04} & +4 & +18 & -1.2 $\pm$ 0.3 & +0.4 $\pm$ 0.3 & 1 & 0 & 1 & 0 \\
43726 & 27jan07 & 01 & +32.2 & +32.002 & 32845 & 9722 & D & 0.000 $\times$ 10\sups{+00} & +27 & +38 & -3.7 $\pm$ 0.2 & -0.1 $\pm$ 0.2 & 30 & 16 & 4 & 10 \\
44897 & 10nov11 & 01 & +26.4 & +26.120 & 30681 & 8728 & D & 0.000 $\times$ 10\sups{+00} & +20 & +34 & -6.6 $\pm$ 0.3 & +0.3 $\pm$ 0.3 & 38 & 30 & 1 & 7 \\
44997 & 05jan10 & 01 & -11.7 & - & 25147 & 9977 & N & 4.314 $\times$ 10\sups{-02} & -18 & -5 & +1.2 $\pm$ 0.4 & -0.5 $\pm$ 0.4 & 1 & 0 & 0 & 1 \\
46066 & 02jan10 & 01 & +4.6 & - & 17515 & 9970 & N & 4.921 $\times$ 10\sups{-01} & -13 & +22 & +4.4 $\pm$ 1.7 & -0.5 $\pm$ 1.7 & 1 & 0 & 0 & 1 \\
46580 & 10may13 & 01 & +30.1 & +29.836 & 31822 & 13267 & D & 0.000 $\times$ 10\sups{+00} & +23 & +38 & +13.5 $\pm$ 0.3 & +0.0 $\pm$ 0.3 & 1 & 1 & 0 & 0 \\
46903 & 06jan10 & 01 & +12.1 & - & 24878 & 9973 & N & 9.994 $\times$ 10\sups{-02} & +0 & +23 & -2.9 $\pm$ 0.7 & +1.0 $\pm$ 0.7 & 1 & 0 & 0 & 1 \\
49081 & 19jan08 & 01 & +56.1 & +55.956 & 30891 & 11082 & N & 6.495 $\times$ 10\sups{-01} & +47 & +67 & -0.8 $\pm$ 0.5 & +0.3 $\pm$ 0.5 & 1 & 0 & 0 & 1 \\
49350 & 23jan10 & 01 & +0.2 & -0.260 & 27673 & 9970 & N & 9.249 $\times$ 10\sups{-01} & -9 & +11 & -0.5 $\pm$ 0.5 & +0.2 $\pm$ 0.5 & 1 & 0 & 0 & 1 \\
49580 & 07jan10 & 01 & -26.7 & - & 26456 & 9956 & N & 5.012 $\times$ 10\sups{-01} & -32 & -20 & +0.6 $\pm$ 0.4 & +0.0 $\pm$ 0.4 & 1 & 0 & 0 & 1 \\
49728 & 23jan10 & 01 & -23.4 & - & 32076 & 9958 & N & 8.897 $\times$ 10\sups{-01} & -29 & -18 & +0.5 $\pm$ 0.3 & -0.2 $\pm$ 0.3 & 1 & 0 & 0 & 1 \\
49756 & 16mar08 & 01 & -17.7 & -17.886 & 21262 & 10025 & N & 7.181 $\times$ 10\sups{-01} & -31 & -4 & -2.1 $\pm$ 1.0 & -13.9 $\pm$ 1.2 & 2 & 0 & 0 & 2 \\
49908 & 11jan13 & 01 & -26.0 & -25.729 & 29082 & 11927 & D & 0.000 $\times$ 10\sups{+00} & -32 & -18 & +3.4 $\pm$ 0.3 & +0.4 $\pm$ 0.3 & 1 & 1 & 0 & 0 \\
50316 & 21jan07 & 01 & +29.2 & +29.019 & 6422 & 10302 & N & 1.338 $\times$ 10\sups{-01} & +23 & +36 & -3.1 $\pm$ 1.4 & +0.3 $\pm$ 1.4 & 1 & 0 & 0 & 1 \\
50505 & 29nov06 & 01 & -7.2 & -7.551 & 15861 & 9671 & N & 7.415 $\times$ 10\sups{-01} & -22 & +7 & -3.4 $\pm$ 1.5 & +0.8 $\pm$ 1.5 & 2 & 0 & 0 & 2 \\
53721 & 31dec09 & 01 & +11.5 & +11.235 & 37281 & 9006 & N & 4.988 $\times$ 10\sups{-01} & +5 & +18 & +0.6 $\pm$ 0.2 & +0.2 $\pm$ 0.2 & 1 & 0 & 0 & 1 \\
54952 & 11may13 & 01 & +9.2 & - & 25294 & 13296 & D & 0.000 $\times$ 10\sups{+00} & +2 & +16 & -8.0 $\pm$ 0.4 & -0.3 $\pm$ 0.4 & 1 & 1 & 0 & 0 \\
\hline
\end{tabular}
\end{center}
\end{table}
\end{landscape}

\addtocounter {table} {-1}

\begin{landscape}
\begin{table}
\scriptsize
\begin{center}
\caption{continued.} 
\begin{tabular}{lcccccccccccccccc}
\hline
HIP & obs. & obs. & rad. vel. (\kmsn) & rad. vel. (\kmsn) & SNR\subs{LSD} & \# lines & Det? & FAP & v\subs{min} & v\subs{max} & \Bl & N$_{l}$ & \# obs. & \# def. & \# mar. & \# non. \\
no. & date & \# & (This Work) &  \citep{NideverDL:2002} & & used & & & (\kmsn) & (\kmsn) & (Gauss) & (Gauss) & tot. & det. & det. & det. \\
\hline
55459 & 31dec09 & 01 & +7.3 & +7.058 & 29014 & 9969 & N & 5.715 $\times$ 10\sups{-01} & -13 & +27 & -1.6 $\pm$ 1.2 & -0.5 $\pm$ 1.2 & 5 & 0 & 0 & 5 \\
56242 & 26jan07 & 01 & -4.7 & -4.854 & 21926 & 8783 & N & 7.598 $\times$ 10\sups{-01} & -16 & +7 & -1.7 $\pm$ 0.8 & +0.6 $\pm$ 0.8 & 1 & 0 & 0 & 1 \\
56948 & 01jan10 & 01 & +5.7 & - & 15643 & 9972 & N & 4.466 $\times$ 10\sups{-01} & +0 & +13 & -1.3 $\pm$ 0.6 & -0.2 $\pm$ 0.6 & 2 & 0 & 0 & 2 \\
56997 & 12feb10 & 01 & -5.5 & -5.565 & 25391 & 10783 & D & 0.000 $\times$ 10\sups{+00} & -13 & +2 & -14.6 $\pm$ 0.4 & -0.1 $\pm$ 0.4 & 41 & 35 & 2 & 4 \\
57939 & 25jan07 & 01 & -98.0 & -98.073 & 15043 & 11128 & N & 4.264 $\times$ 10\sups{-01} & -117 & -79 & -6.6 $\pm$ 4.1 & -2.5 $\pm$ 4.1 & 1 & 0 & 0 & 1 \\
58708\sups{SG} & 24jan12 & 01 & -10.5 & -10.678 & 35868 & 12147 & N & 8.731 $\times$ 10\sups{-01} & -20 & -2 & +0.4 $\pm$ 0.4 & -0.5 $\pm$ 0.4 & 1 & 0 & 0 & 1 \\
60098 & 25jan12 & 01 & -9.3 & -9.3\sups{VF} & 41026 & 9592 & N & 9.034 $\times$ 10\sups{-01} & -20 & +2 & +0.9 $\pm$ 0.4 & -0.5 $\pm$ 0.4 & 1 & 0 & 0 & 1 \\
60353 & 25jan12 & 01 & +5.0 & +4.759 & 35507 & 7878 & N & 4.519 $\times$ 10\sups{-01} & -5 & +14 & +0.7 $\pm$ 0.4 & -0.3 $\pm$ 0.4 & 1 & 0 & 0 & 1 \\
61901 & 23mar13 & 01 & +24.6 & +24.604 & 12892 & 12384 & N & 9.809 $\times$ 10\sups{-01} & +5 & +43 & +1.9 $\pm$ 2.3 & -1.9 $\pm$ 2.3 & 1 & 0 & 0 & 1 \\
62523 & 28jan07 & 01 & -8.8 & -6.0\sups{VF} & 34523 & 9723 & D & 0.000 $\times$ 10\sups{+00} & -16 & -2 & -3.0 $\pm$ 0.3 & -0.3 $\pm$ 0.3 & 1 & 1 & 0 & 0 \\
64797 & 23mar13 & 01 & +7.9 & - & 21612 & 11198 & N & 6.516 $\times$ 10\sups{-01} & -2 & +18 & -1.4 $\pm$ 0.6 & -0.5 $\pm$ 0.6 & 1 & 0 & 0 & 1 \\
65347 & 19feb11 & 01 & +33.3 & - & 16349 & 8110 & D & 0.000 $\times$ 10\sups{+00} & +22 & +45 & +1.7 $\pm$ 1.1 & +0.7 $\pm$ 1.1 & 1 & 1 & 0 & 0 \\
66147 & 15apr13 & 01 & -5.8 & -6.086 & 19246 & 13283 & D & 0.000 $\times$ 10\sups{+00} & -13 & +0 & -7.3 $\pm$ 0.4 & +0.0 $\pm$ 0.4 & 1 & 1 & 0 & 0 \\
66774\sups{SG} & 16mar11 & 01 & -19.3 & - & 27033 & 12047 & N & 2.556 $\times$ 10\sups{-01} & -36 & -4 & -2.1 $\pm$ 1.0 & +0.6 $\pm$ 1.0 & 1 & 0 & 0 & 1 \\
67275 & 13may13 & 01 & -16.0 & -16.542 $\pm$ 0.340 & 75344 & 8271 & D & 6.090 $\times$ 10\sups{-08} & -38 & +5 & +3.2 $\pm$ 0.5 & -0.1 $\pm$ 0.5 & 8 & 3 & 2 & 3 \\
67422 & 17apr13 & 01 & -20.9 & -20.380 & 18090 & 11838 & M\sups{a} & 4.315 $\times$ 10\sups{-04} & -36 & -7 & +4.8 $\pm$ 1.1 & -2.3 $\pm$ 1.1 & 1 & 0 & 1 & 0 \\
68184 & 16apr13 & 01 & -26.3 & -26.471 & 41068 & 13268 & D & 0.000 $\times$ 10\sups{+00} & -32 & -20 & +3.7 $\pm$ 0.2 & +0.1 $\pm$ 0.2 & 1 & 1 & 0 & 0 \\
71181 & 21mar13 & 01 & +11.5 & +12.4\sups{VF} & 18618 & 13313 & D & 0.000 $\times$ 10\sups{+00} & +5 & +18 & +5.1 $\pm$ 0.5 & +0.0 $\pm$ 0.5 & 1 & 1 & 0 & 0 \\
71631 & 13jan12 & 01 & -20.5 & -20.3\sups{VF} & 10747 & 9661 & D & 0.000 $\times$ 10\sups{+00} & -38 & -4 & +45.3 $\pm$ 2.8 & +4.4 $\pm$ 2.8 & 40 & 39 & 0 & 1 \\
72848 & 17apr13 & 01 & -21.8 & -31.808 $\pm$ 8.832 & 55793 & 13103 & D & 0.000 $\times$ 10\sups{+00} & -31 & -13 & +7.6 $\pm$ 0.2 & +0.4 $\pm$ 0.2 & 1 & 1 & 0 & 0 \\
74432 & 25jan07 & 01 & -38.8 & -38.921 & 17786 & 9687 & N & 9.134 $\times$ 10\sups{-01} & -45 & -32 & +0.7 $\pm$ 0.5 & -0.3 $\pm$ 0.5 & 1 & 0 & 0 & 1 \\
76114 & 16mar08 & 01 & -35.5 & -35.691 & 25605 & 9993 & N & 8.521 $\times$ 10\sups{-01} & -50 & -20 & +1.9 $\pm$ 1.0 & -9.9 $\pm$ 1.3 & 2 & 0 & 0 & 2 \\
79578 & 16mar08 & 01 & -20.5 & -22.8\sups{VF} & 17942 & 10026 & D\sups{a} & 8.489 $\times$ 10\sups{-11} & -27 & -14 & -3.4 $\pm$ 0.5 & -7.1 $\pm$ 0.6 & 4 & 2 & 0 & 2 \\
79672 & 25jul07 & 01 & +12.0 & +11.748 & 21510 & 9570 & D & 5.446 $\times$ 10\sups{-05} & +5 & +18 & -2.3 $\pm$ 0.4 & -0.5 $\pm$ 0.4 & 57 & 6 & 6 & 45 \\
81300 & 27jan07 & 01 & -12.7 & -12.857 & 26382 & 11723 & N & 6.555 $\times$ 10\sups{-02} & -22 & -5 & +0.6 $\pm$ 0.4 & -0.9 $\pm$ 0.4 & 1 & 0 & 0 & 1 \\
82588 & 26jan07 & 01 & +45.3 & +45.066 & 19091 & 10767 & D\sups{a} & 1.238 $\times$ 10\sups{-13} & +25 & +65 & +8.5 $\pm$ 1.9 & -16.1 $\pm$ 1.9 & 1 & 1 & 0 & 0 \\
86193 & 02jul08 & 01 & -76.6 & -76.809 & 28259 & 9933 & D & 4.864 $\times$ 10\sups{-05} & -83 & -70 & -1.5 $\pm$ 0.3 & +0.3 $\pm$ 0.3 & 3 & 1 & 0 & 2 \\
86400 & 01aug07 & 01 & +26.7 & +15.947 $\pm$ 3.724\sups{NS} & 23973 & 12670 & D & 0.000 $\times$ 10\sups{+00} & +20 & +32 & -6.3 $\pm$ 0.4 & -0.3 $\pm$ 0.4 & 1 & 1 & 0 & 0 \\
86974 & 08jun11 & 01 & -17.6 & -17.004 $\pm$ 0.210\sups{NS} & 41378 & 13228 & N & 7.275 $\times$ 10\sups{-01} & -27 & -7 & -0.5 $\pm$ 0.3 & -0.1 $\pm$ 0.3 & 1 & 0 & 0 & 1 \\
88194 & 28jun08 & 01 & +2.2 & +2.020 & 24134 & 10031 & N & 5.087 $\times$ 10\sups{-01} & -9 & +14 & +1.0 $\pm$ 0.8 & -0.8 $\pm$ 0.8 & 3 & 0 & 0 & 3 \\
88945 & 07aug12 & 01 & -14.3 & -14.403 $\pm$ 0.105\sups{NS} & 24965 & 9710 & D & 0.000 $\times$ 10\sups{+00} & -23 & -5 & +11.2 $\pm$ 0.6 & -0.4 $\pm$ 0.6 & 37 & 34 & 0 & 3 \\
88972 & 25jul07 & 01 & -19.2 & -19.418 & 26087 & 12660 & N & 6.062 $\times$ 10\sups{-01} & -34 & -4 & +1.5 $\pm$ 1.0 & +1.1 $\pm$ 1.0 & 1 & 0 & 0 & 1 \\
90729\sups{SG} & 06jul11 & 01 & -53.0 & -75.3\sups{VF} & 50302 & 11765 & N & 6.095 $\times$ 10\sups{-01} & -70 & -36 & +0.8 $\pm$ 0.5 & +0.7 $\pm$ 0.5 & 1 & 0 & 0 & 1 \\
91043 & 28may08 & 06 & -23.1 & - & 11923 & 8406 & D & 0.000 $\times$ 10\sups{+00} & -83 & +36 & -89.7 $\pm$ 12.2 & +29.5 $\pm$ 12.2 & 53 & 47 & 1 & 5 \\
92984 & 17jul08 & 03 & +10.6 & +10.277 & 30904 & 8774 & D & 0.000 $\times$ 10\sups{+00} & -7 & +29 & +11.3 $\pm$ 1.1 & +1.0 $\pm$ 1.1 & 46 & 30 & 7 & 9 \\
95253 & 25jul07 & 01 & -11.7 & - & 18578 & 6856 & N & 1.811 $\times$ 10\sups{-01} & -40 & +16 & +7.6 $\pm$ 3.4 & +7.8 $\pm$ 3.4 & 2 & 0 & 0 & 2 \\
95962 & 05jul08 & 01 & +58.4 & +58.176 & 33135 & 9983 & N & 5.148 $\times$ 10\sups{-02} & +52 & +65 & -1.0 $\pm$ 0.3 & -0.1 $\pm$ 0.3 & 4 & 0 & 0 & 4 \\
96085 & 05jul13 & 01 & -48.7 & -48.864 & 16167 & 12578 & N & 2.188 $\times$ 10\sups{-02} & -56 & -41 & -1.8 $\pm$ 0.6 & -0.2 $\pm$ 0.6 & 1 & 0 & 0 & 1 \\
96100 & 14aug07 & 01 & +27.0 & +26.691 & 66739 & 10340 & D & 0.000 $\times$ 10\sups{+00} & +20 & +34 & -3.1 $\pm$ 0.1 & +0.0 $\pm$ 0.1 & 11 & 7 & 0 & 4 \\
96895 & 20apr13 & 01 & -27.7 & -27.377 & 33423 & 9663 & N & 5.969 $\times$ 10\sups{-01} & -45 & -11 & +2.7 $\pm$ 0.8 & -1.4 $\pm$ 0.8 & 4 & 0 & 0 & 4 \\
96901 & 17jul11 & 01 & -27.7 & -27.871 & 35324 & 9669 & N & 5.381 $\times$ 10\sups{-01} & -47 & -9 & -1.7 $\pm$ 0.9 & -0.4 $\pm$ 0.9 & 4 & 0 & 0 & 4 \\
98921 & 12jul10 & 01 & -26.2 & -25.063 $\pm$ 0.136\sups{NS} & 26320 & 11111 & D & 0.000 $\times$ 10\sups{+00} & -34 & -18 & -9.8 $\pm$ 0.5 & +0.3 $\pm$ 0.5 & 97 & 72 & 5 & 18 \\
100511 & 17aug07 & 01 & +4.8 & - & 23046 & 6864 & N & 5.755 $\times$ 10\sups{-01} & -5 & +14 & +0.6 $\pm$ 0.7 & -0.2 $\pm$ 0.7 & 1 & 0 & 0 & 1 \\
100970 & 21jun11 & 01 & -91.1 & -91.582 $\pm$ 0.188\sups{NS} & 36392 & 10466 & N & 4.175 $\times$ 10\sups{-01} & -104 & -77 & +1.7 $\pm$ 0.6 & -0.6 $\pm$ 0.6 & 1 & 0 & 0 & 1 \\
101875 & 17aug07 & 01 & -20.8 & -21.045 & 14268 & 8462 & N & 8.522 $\times$ 10\sups{-01} & -27 & -14 & +0.9 $\pm$ 0.6 & +0.3 $\pm$ 0.6 & 2 & 0 & 0 & 2 \\
104214 & 26jul07 & 01 & -65.4 & -65.726 & 35154 & 12357 & D & 0.000 $\times$ 10\sups{+00} & -74 & -58 & -8.9 $\pm$ 0.3 & -0.4 $\pm$ 0.3 & 68 & 45 & 4 & 19 \\
107350 & 21jun10 & 01 & -16.6 & -16.833 & 21322 & 8764 & D & 0.000 $\times$ 10\sups{+00} & -29 & -5 & +14.8 $\pm$ 0.9 & +0.0 $\pm$ 0.9 & 91 & 68 & 7 & 16 \\
108473 & 04dec10 & 01 & +29.9 & +31.491 $\pm$ 0.601\sups{NS} & 16028 & 9315 & N & 7.885 $\times$ 10\sups{-01} & +14 & +45 & -2.2 $\pm$ 1.7 & -2.3 $\pm$ 1.7 & 1 & 0 & 0 & 1 \\
108506\sups{SG} & 26nov10 & 01 & -50.0 & -50.179 & 43451 & 13547 & N & 9.392 $\times$ 10\sups{-01} & -61 & -40 & +0.3 $\pm$ 0.3 & -0.4 $\pm$ 0.3 & 1 & 0 & 0 & 1 \\
109378 & 26nov10 & 01 & -20.7 & -20.873 & 37867 & 12151 & N &  4.703 $\times$ 10\sups{-01} & -32 & -9 & -0.9 $\pm$ 0.5 & +0.2 $\pm$ 0.5 & 1 & 0 & 0 & 1 \\
109439\sups{SG} & 14oct10 & 01 & +20.7 & +20.429 & 43917 & 9064 & D & 0.000 $\times$ 10\sups{+00} & +14 & +27 & -3.7 $\pm$ 0.2 & -0.3 $\pm$ 0.2 & 1 & 1 & 0 & 0 \\
109572 & 24aug13 & 01 & -20.3 & - & 57024 & 9591 & N & 3.054 $\times$ 10\sups{-01} & -34 & -5 & +0.8 $\pm$ 0.5 & +1.0 $\pm$ 0.5 & 1 & 0 & 0 & 1 \\
109674\sups{SG,B} & 22aug13 & 01 & +59.2 & - & 30074 & 7436 & N & 6.015 $\times$ 10\sups{-01} & +41 & +76 & +0.7 $\pm$ 1.3 & +0.2 $\pm$ 1.3 & 1 & 0 & 0 & 1 \\
\hline
\end{tabular}
\end{center}
\end{table}
\end{landscape}

\addtocounter {table} {-1}

\begin{landscape}
\begin{table}
\scriptsize
\begin{center}
\caption{continued.} 
\begin{tabular}{lcccccccccccccccc}
\hline
HIP & obs. & obs. & rad. vel. (\kmsn) & rad. vel. (\kmsn) & SNR\subs{LSD} & \# lines & Det? & FAP & v\subs{min} & v\subs{max} & \Bl & N$_{l}$ & \# obs. & \# def. & \# mar. & \# non. \\
no. & date & \# & (This Work) &  \citep{NideverDL:2002} & & used & & & (\kmsn) & (\kmsn) & (Gauss) & (Gauss) & tot. & det. & det. & det. \\
\hline
111274 & 20oct10 & 01 & -21.4 & -21.544 & 17523 & 10304 & N & 8.556 $\times$ 10\sups{-01} & -31 & -13 & +1.2 $\pm$ 0.8 & -0.4 $\pm$ 0.8 & 1 & 0 & 0 & 1 \\
113357 & 15dec10 & 01 & -33.1 & -33.225 & 20777 & 11111 & N & 7.351 $\times$ 10\sups{-01} & -40 & -27 & +0.6 $\pm$ 0.4 & +0.1 $\pm$ 0.4 & 3 & 0 & 0 & 3 \\
113421 & 20oct10 & 01 & -13.3 & -13.399 & 21447 & 11135 & N & 5.690 $\times$ 10\sups{-01} & -31 & +4 & -1.9 $\pm$ 1.2 & -0.5 $\pm$ 1.2 & 1 & 0 & 0 & 1 \\
113829 & 15oct10 & 01 & +2.2 & +2.030 & 28847 & 8781 & D & 0.000 $\times$ 10\sups{+00} & -4 & +9 & +2.7 $\pm$ 0.3 & +0.0 $\pm$ 0.3 & 1 & 1 & 0 & 0 \\
113896 & 18oct10 & 01 & -12.5 & -12.676 & 13046 & 8784 & N & 8.051 $\times$ 10\sups{-01} & -25 & +2 & -2.5 $\pm$ 1.8 & -2.0 $\pm$ 1.8 & 1 & 0 & 0 & 1 \\
113994\sups{SG} & 15oct10 & 01 & -32.1 & -31.3\sups{VF} & 39724 & 12513 & D & 2.243 $\times$ 10\sups{-14} & -40 & -25 & -1.5 $\pm$ 0.2 & +0.3 $\pm$ 0.2 & 1 & 1 & 0 & 0 \\
114378 & 12dec10 & 01 & -7.6 & -23.2\sups{VF} & 27114 & 8785 & D & 0.000 $\times$ 10\sups{+00} & -32 & +16 & -11.1 $\pm$ 1.8 & -0.2 $\pm$ 1.8 & 2 & 2 & 0 & 0 \\
114456 & 04oct10 & 01 & -30.4 & -30.622 & 31241 & 12122 & N\sups{D} & 3.843 $\times$ 10\sups{-01} & -38 & -22 & -1.2 $\pm$ 0.4 & +0.1 $\pm$ 0.4 & 2 & 1 & 0 & 1 \\
114622 & 29sep10 & 01 & -18.3 & -18.558 & 62098 & 13325 & D & 0.000 $\times$ 10\sups{+00} & -25 & -13 & +1.1 $\pm$ 0.1 & -0.1 $\pm$ 0.1 & 1 & 1 & 0 & 0 \\
115951 & 14oct10 & 01 & -14.8 & -15.137 & 32164 & 9329 & N & 2.360 $\times$ 10\sups{-01} & -29 & -2 & -1.1 $\pm$ 0.6 & +0.4 $\pm$ 0.6 & 1 & 0 & 0 & 1 \\
116085 & 17oct10 & 01 & -24.9 & -25.113 & 22588 & 11713 & N & 9.082 $\times$ 10\sups{-01} & -43 & -5 & +2.3 $\pm$ 1.3 & -0.7 $\pm$ 1.3 & 2 & 0 & 0 & 2 \\
116106 & 17oct10 & 01 & -12.5 & -12.713 & 16434 & 7610 & N & 7.958 $\times$ 10\sups{-01} & -31 & +5 & -2.0 $\pm$ 2.1 & +0.3 $\pm$ 2.1 & 1 & 0 & 0 & 1 \\
116421 & 28sep10 & 01 & -112.1 & -112.299 & 26538 & 8928 & N & 8.160 $\times$ 10\sups{-01} & -122 & -103 & -1.0 $\pm$ 0.6 & +0.7 $\pm$ 0.6 & 1 & 0 & 0 & 1 \\
116613 & 19oct10 & 01 & +0.1 & -0.169 & 29842 & 11139 & D & 0.000 $\times$ 10\sups{+00} & -7 & +7 & -5.2 $\pm$ 0.4 & +0.4 $\pm$ 0.4 & 1 & 1 & 0 & 0 \\
hd131156a & 02aug07 & 01 & +1.9 & +1.303 & 33578 & 10803 & D & 0.000 $\times$ 10\sups{+00} & -5 & +9 & +18.4 $\pm$ 0.3 & +0.4 $\pm$ 0.3 & 112 & 101 & 3 & 8 \\
hd131156b & 08feb12 & 02 & +2.9 & +3.274 & 22022 & 14055 & D & 0.000 $\times$ 10\sups{+00} & -4 & +11 & -18.9 $\pm$ 0.5 & -0.9 $\pm$ 0.5 & 43 & 38 & 3 & 2 \\
hd179958 & 28jun08 & 01 & -41.6 & -41.158 & 19052 & 9989 & N & 8.814 $\times$ 10\sups{-01} & -50 & -34 & +0.6 $\pm$ 0.6 & -0.4 $\pm$ 0.6 & 3 & 0 & 0 & 3 \\
moon & 15jul11 & 01 & +0.3 & -0.1\sups{VF} & 28555 & 10004 & N & 1.420 $\times$ 10\sups{-02} & -5 & +5 & +0.5 $\pm$ 0.3 & -0.5 $\pm$ 0.4 & 1 & 0 & 0 & 1 \\
\hline
\end{tabular}
\end{center}
\sups{a}: Magnetic detection uncertain as a significant signal is also evidenced in the Null profile (N$_{l}$).
\end{table}
\end{landscape}

\begin{table}
\begin{center}
\caption{Table of the normalisation parameters used to create the LSD profiles for every 250 K step in T\subs{eff}.}
\label{Weight_params}
\begin{tabular}{lccc}
\hline
T\subs{eff} (K) & $d_{0}$ & $\lambda_{0}$ & $g_{0}$\\
\hline
4000 & 0.55 & 650.0 & 1.22\\
4250 & 0.55 & 640.0 & 1.22\\
4500 & 0.55 & 630.0 & 1.22\\
4750 & 0.55 & 620.0 & 1.22\\
5000 & 0.54 & 610.0 & 1.22\\
5250 & 0.54 & 600.0 & 1.22\\
5500 & 0.53 & 590.0 & 1.22\\
5750 & 0.52 & 580.0 & 1.22\\
6000 & 0.51 & 570.0 & 1.22\\
6250 & 0.50 & 570.0 & 1.21\\
6500 & 0.49 & 560.0 & 1.21\\
\hline
\end{tabular}
\end{center}
\end{table}

For each Stokes V and I LSD profile, the line mask that matches most closely to each star's parameters (T\subs{eff}, Log(g) and Log(M/H)) from Table~\ref{Bcool_params} is used. For several stars without a Log(g) or Log(M/H) values, Log(g) = 4.5 cm s\sups{-2} and Log(M/H) = 0.00 was used and Log(Fe/H) was substituted for Log(M/H) where a star had only a Log(Fe/H) measurement.  In the Bcool snapshot sample, there are four stars that have metallicities far outside the range shown in Table~\ref{Mask_params}. For these stars (HIP 17147, HIP 32851, HIP 57939, HIP 91043) individual line masks were created. However, the use of different metallicities was found to be of no significance. Examples of the LSD profiles of some of the more widely known targets are given in Appendix~\ref{Appendix_LSD} along with more detailed notes on these stars.

\section{The longitudinal magnetic field, B\subs{\lowercase{l}}}\label{Sec_Bl}

A star's surface magnetic field can be measured directly from the stellar Stokes V and I LSD profiles.  This results in a value for the mean longitudinal magnetic field ($B_{l}$, which can also be referred to as $<$B\subs{z}$>$), where $B_{l}$ is the line-of-sight component of the stellar magnetic field integrated over the visible stellar disc. Because the LSD profiles are in velocity space ($v$), we can determine \Bl from the first-order moment of the Stokes V LSD profile using Equation~\ref{Blong_eqn} \citep{MathysG:1989, DonatiJF:1997}:

\BE
B_{l} = -2.14 \times 10^{11} \frac{\int vV(v)dv}{\lambda_{0} g_{0} c \int [I_{c} - I(v)]dv}, \label{Blong_eqn}
\EE
\label{eq:bl}

where \Bl is in gauss, $\lambda_{0}$ and $g_{0}$ are given in Table~\ref{Weight_params} and $c$ is the speed of light in \kmsn. $I_{c}$ is the continuum level of the intensity profile and as this is normalised to 1.0, $I_{c}$ is often replaced by 1.0 in some formulations of Equation~\ref{Blong_eqn}.  The error in the measure of \Bl ($B_{\rm err}$) is determined by propagating the uncertainties computed by the reduction pipeline for each spectral bin of the normalised spectrum through Equation~\ref{Blong_eqn}. 

\begin{figure}
  \centering
  \includegraphics[trim =0 15 0 15, clip, angle=-90, width=\columnwidth]{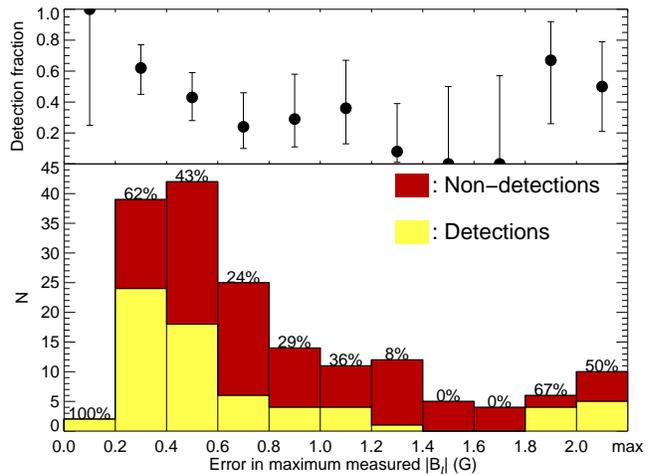}
  \caption{Histogram of detections/non-detections of surface magnetic fields against the error bar in $B_{l}$, using bins of 0.2~G. The detection rate (including both marginal and definite detections) for each bin is given above each column as a percentage, with the detection fraction (along with $\pm$2-sigma error bars) given in the upper panel.} 
  \label{Fig_Hist_Err}
\end{figure}

We show in Figure~\ref{Fig_Hist_Err} the distribution of our observations as a function of the error bars in \Bl estimates. The main factors affecting the uncertainties are the stellar \vsini, the signal-to-noise ratio of the observation and the number of lines included in the LSD analysis as discussed in \citet{ShorlinSLS:2002}. The majority of our measurements have error bars lower than 1~G, and the detection rates tend to be higher for measurements with the lowest uncertainties.  

An additional source of uncertainly in measuring \Bl is the determination of the velocity range over which to integrate Equation~\ref{Blong_eqn}. Calculations using a velocity range that is too narrow may result in the exclusion of polarised signal, while using a velocity range that is too wide could result in the introduction of additional noise, and possibly spurious polarisation signal, to the measurement.  For stars with definite strong magnetic detections, the width over which to integrate Equation~\ref{Blong_eqn} can be determined from the presence of a well defined Stokes V profile. However, most of the stars in our sample have no magnetic detection in their Stokes V LSD profiles, so it was necessary to use a more robust technique for the entire dataset. We initially calculated \Bl over a range of velocity widths and chose the width for which the ratio of $|B_{l}|$/$B_{\rm err}$ is maximised as this should minimise the amount of noise in our measurement of $B_{l}$.  This was performed on each observation of the Bcool sample.  For stars with multiple observations, only the observation with the largest value of \Bla was chosen to be analysed and to be presented in this paper.  This is  because the upper envelope of \Bla values are the most relevant, see Section~\ref{Sec_Resvar}. For stars with no strong magnetic detection the measure of \Bl is likely to be an upper estimate due to the noise contained in the Stokes V profile.

For each star, the value of $B_{l}$, the velocity ranges over which \Bl was calculated and the magnetic field measured from the Null profile (N$_{l}$) are presented in Table~\ref{Bcool_LSD}. For most stars, the value of N$_{l}$ is near zero, which is consistent with a magnetic field measurement unaffected by spurious polarisation signals. However, in a number of observations the N$_{l}$ value diverges from zero by more than 3$\sigma$. In these cases, the magnetic field detection can still be considered as robust if $|N_{l}|$ is small compared to $|B_{l}|$, although the statistical error bar given on $B_{l}$ is then likely underestimating the true uncertainty on the measurement. If $|N_{l}|$ is significantly greater than zero and is significant compared to $|B_{l}|$, the magnetic field estimate should be considered with great care.  We note that other activity indicators deduced from the same observation and listed in the present study are unaffected by this issue.

Table~\ref{Bcool_LSD} also lists the observation date (of the selected observations) and the radial velocity of the star as measured by this work and previous works \citep{NideverDL:2002, ValentiJA:2005}.  Additionally, Table~\ref{Bcool_LSD} shows the signal-to-noise ratio of the Stokes V LSD profile for the chosen observation, the number of spectral lines used in the computation of the LSD profile, if there was a magnetic field detected on the star and the false alarm probability (FAP) of the magnetic field detection.  Finally, Table~\ref{Bcool_LSD} also lists the total number of observations for each star, the number of definite detections, marginal detections and null detections of a magnetic field, see Section~\ref{Sec_fap} for the definition of definite and marginal detections.

\subsection{False Alarm Probability}\label{Sec_fap}

In order to determine if any variations in a star's Stokes V LSD profile are due to a magnetic field or just noise we determine the False Alarm Probability (FAP) for each observation. The FAP is based on a $\chi^{2}$ probability function measured both inside and outside of the stellar spectral lines (as defined by the position of the Stokes I LSD profile). The FAP is determined for both the Stokes V and Null LSD profiles with a definite detection (D) being defined as having a FAP smaller than 10\sups{-5} (i.e. a $\chi^{2}$ probability of larger than 99.999\%), while a marginal detection (M) is defined as having a FAP less than 10\sups{-3} but greater than 10\sups{-5} (i.e. a $\chi^{2}$ probability between 99.999\% and 99.9\%) otherwise the detection is classified as a non detection (N). For a definite or marginal detection, the signal must also be found within the line profile of the star (i.e. between $v_{\rm rad}$ $\pm$ \vsini). Further information on the evaluation of the FAP can be found in \citet*{DonatiJF:1992} and \citet{DonatiJF:1997}.

\subsection{Radial Velocities}

The radial velocities determined from this work (given in Table~\ref{Bcool_LSD}) were calculated by fitting a Gaussian profile to the unpolarised Stokes I LSD profile of each star and taking the centroid of the Gaussian as the radial velocity. As can be seen from Table~\ref{Bcool_LSD}, for those stars with previously measured radial velocities and known to be stable (see \citealt{NideverDL:2002}) our values are mostly within $\pm$ 0.3 \kmss of those of \citet{NideverDL:2002}.

\section{Stellar activity proxies}\label{Sec_activity}

In addition to the measurement of the magnetic field on the visible surface of each star we also analysed more traditional measures of stellar activity, namely the emission in the \Ca H \& K, \Ca infrared triplet and H$\alpha$ lines. 

\subsection{Calcium HK emission}\label{Sec_CaHK}

The most widely used activity proxy is a star's \Ca H \& K emission. This has been used by a large number of authors and the Mount Wilson survey's S-index \citep[cf.,][]{BaliunasSL:1995} is a de facto standard for measuring \Ca H \& K emission of solar-type stars. In order for our measurements to match the Mount Wilson S-index we follow a similar method to that of \citet{WrightJT:2004}.

By early 2013, we had 113 stars in our sample in common with those of \citet{WrightJT:2004} for which they had calculated the S-index. Of these 113 stars we had 23 stars that were observed using ESPaDOnS on the CFHT and 94 stars that were observed using NARVAL on the TBL (with 4 stars being observed on both telescopes). For every intensity spectrum of these 113 stars we removed all the overlapping orders from the reduced spectra from \mbox{\scriptsize{LIBRE-ESPRIT}} so that only the order containing the \Ca H \& K lines (at 3933.663 \AA\/ and 3968.469 \AA) remained. This order spanned from $\sim$3870 to $\sim$4020 \AA\/ for both spectrographs, thus easily covering the wavelength range required to calculate the S-index \citep[see][Figure 1]{WrightJT:2004}. Unlike \citet{MorgenthalerA:2012} and Waite et al. (in prep.) we did not re-normalise the spectra of each star before calculating the S-index as tests showed only minor differences in the calculation of the S-index when this was done. In order to determine the S-index for our stars we followed the method of \citet{WrightJT:2004} and determined the S-index by fitting:
\BE
{\rm S-index = Ca_{HK}-index} = \frac{aF_{H} + bF_{K}}{cF_{R_{HK}} + dF_{V_{HK}}} + e, \label{Sindex_eqn}
\EE
where $F_{H}$ and $F_{K}$ are the fluxes in 2.18 \AA\/ triangular bandpasses centred on the cores of the \Ca H \& K lines respectively, $F_{R_{HK}}$ and $F_{V_{HK}}$ are the fluxes in two rectangular 20 \AA\/ bandpasses centered on the continuum either side of the HK lines at 3901.07 \AA\/ and 4001.07 \AA, see \citet[][Figure 1]{WrightJT:2004} and $a$, $b$, $c$, $d$ and $e$ are coefficients to be determined.

\begin{figure}
  \centering
  \includegraphics[trim =30 35 10 15, clip, angle=-90, width=\columnwidth]{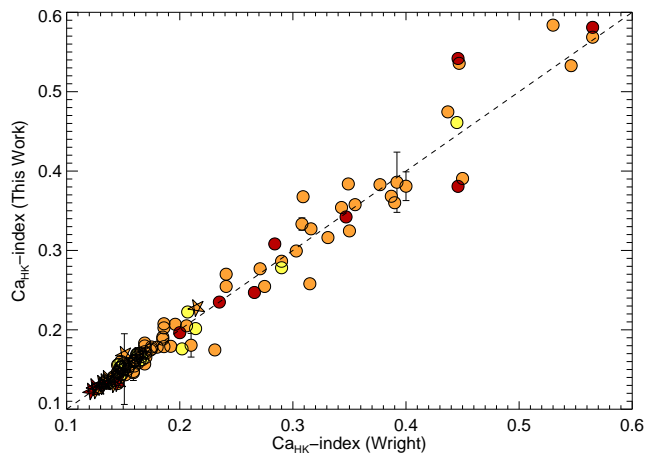}
  \caption{Plot of the S-index found in this work against that from Wright et al. (2004). The symbols are the same as in Figure~\ref{Fig_HR}.} 
  \label{Fig_Sindex}
\end{figure}

In order to determine the coefficients from Equation~\ref{Sindex_eqn} we took every intensity spectrum of our 113 stars and calculated the flux through the $H$, $K$, $R$ and $V$ bandpasses. Then the mean values of $F_{H}$, $F_{K}$, $F_{R}$ and $F_{V}$ were calculated for each star (so as to minimise errors associated with temporal variations in stellar activity levels) and the coefficients of Equation~\ref{Sindex_eqn} were then adjusted, using a least-squares fit, so that the S-indexes of our stars matched (as closely as possible given the different dates of observation) those from \citet{WrightJT:2004}. This was done separately for the NARVAL and ESPaDOnS observations with the results given in Table~\ref{Tab_Sindex_params}. A calculation of the coefficients for NARVAL was previously done by \citet{MorgenthalerA:2012}. However, this used a much smaller number of stars (31) and only used the \Ca H line. A fit of the S-index found from this project (using only the single observation for each star given in Table~\ref{Bcool_LSD}) to those from \citet{WrightJT:2004} is given in Figure~\ref{Fig_Sindex}.

The errors in the S-indices are an empirical estimate of the uncertainty in this single measurement and are calculated as the sample standard deviation of the S-indices of each of the four exposures (see Section~\ref{Sec_polar}) of each observation. The S-indices were then converted to Log(R$^{\prime}$\subs{HK}) using the formulations of \citet[][Equations 9 to 12]{WrightJT:2004}, with the B-V values for the stars coming from HIPPARCOS \citep{PerrymanMAC:1997} and \citet{WrightJT:2004}. The range in the Log(R$^{\prime}$\subs{HK}) values given in column 6 of Table~\ref{Bcool_Activity}, gives the spread in Log(R$^{\prime}$\subs{HK}) found from all observations of the star (ignoring the small error bars). 

\begin{landscape}
\begin{table}
\scriptsize
\begin{center}
\caption{The chromospheric activity of the Bcool solar-type stars sample. Note that the chromospheric ages and rotation periods have been determined from the equations of \citet{WrightJT:2004} from the individual Log(R$^{\prime}_{\rm{HK}}$) measures rather than the range of values. \sups{SG}: indicates a subgiant, see Figure~\ref{Fig_HR}. For the (B-V) values, \sups{W}: indicates a (B-V) value from \citet{WrightJT:2004}.} \label{Bcool_Activity} 
\begin{tabular}{lccccccccccc}
\hline
HIP & (B-V) & S-index & S-index & Log(R$^{\prime}_{\rm{HK}}$) & Chromospheric & Chromospheric & Log(R$^{\prime}_{\rm{HK}}$) & Ca\subs{IRT}-index & Ca\subs{IRT}-index & H\subs{\alpha}-index & H\subs{\alpha}-index\\
no. & (HIPP.) & (Wright) & (This work) & & age (Gyr) & period (d) & range & & range & & range\\
\hline
400 & 0.755 & 0.174 & 0.1744 $\pm$ 0.0053 & -4.95$^{+0.02}_{-0.03}$ & 4.664$^{+0.537}_{-0.330}$ & 35.3$^{+1.1}_{-0.7}$ & -4.95 & 0.7976 $\pm$ 0.0045 & 0.7976 & 0.3046 $\pm$ 0.0004 & 0.3046 \\[0.9mm]
544 & 0.752 & - & 0.4591 $\pm$ 0.0015 & -4.36$^{+0.00}_{-0.00}$ & 0.221$^{+0.000}_{-0.000}$ & 7.0$^{+0.0}_{-0.0}$ & -4.36 & 0.8977 $\pm$ 0.0011 & 0.8977 & 0.3269 $\pm$ 0.0002 & 0.3269 \\[0.9mm]
682 & 0.626 & 0.377 & 0.3829 $\pm$ 0.0025 & -4.36$^{+0.01}_{-0.00}$ & 0.221$^{+0.000}_{-0.037}$ & 4.3$^{+0.0}_{-0.2}$ & -4.36 & 0.9238 $\pm$ 0.0009 & 0.9238 & 0.3183 $\pm$ 0.0004 & 0.3183 \\[0.9mm]
1499 & 0.674 & 0.156 & 0.1567 $\pm$ 0.0014 & -5.03$^{+0.01}_{-0.01}$ & 6.214$^{+0.221}_{-0.215}$ & 29.0$^{+0.3}_{-0.3}$ & -5.03 & 0.7149 $\pm$ 0.0016 & 0.7149 & 0.2908 $\pm$ 0.0001 & 0.2908 \\[0.9mm]
1813 & 0.639 & 0.170 & 0.1705 $\pm$ 0.0008 & -4.92$^{+0.00}_{-0.01}$ & 4.178$^{+0.157}_{-0.000}$ & 22.1$^{+0.2}_{-0.0}$ & -4.92 & 0.7372 $\pm$ 0.0013 & 0.7372 & 0.2867 $\pm$ 0.0002 & 0.2867 \\[0.9mm]
3093 & 0.850 & 0.169 & 0.1724 $\pm$ 0.0007 & -5.00$^{+0.00}_{-0.01}$ & 5.589$^{+0.203}_{-0.000}$ & 44.0$^{+0.4}_{-0.0}$ & -5.02 --- -4.93 & 0.7065 $\pm$ 0.0007 & 0.6973 --- 0.7240 & 0.3122 $\pm$ 0.0002 & 0.3122 --- 0.3153 \\[0.9mm]
3203 & 0.620 & 0.303 & 0.2993 $\pm$ 0.0029 & -4.50$^{+0.01}_{-0.01}$ & 0.873$^{+0.046}_{-0.047}$ & 7.9$^{+0.3}_{-0.3}$ & -4.50 & 0.8785 $\pm$ 0.0034 & 0.8785 & 0.3115 $\pm$ 0.0007 & 0.3115 \\[0.9mm]
3206 & 0.937 & 0.210 & 0.1806 $\pm$ 0.0149 & -5.03$^{+0.04}_{-0.05}$ & 6.214$^{+1.165}_{-0.822}$ & 48.5$^{+2.2}_{-1.7}$ & -5.03 & 0.7166 $\pm$ 0.0030 & 0.7166 & 0.3289 $\pm$ 0.0003 & 0.3289 \\[0.9mm]
3765 & 0.890 & - & 0.2042 $\pm$ 0.0056 & -4.93$^{+0.02}_{-0.01}$ & 4.334$^{+0.162}_{-0.308}$ & 42.8$^{+0.4}_{-0.9}$ & -4.93 --- -4.92 & 0.7734 $\pm$ 0.0016 & 0.7375 --- 0.7734 & 0.3220 $\pm$ 0.0007 & 0.3220 --- 0.3222 \\[0.9mm]
3821 & 0.587 & 0.165 & 0.1616 $\pm$ 0.0006 & -4.97$^{+0.00}_{-0.00}$ & 5.016$^{+0.000}_{-0.000}$ & 17.2$^{+0.0}_{-0.0}$ & -4.97 & 0.7741 $\pm$ 0.0021 & 0.7741 & 0.2820 $\pm$ 0.0003 & 0.2820 \\[0.9mm]
3979 & 0.663 & 0.186 & 0.2025 $\pm$ 0.0043 & -4.80$^{+0.01}_{-0.02}$ & 2.686$^{+0.203}_{-0.096}$ & 21.2$^{+0.6}_{-0.3}$ & -4.80 & 0.8077 $\pm$ 0.0003 & 0.8077 & 0.2957 $\pm$ 0.0005 & 0.2957 \\[0.9mm]
4127 & 0.595 & 0.141 & 0.1403 $\pm$ 0.0003 & -5.12$^{+0.00}_{-0.00}$ & 8.421$^{+0.000}_{-0.000}$ & 20.7$^{+0.0}_{-0.0}$ & -5.12 & 0.7483 $\pm$ 0.0004 & 0.7483 & 0.2802 $\pm$ 0.0003 & 0.2802 \\[0.9mm]
5315\sups{SG} & 0.847 & 0.131 & 0.1326 $\pm$ 0.0023 & -5.19$^{+0.01}_{-0.01}$ & 10.470$^{+0.315}_{-0.310}$ & 52.1$^{+0.5}_{-0.5}$ & -5.19 & 0.7166 $\pm$ 0.0019 & 0.7166 & 0.3013 $\pm$ 0.0004 & 0.3013 \\[0.9mm]
5493\sups{SG} & 0.603 & - & 0.1982 $\pm$ 0.0006 & -4.79$^{+0.01}_{-0.00}$ & 2.590$^{+0.000}_{-0.092}$ & 15.2$^{+0.0}_{-0.2}$ & -4.79 --- -4.77 & 0.7522 $\pm$ 0.0005 & 0.7522 --- 0.7572 & 0.2854 $\pm$ 0.0001 & 0.2818 --- 0.2854 \\[0.9mm]
5985 & 0.563 & 0.162 & 0.1694 $\pm$ 0.0011 & -4.91$^{+0.00}_{-0.01}$ & 4.026$^{+0.151}_{-0.000}$ & 13.6$^{+0.2}_{-0.0}$ & -4.91 & 0.7722 $\pm$ 0.0018 & 0.7722 & 0.2818 $\pm$ 0.0008 & 0.2818 \\[0.9mm]
6405 & 0.627 & 0.170 & 0.1653 $\pm$ 0.0059 & -4.95$^{+0.03}_{-0.04}$ & 4.664$^{+0.728}_{-0.486}$ & 21.4$^{+0.8}_{-0.7}$ & -4.95 & 0.7958 $\pm$ 0.0028 & 0.7958 & 0.2935 $\pm$ 0.0014 & 0.2935 \\[0.9mm]
7244 & 0.666 & 0.308 & 0.3333 $\pm$ 0.0078 & -4.47$^{+0.01}_{-0.02}$ & 0.731$^{+0.095}_{-0.048}$ & 8.8$^{+0.8}_{-0.4}$ & -4.47 & 0.8722 $\pm$ 0.0014 & 0.8722 & 0.3124 $\pm$ 0.0004 & 0.3124 \\[0.9mm]
7276 & 0.639 & 0.135 & 0.1355 $\pm$ 0.0004 & -5.17$^{+0.00}_{-0.01}$ & 9.856$^{+0.304}_{-0.000}$ & 27.9$^{+0.3}_{-0.0}$ & -5.17 & 0.6936 $\pm$ 0.0010 & 0.6936 & 0.2816 $\pm$ 0.0002 & 0.2816 \\[0.9mm]
7339 & 0.686 & 0.160 & 0.1601 $\pm$ 0.0007 & -5.01$^{+0.00}_{-0.01}$ & 5.791$^{+0.209}_{-0.000}$ & 29.9$^{+0.3}_{-0.0}$ & -5.01 & 0.7511 $\pm$ 0.0013 & 0.7511 & 0.2917 $\pm$ 0.0002 & 0.2917 \\[0.9mm]
7513 & 0.536 & 0.146 & 0.1565 $\pm$ 0.0009 & -4.98$^{+0.01}_{-0.00}$ & 5.201$^{+0.000}_{-0.185}$ & 11.9$^{+0.0}_{-0.1}$ & -5.02 --- -4.98 & 0.7498 $\pm$ 0.0014 & 0.7442 --- 0.7498 & 0.2768 $\pm$ 0.0001 & 0.2768 --- 0.2777 \\[0.9mm]
7585 & 0.648 & 0.175 & 0.1784 $\pm$ 0.0005 & -4.89$^{+0.00}_{-0.00}$ & 3.740$^{+0.000}_{-0.000}$ & 22.4$^{+0.0}_{-0.0}$ & -4.96 --- -4.87 & 0.7364 $\pm$ 0.0010 & 0.7330 --- 0.7840 & 0.2877 $\pm$ 0.0002 & 0.2865 --- 0.2907 \\[0.9mm]
7734 & 0.690 & 0.275 & 0.2546 $\pm$ 0.0013 & -4.65$^{+0.01}_{-0.00}$ & 1.586$^{+0.000}_{-0.054}$ & 17.7$^{+0.0}_{-0.4}$ & -4.65 & 0.7855 $\pm$ 0.0011 & 0.7855 & 0.2980 $\pm$ 0.0002 & 0.2980 \\[0.9mm]
7918 & 0.618 & - & 0.1552 $\pm$ 0.0002 & -5.02$^{+0.00}_{-0.00}$ & 6.000$^{+0.000}_{-0.000}$ & 21.7$^{+0.0}_{-0.0}$ & -5.03 --- -5.02 & 0.7351 $\pm$ 0.0009 & 0.7351 --- 0.7466 & 0.2855 $\pm$ 0.0001 & 0.2844 --- 0.2855 \\[0.9mm]
7981 & 0.836 & 0.186 & 0.2077 $\pm$ 0.0022 & -4.88$^{+0.01}_{-0.01}$ & 3.604$^{+0.136}_{-0.131}$ & 38.2$^{+0.5}_{-0.5}$ & -4.91 --- -4.84 & 0.7473 $\pm$ 0.0045 & 0.7341 --- 0.7509 & 0.3107 $\pm$ 0.0002 & 0.3094 --- 0.3124 \\[0.9mm]
8159 & 0.720 & 0.149 & 0.1445 $\pm$ 0.0013 & -5.11$^{+0.01}_{-0.01}$ & 8.151$^{+0.269}_{-0.263}$ & 36.9$^{+0.3}_{-0.3}$ & -5.11 & 0.6900 $\pm$ 0.0012 & 0.6900 & 0.2889 $\pm$ 0.0002 & 0.2889 \\[0.9mm]
8362 & 0.804 & 0.271 & 0.2769 $\pm$ 0.0002 & -4.68$^{+0.00}_{-0.00}$ & 1.758$^{+0.000}_{-0.000}$ & 25.9$^{+0.0}_{-0.0}$ & -4.68 & 0.7650 $\pm$ 0.0009 & 0.7650 & 0.3092 $\pm$ 0.0005 & 0.3092 \\[0.9mm]
8486 & 0.654 & - & 0.3213 $\pm$ 0.0022 & -4.48$^{+0.00}_{-0.00}$ & 0.779$^{+0.000}_{-0.000}$ & 8.7$^{+0.0}_{-0.0}$ & -4.48 & 0.8760 $\pm$ 0.0008 & 0.8760 & 0.3166 $\pm$ 0.0006 & 0.3166 \\[0.9mm]
9349 & 0.660 & - & 0.2414 $\pm$ 0.0058 & -4.67$^{+0.02}_{-0.01}$ & 1.698$^{+0.059}_{-0.113}$ & 16.3$^{+0.4}_{-0.8}$ & -4.67 & 0.8016 $\pm$ 0.0003 & 0.8016 & 0.2941 $\pm$ 0.0002 & 0.2941 \\[0.9mm]
9406\sups{SG} & 0.960 & 0.133 & 0.1306 $\pm$ 0.0013 & -5.24$^{+0.01}_{-0.00}$ & 12.097$^{+0.000}_{-0.336}$ & 59.9$^{+0.0}_{-0.7}$ & -5.24 & 0.6440 $\pm$ 0.0005 & 0.6440 & 0.3092 $\pm$ 0.0001 & 0.3092 \\[0.9mm]
9829 & 0.662 & 0.163 & 0.1670 $\pm$ 0.0012 & -4.96$^{+0.01}_{-0.00}$ & 4.837$^{+0.000}_{-0.173}$ & 25.7$^{+0.0}_{-0.3}$ & -4.96 & 0.7839 $\pm$ 0.0026 & 0.7839 & 0.2902 $\pm$ 0.0007 & 0.2902 \\[0.9mm]
9911 & 0.624 & 0.150 & 0.1573 $\pm$ 0.0011 & -5.00$^{+0.01}_{-0.00}$ & 5.589$^{+0.000}_{-0.197}$ & 22.1$^{+0.0}_{-0.2}$ & -5.00 & 0.7388 $\pm$ 0.0007 & 0.7388 & 0.2847 $\pm$ 0.0003 & 0.2847 \\[0.9mm]
10339 & 0.700 & 0.390 & 0.3601 $\pm$ 0.0060 & -4.44$^{+0.00}_{-0.01}$ & 0.586$^{+0.049}_{-0.000}$ & 8.9$^{+0.4}_{-0.0}$ & -4.44 & 0.8810 $\pm$ 0.0005 & 0.8810 & 0.3174 $\pm$ 0.0000 & 0.3174 \\[0.9mm]
10505 & 0.690 & 0.169 & 0.1570 $\pm$ 0.0028 & -5.03$^{+0.02}_{-0.01}$ & 6.214$^{+0.221}_{-0.423}$ & 30.9$^{+0.3}_{-0.6}$ & -5.03 & 0.7200 $\pm$ 0.0013 & 0.7200 & 0.2926 $\pm$ 0.0005 & 0.2926 \\[0.9mm]
11548 & 0.591 & 0.142 & 0.1409 $\pm$ 0.0008 & -5.12$^{+0.00}_{-0.02}$ & 8.421$^{+0.557}_{-0.000}$ & 20.2$^{+0.4}_{-0.0}$ & -5.12 & 0.7584 $\pm$ 0.0003 & 0.7584 & 0.2790 $\pm$ 0.0001 & 0.2790 \\[0.9mm]
12048 & 0.670 & 0.145 & 0.1461 $\pm$ 0.0019 & -5.10$^{+0.01}_{-0.01}$ & 7.888$^{+0.263}_{-0.257}$ & 30.3$^{+0.3}_{-0.3}$ & -5.10 & 0.7067 $\pm$ 0.0013 & 0.7067 & 0.2860 $\pm$ 0.0003 & 0.2860 \\[0.9mm]
12114 & 0.918 & 0.266 & 0.2470 $\pm$ 0.0014 & -4.85$^{+0.01}_{-0.00}$ & 3.226$^{+0.000}_{-0.117}$ & 39.7$^{+0.0}_{-0.5}$ & -4.85 & 0.7167 $\pm$ 0.0013 & 0.7167 & 0.3357 $\pm$ 0.0002 & 0.3357 \\[0.9mm]
13702 & 0.471 & - & 0.1808 $\pm$ 0.0001 & -4.83$^{+0.00}_{-0.00}$ & 2.997$^{+0.000}_{-0.000}$ & 5.4$^{+0.0}_{-0.0}$ & -4.83 & 0.8029 $\pm$ 0.0012 & 0.8029 & 0.2848 $\pm$ 0.0001 & 0.2848 \\[0.9mm]
14150 & 0.696 & 0.185 & 0.1883 $\pm$ 0.0009 & -4.87$^{+0.01}_{-0.00}$ & 3.473$^{+0.000}_{-0.126}$ & 26.8$^{+0.0}_{-0.3}$ & -4.87 & 0.7347 $\pm$ 0.0007 & 0.7347 & 0.2938 $\pm$ 0.0005 & 0.2938 \\[0.9mm]
\hline
\end{tabular}
\end{center}
\end{table}
\end{landscape}

\addtocounter {table} {-1}

\begin{landscape}
\begin{table}
\scriptsize
\begin{center}
\caption{continued.} 
\begin{tabular}{lccccccccccc}
\hline
HIP & (B-V) & S-index & S-index & Log(R$^{\prime}_{\rm{HK}}$) & Chromospheric & Chromospheric & Log(R$^{\prime}_{\rm{HK}}$) & Ca\subs{IRT}-index & Ca\subs{IRT}-index & H\subs{\alpha}-index & H\subs{\alpha}-index\\
no. & (HIPP.) & (Wright) & (This work) & & age (Gyr) & period (d) & range & & range & & range\\
\hline
15457 & 0.681 & - & 0.3263 $\pm$ 0.0013 & -4.49$^{+0.01}_{-0.00}$ & 0.826$^{+0.000}_{-0.047}$ & 10.2$^{+0.0}_{-0.4}$ & -4.55 --- -4.45 & 0.8571 $\pm$ 0.0013 & 0.8214 --- 0.8649 & 0.3082 $\pm$ 0.0002 & 0.3037 --- 0.3103 \\[0.9mm]
15776\sups{SG} & 0.702 & 0.147 & 0.1472 $\pm$ 0.0002 & -5.09$^{+0.00}_{-0.01}$ & 7.630$^{+0.257}_{-0.000}$ & 34.1$^{+0.3}_{-0.0}$ & -5.09 & 0.7727 $\pm$ 0.0015 & 0.7727 & 0.2878 $\pm$ 0.0002 & 0.2878 \\[0.9mm]
16537 & 0.881 & 0.447 & 0.5357 $\pm$ 0.0019 & -4.42$^{+0.00}_{-0.00}$ & 0.489$^{+0.000}_{-0.000}$ & 11.8$^{+0.0}_{-0.0}$ & -4.55 --- -4.40 & 0.8566 $\pm$ 0.0017 & 0.8044 --- 0.8686 & 0.3446 $\pm$ 0.0004 & 0.3293 --- 0.3478 \\[0.9mm]
16641\sups{SG} & 0.891 & 0.132 & 0.1297 $\pm$ 0.0009 & -5.22$^{+0.01}_{-0.00}$ & 11.431$^{+0.000}_{-0.326}$ & 56.2$^{+0.0}_{-0.6}$ & -5.22 & 0.6744 $\pm$ 0.0004 & 0.6744 & 0.3019 $\pm$ 0.0001 & 0.3019 \\[0.9mm]
17027\sups{SG} & 0.921 & 0.142 & 0.1320 $\pm$ 0.0004 & -5.21$^{+0.00}_{-0.00}$ & 11.105$^{+0.000}_{-0.000}$ & 56.8$^{+0.0}_{-0.0}$ & -5.21 & 0.6556 $\pm$ 0.0002 & 0.6556 & 0.3047 $\pm$ 0.0003 & 0.3047 \\[0.9mm]
17147 & 0.554 & 0.163 & 0.1581 $\pm$ 0.0004 & -4.98$^{+0.00}_{-0.00}$ & 5.201$^{+0.000}_{-0.000}$ & 13.7$^{+0.0}_{-0.0}$ & -4.98 & 0.8453 $\pm$ 0.0009 & 0.8453 & 0.2873 $\pm$ 0.0001 & 0.2873 \\[0.9mm]
17183\sups{SG} & 0.954 & 0.145 & 0.1335 $\pm$ 0.0018 & -5.22$^{+0.01}_{-0.01}$ & 11.431$^{+0.331}_{-0.326}$ & 58.4$^{+0.6}_{-0.6}$ & -5.22 & 0.6623 $\pm$ 0.0013 & 0.6623 & 0.3160 $\pm$ 0.0003 & 0.3160 \\[0.9mm]
17378\sups{SG} & 0.915 & 0.136 & 0.1359 $\pm$ 0.0005 & -5.19$^{+0.00}_{-0.00}$ & 10.470$^{+0.000}_{-0.000}$ & 55.4$^{+0.0}_{-0.0}$ & -5.19 & 0.6396 $\pm$ 0.0040 & 0.6396 & 0.3075 $\pm$ 0.0005 & 0.3075 \\[0.9mm]
18106 & 0.583 & 0.148 & 0.1438 $\pm$ 0.0019 & -5.10$^{+0.01}_{-0.02}$ & 7.888$^{+0.533}_{-0.257}$ & 18.8$^{+0.3}_{-0.2}$ & -5.10 --- -5.08 & 0.7268 $\pm$ 0.0027 & 0.7240 --- 0.7337 & 0.2804 $\pm$ 0.0005 & 0.2794 --- 0.2810 \\[0.9mm]
18267 & 0.719 & 0.192 & 0.1792 $\pm$ 0.0037 & -4.92$^{+0.02}_{-0.02}$ & 4.178$^{+0.319}_{-0.297}$ & 30.8$^{+0.7}_{-0.7}$ & -4.92 & 0.7409 $\pm$ 0.0024 & 0.7409 & 0.2976 $\pm$ 0.0009 & 0.2976 \\[0.9mm]
18606\sups{SG} & 1.001 & 0.122 & 0.1229 $\pm$ 0.0018 & -5.29$^{+0.00}_{-0.01}$ & 13.844$^{+0.362}_{-0.000}$ & 64.7$^{+0.9}_{-0.0}$ & -5.29 & 0.6177 $\pm$ 0.0020 & 0.6177 & 0.3078 $\pm$ 0.0006 & 0.3078 \\[0.9mm]
19076 & 0.620 & - & 0.3002 $\pm$ 0.0007 & -4.50$^{+0.00}_{-0.00}$ & 0.873$^{+0.000}_{-0.000}$ & 7.9$^{+0.0}_{-0.0}$ & -4.50 & 0.8365 $\pm$ 0.0012 & 0.8365 & 0.3029 $\pm$ 0.0002 & 0.3029 \\[0.9mm]
19849 & 0.820 & 0.196 & 0.2070 $\pm$ 0.0005 & -4.87$^{+0.00}_{-0.00}$ & 3.473$^{+0.000}_{-0.000}$ & 36.8$^{+0.0}_{-0.0}$ & -4.87 & 0.7640 $\pm$ 0.0010 & 0.7640 & 0.3154 $\pm$ 0.0002 & 0.3154 \\[0.9mm]
19925 & 0.669 & - & 0.2111 $\pm$ 0.0157 & -4.77$^{+0.05}_{-0.06}$ & 2.410$^{+0.587}_{-0.391}$ & 20.8$^{+2.0}_{-1.8}$ & -4.77 & 0.7641 $\pm$ 0.0006 & 0.7641 & 0.2920 $\pm$ 0.0003 & 0.2920 \\[0.9mm]
20800 & 0.711 & 0.151 & 0.1510 $\pm$ 0.0020 & -5.08$^{+0.01}_{-0.01}$ & 7.379$^{+0.251}_{-0.245}$ & 34.9$^{+0.3}_{-0.3}$ & -5.08 & 0.6984 $\pm$ 0.0012 & 0.6984 & 0.2898 $\pm$ 0.0006 & 0.2898 \\[0.9mm]
22319\sups{SG} & 0.846 & 0.151 & 0.1692 $\pm$ 0.0010 & -5.02$^{+0.01}_{-0.00}$ & 6.000$^{+0.000}_{-0.209}$ & 44.6$^{+0.0}_{-0.4}$ & -5.02 & 0.7091 $\pm$ 0.0005 & 0.7091 & 0.2958 $\pm$ 0.0002 & 0.2958 \\[0.9mm]
22336 & 0.631 & 0.152 & 0.1434 $\pm$ 0.0006 & -5.11$^{+0.00}_{-0.01}$ & 8.151$^{+0.269}_{-0.000}$ & 25.3$^{+0.2}_{-0.0}$ & -5.11 & 0.6830 $\pm$ 0.0009 & 0.6830 & 0.2830 $\pm$ 0.0003 & 0.2830 \\[0.9mm]
22449 & 0.453 & 0.214 & 0.2016 $\pm$ 0.0003 & -4.72$^{+0.00}_{-0.01}$ & 2.019$^{+0.072}_{-0.000}$ & 3.7$^{+0.1}_{-0.0}$ & -4.72 & 0.8218 $\pm$ 0.0009 & 0.8218 & 0.2881 $\pm$ 0.0001 & 0.2881 \\
22633\sups{SG} & 0.875 & 0.134 & 0.1347 $\pm$ 0.0005 & -5.19$^{+0.00}_{-0.00}$ & 10.470$^{+0.000}_{-0.000}$ & 53.7$^{+0.0}_{-0.0}$ & -5.19 & 0.6889 $\pm$ 0.0003 & 0.6889 & 0.2962 $\pm$ 0.0004 & 0.2962 \\[0.9mm]
23311 & 1.049 & - & 0.3391 $\pm$ 0.0020 & -4.85$^{+0.00}_{-0.00}$ & 3.226$^{+0.000}_{-0.000}$ & 41.9$^{+0.0}_{-0.0}$ & -4.85 & 0.6882 $\pm$ 0.0005 & 0.6882 & 0.3419 $\pm$ 0.0002 & 0.3419 \\[0.9mm]
24813 & 0.630 & 0.151 & 0.1505 $\pm$ 0.0445 & -5.06$^{+0.25}_{-0.62}$ & 6.895$^{+22.188}_{-4.110}$ & 24.1$^{+58.4}_{-5.8}$ & -5.06 & 0.7222 $\pm$ 0.0480 & 0.7222 & 0.2839 $\pm$ 0.0141 & 0.2839 \\[0.9mm]
25278 & 0.544 & - & 0.3138 $\pm$ 0.0006 & -4.43$^{+0.00}_{-0.00}$ & 0.538$^{+0.000}_{-0.000}$ & 3.5$^{+0.0}_{-0.0}$ & -4.47 --- -4.41 & 0.8936 $\pm$ 0.0010 & 0.8734 --- 0.9029 & 0.3104 $\pm$ 0.0003 & 0.3063 --- 0.3125 \\[0.9mm]
25486 & 0.538 & 0.445 & 0.4611 $\pm$ 0.0023 & -4.20$^{+0.00}_{-0.01}$ & 0.000$^{+0.000}_{-0.000}$ & 0.8$^{+0.1}_{-0.0}$ & -4.24 --- -4.18 & 1.0596 $\pm$ 0.0011 & 1.0344 --- 1.0658 & 0.3624 $\pm$ 0.0007 & 0.3408 --- 0.3624 \\[0.9mm]
27913 & 0.594 & - & 0.3193 $\pm$ 0.0005 & -4.44$^{+0.00}_{-0.00}$ & 0.586$^{+0.000}_{-0.000}$ & 5.3$^{+0.0}_{-0.0}$ & -4.47 --- -4.38 & 0.8777 $\pm$ 0.0007 & 0.8751 --- 0.9023 & 0.3094 $\pm$ 0.0002 & 0.3078 --- 0.3163 \\[0.9mm]
29568 & 0.713 & 0.400 & 0.3809 $\pm$ 0.0181 & -4.42$^{+0.02}_{-0.03}$ & 0.489$^{+0.146}_{-0.095}$ & 8.5$^{+1.3}_{-0.8}$ & -4.42 & 0.9000 $\pm$ 0.0012 & 0.9000 & 0.3214 $\pm$ 0.0003 & 0.3214 \\[0.9mm]
30476 & 0.673 & - & 0.1507 $\pm$ 0.0071 & -5.07$^{+0.05}_{-0.05}$ & 7.134$^{+1.287}_{-1.134}$ & 29.9$^{+1.3}_{-1.3}$ & -5.07 & 0.7294 $\pm$ 0.0003 & 0.7294 & 0.2886 $\pm$ 0.0001 & 0.2886 \\[0.9mm]
31965 & 0.672 & - & 0.1513 $\pm$ 0.0074 & -5.07$^{+0.05}_{-0.05}$ & 7.134$^{+1.287}_{-1.134}$ & 29.7$^{+1.3}_{-1.3}$ & -5.07 & 0.7408 $\pm$ 0.0004 & 0.7408 & 0.2853 $\pm$ 0.0003 & 0.2853 \\[0.9mm]
32673 & 0.677 & - & 0.1852 $\pm$ 0.0067 & -4.87$^{+0.03}_{-0.03}$ & 3.473$^{+0.407}_{-0.364}$ & 24.9$^{+0.9}_{-1.0}$ & -4.87 & 0.7529 $\pm$ 0.0028 & 0.7529 & 0.2908 $\pm$ 0.0003 & 0.2908 \\[0.9mm]
32851 & 0.396 & - & 0.2269 $\pm$ 0.0005 & -4.65$^{+0.00}_{-0.00}$ & 1.586$^{+0.000}_{-0.000}$ & 1.6$^{+0.0}_{-0.0}$ & -4.65 & 0.9142 $\pm$ 0.0008 & 0.9142 & 0.3005 $\pm$ 0.0001 & 0.3005 \\[0.9mm]
33277 & 0.573 & 0.162 & 0.1648 $\pm$ 0.0024 & -4.94$^{+0.02}_{-0.01}$ & 4.496$^{+0.168}_{-0.319}$ & 15.1$^{+0.2}_{-0.3}$ & -4.94 --- -4.92 & 0.7781 $\pm$ 0.0037 & 0.7781 --- 0.7852 & 0.2845 $\pm$ 0.0011 & 0.2831 --- 0.2845 \\[0.9mm]
35185 & 0.641 & - & 0.3344 $\pm$ 0.0066 & -4.45$^{+0.02}_{-0.01}$ & 0.635$^{+0.048}_{-0.097}$ & 7.2$^{+0.3}_{-0.6}$ & -4.45 & 0.8773 $\pm$ 0.0022 & 0.8773 & 0.3110 $\pm$ 0.0003 & 0.3110 \\[0.9mm]
35265 & 0.631 & 0.175 & 0.1758 $\pm$ 0.0010 & -4.90$^{+0.00}_{-0.01}$ & 3.880$^{+0.146}_{-0.000}$ & 20.7$^{+0.2}_{-0.0}$ & -4.93 --- -4.86 & 0.7616 $\pm$ 0.0009 & 0.7557 --- 0.7763 & 0.2865 $\pm$ 0.0002 & 0.2849 --- 0.2892 \\[0.9mm]
36704 & 0.863 & 0.565 & 0.5686 $\pm$ 0.0037 & -4.37$^{+0.00}_{-0.00}$ & 0.261$^{+0.000}_{-0.000}$ & 9.0$^{+0.0}_{-0.0}$ & -4.37 & 0.8660 $\pm$ 0.0025 & 0.8660 & 0.3443 $\pm$ 0.0004 & 0.3443 \\[0.9mm]
38018 & 0.712 & 0.241 & 0.2547 $\pm$ 0.0021 & -4.66$^{+0.00}_{-0.01}$ & 1.641$^{+0.057}_{-0.000}$ & 19.6$^{+0.5}_{-0.0}$ & -4.66 & 0.7873 $\pm$ 0.0008 & 0.7873 & 0.3074 $\pm$ 0.0004 & 0.3074 \\[0.9mm]
38228 & 0.682 & 0.387 & 0.3683 $\pm$ 0.0015 & -4.42$^{+0.01}_{-0.00}$ & 0.489$^{+0.000}_{-0.048}$ & 7.6$^{+0.0}_{-0.4}$ & -4.42 & 0.8838 $\pm$ 0.0014 & 0.8838 & 0.3170 $\pm$ 0.0002 & 0.3170 \\[0.9mm]
38647 & 0.659 & 0.241 & 0.2701 $\pm$ 0.0040 & -4.59$^{+0.01}_{-0.01}$ & 1.284$^{+0.047}_{-0.047}$ & 13.1$^{+0.4}_{-0.4}$ & -4.59 & 0.8160 $\pm$ 0.0013 & 0.8160 & 0.2983 $\pm$ 0.0002 & 0.2983 \\[0.9mm]
38747 & 0.671 & - & 0.3717 $\pm$ 0.0187 & -4.40$^{+0.02}_{-0.03}$ & 0.393$^{+0.144}_{-0.091}$ & 6.6$^{+1.0}_{-0.6}$ & -4.40 & 0.8921 $\pm$ 0.0016 & 0.8921 & 0.3149 $\pm$ 0.0003 & 0.3149 \\[0.9mm]
41484 & 0.624 & 0.165 & 0.1704 $\pm$ 0.0035 & -4.92$^{+0.02}_{-0.02}$ & 4.178$^{+0.319}_{-0.297}$ & 20.4$^{+0.4}_{-0.4}$ & -4.97 --- -4.91 & 0.7401 $\pm$ 0.0002 & 0.7398 --- 0.7655 & 0.2850 $\pm$ 0.0002 & 0.2850 --- 0.2861 \\[0.9mm]
41526 & 0.639 & - & 0.1575 $\pm$ 0.0031 & -5.00$^{+0.02}_{-0.02}$ & 5.589$^{+0.411}_{-0.387}$ & 23.9$^{+0.4}_{-0.4}$ & -5.00 & 0.7598 $\pm$ 0.0006 & 0.7598 & 0.2844 $\pm$ 0.0002 & 0.2844 \\[0.9mm]
41844 & 0.630 & 0.159 & 0.1476 $\pm$ 0.0114 & -5.08$^{+0.08}_{-0.10}$ & 7.379$^{+2.781}_{-1.791}$ & 24.5$^{+2.3}_{-1.7}$ & -5.08 & 0.7555 $\pm$ 0.0004 & 0.7555 & 0.2824 $\pm$ 0.0001 & 0.2824 \\[0.9mm]
\hline
\end{tabular}
\end{center}
\end{table}
\end{landscape}

\addtocounter {table} {-1}

\begin{landscape}
\begin{table}
\scriptsize
\begin{center}
\caption{continued.} 
\begin{tabular}{lccccccccccc}
\hline
HIP & (B-V) & S-index & S-index & Log(R$^{\prime}_{\rm{HK}}$) & Chromospheric & Chromospheric & Log(R$^{\prime}_{\rm{HK}}$) & Ca\subs{IRT}-index & Ca\subs{IRT}-index & H\subs{\alpha}-index & H\subs{\alpha}-index\\
no. & (HIPP.) & (Wright) & (This work) & & age (Gyr) & period (d) & range & & range & & range\\
\hline
42333 & 0.655 & 0.315 & 0.2580 $\pm$ 0.0008 & -4.62$^{+0.00}_{-0.00}$ & 1.429$^{+0.000}_{-0.000}$ & 14.0$^{+0.0}_{-0.0}$ & -4.64 --- -4.52 & 0.7971 $\pm$ 0.0021 & 0.7791 --- 0.8278 & 0.2965 $\pm$ 0.0006 & 0.2947 --- 0.3019 \\[0.9mm]
42403 & 0.547 & 0.207 & 0.2225 $\pm$ 0.0014 & -4.66$^{+0.00}_{-0.01}$ & 1.641$^{+0.057}_{-0.000}$ & 7.9$^{+0.2}_{-0.0}$ & -4.66 & 0.7795 $\pm$ 0.0028 & 0.7795 & 0.2877 $\pm$ 0.0019 & 0.2877 \\[0.9mm]
42438 & 0.618 & 0.349 & 0.3838 $\pm$ 0.0015 & -4.35$^{+0.00}_{-0.00}$ & 0.184$^{+0.000}_{-0.000}$ & 3.9$^{+0.0}_{-0.0}$ & -4.38 --- -4.34 & 0.9297 $\pm$ 0.0010 & 0.9206 --- 0.9347 & 0.3217 $\pm$ 0.0006 & 0.3178 --- 0.3231 \\[0.9mm]
43410 & 0.549 & 0.290 & 0.2782 $\pm$ 0.0019 & -4.51$^{+0.01}_{-0.00}$ & 0.919$^{+0.000}_{-0.000}$ & 5.1$^{+0.0}_{-0.2}$ & -4.55 --- -4.49 & 0.8737 $\pm$ 0.0013 & 0.8284 --- 0.8779 & 0.3009 $\pm$ 0.0003 & 0.2957 --- 0.3024 \\[0.9mm]
43557 & 0.640 & - & 0.2195 $\pm$ 0.0029 & -4.72$^{+0.01}_{-0.02}$ & 2.019$^{+0.147}_{-0.069}$ & 16.5$^{+0.7}_{-0.3}$ & -4.72 & 0.8006 $\pm$ 0.0012 & 0.8006 & 0.2944 $\pm$ 0.0002 & 0.2944 \\[0.9mm]
43726 & 0.661 & - & 0.2202 $\pm$ 0.0019 & -4.73$^{+0.00}_{-0.01}$ & 2.091$^{+0.075}_{-0.000}$ & 18.6$^{+0.4}_{-0.0}$ & -4.75 --- -4.63 & 0.7618 $\pm$ 0.0010 & 0.7447 --- 0.7917 & 0.2924 $\pm$ 0.0003 & 0.2904 --- 0.2959 \\[0.9mm]
44897 & 0.585 & - & 0.2736 $\pm$ 0.0009 & -4.54$^{+0.00}_{-0.00}$ & 1.055$^{+0.000}_{-0.000}$ & 7.4$^{+0.0}_{-0.0}$ & -4.63 --- -4.53 & 0.8463 $\pm$ 0.0017 & 0.7990 --- 0.8477 & 0.2992 $\pm$ 0.0001 & 0.2915 --- 0.2993 \\[0.9mm]
44997 & 0.665 & - & 0.1663 $\pm$ 0.0035 & -4.97$^{+0.02}_{-0.02}$ & 5.016$^{+0.376}_{-0.352}$ & 26.4$^{+0.5}_{-0.5}$ & -4.97 & 0.7574 $\pm$ 0.0009 & 0.7574 & 0.2900 $\pm$ 0.0001 & 0.2900 \\[0.9mm]
46066 & 0.678 & - & 0.2043 $\pm$ 0.0128 & -4.80$^{+0.05}_{-0.05}$ & 2.686$^{+0.540}_{-0.442}$ & 22.7$^{+1.7}_{-1.8}$ & -4.80 & 0.7724 $\pm$ 0.0012 & 0.7724 & 0.2933 $\pm$ 0.0004 & 0.2933 \\[0.9mm]
46580 & 1.002 & 0.662 & 0.6574 $\pm$ 0.0033 & -4.49$^{+0.01}_{-0.00}$ & 0.826$^{+0.000}_{-0.047}$ & 17.1$^{+0.0}_{-0.7}$ & -4.49 & 0.8249 $\pm$ 0.0005 & 0.8249 & 0.3543 $\pm$ 0.0001 & 0.3543 \\[0.9mm]
46903 & 0.670 & - & 0.1919 $\pm$ 0.0055 & -4.84$^{+0.02}_{-0.03}$ & 3.109$^{+0.364}_{-0.220}$ & 23.2$^{+0.9}_{-0.6}$ & -4.84 & 0.7709 $\pm$ 0.0011 & 0.7709 & 0.2901 $\pm$ 0.0002 & 0.2901 \\[0.9mm]
49081 & 0.676 & 0.152 & 0.1543 $\pm$ 0.0009 & -5.04$^{+0.00}_{-0.01}$ & 6.435$^{+0.227}_{-0.000}$ & 29.5$^{+0.3}_{-0.0}$ & -5.04 & 0.7055 $\pm$ 0.0019 & 0.7055 & 0.2906 $\pm$ 0.0002 & 0.2906 \\[0.9mm]
49350 & 0.689 & 0.185 & 0.1908 $\pm$ 0.0048 & -4.86$^{+0.02}_{-0.02}$ & 3.347$^{+0.257}_{-0.238}$ & 25.8$^{+0.7}_{-0.7}$ & -4.86 & 0.7571 $\pm$ 0.0008 & 0.7571 & 0.2936 $\pm$ 0.0002 & 0.2936 \\[0.9mm]
49580 & 0.670 & - & 0.1833 $\pm$ 0.0062 & -4.88$^{+0.03}_{-0.03}$ & 3.604$^{+0.423}_{-0.378}$ & 24.5$^{+0.9}_{-0.9}$ & -4.88 & 0.7854 $\pm$ 0.0007 & 0.7854 & 0.2905 $\pm$ 0.0002 & 0.2905 \\[0.9mm]
49728 & 0.649 & - & 0.1572 $\pm$ 0.0057 & -5.01$^{+0.04}_{-0.04}$ & 5.791$^{+0.871}_{-0.775}$ & 25.4$^{+0.9}_{-0.9}$ & -5.01 & 0.7718 $\pm$ 0.0013 & 0.7718 & 0.2878 $\pm$ 0.0001 & 0.2878 \\[0.9mm]
49756 & 0.647 & 0.163 & 0.1701 $\pm$ 0.0037 & -4.94$^{+0.02}_{-0.02}$ & 4.496$^{+0.341}_{-0.319}$ & 23.5$^{+0.5}_{-0.5}$ & -4.95 --- -4.94 & 0.7386 $\pm$ 0.0012 & 0.7386 --- 0.7399 & 0.2872 $\pm$ 0.0002 & 0.2872 --- 0.2878 \\[0.9mm]
49908 & 1.371 & - & 1.4390 $\pm$ 0.0063 & -4.73$^{+0.00}_{-0.00}$ & 2.091$^{+0.000}_{-0.000}$ & 38.3$^{+0.0}_{-0.0}$ & -4.73 & 0.8138 $\pm$ 0.0019 & 0.8138 & 0.3908 $\pm$ 0.0003 & 0.3908 \\[0.9mm]
50316 & 0.635 & 0.148 & 0.1528 $\pm$ 0.0040 & -5.04$^{+0.03}_{-0.03}$ & 6.435$^{+0.699}_{-0.644}$ & 24.3$^{+0.6}_{-0.6}$ & -5.04 & 0.7280 $\pm$ 0.0113 & 0.7280 & 0.2898 $\pm$ 0.0016 & 0.2898 \\[0.9mm]
50505 & 0.653 & 0.169 & 0.1797 $\pm$ 0.0011 & -4.89$^{+0.01}_{-0.00}$ & 3.740$^{+0.000}_{-0.136}$ & 22.9$^{+0.0}_{-0.3}$ & -4.91 --- -4.89 & 0.7985 $\pm$ 0.0014 & 0.7958 --- 0.7985 & 0.2923 $\pm$ 0.0003 & 0.2918 --- 0.2923 \\[0.9mm]
53721 & 0.624 & 0.154 & 0.1501 $\pm$ 0.0014 & -5.05$^{+0.01}_{-0.01}$ & 6.662$^{+0.233}_{-0.227}$ & 23.1$^{+0.2}_{-0.2}$ & -5.05 & 0.7519 $\pm$ 0.0014 & 0.7519 & 0.2838 $\pm$ 0.0001 & 0.2838 \\[0.9mm]
54952 & 1.043 & 0.757 & 0.6596 $\pm$ 0.0005 & -4.55$^{+0.01}_{-0.00}$ & 1.100$^{+0.000}_{-0.045}$ & 21.6$^{+0.0}_{-0.7}$ & -4.55 & 0.8133 $\pm$ 0.0005 & 0.8133 & 0.3586 $\pm$ 0.0003 & 0.3586 \\[0.9mm]
55459 & 0.642 & 0.159 & 0.1460 $\pm$ 0.0033 & -5.09$^{+0.02}_{-0.03}$ & 7.630$^{+0.790}_{-0.496}$ & 26.3$^{+0.7}_{-0.5}$ & -5.09 --- -5.04 & 0.7502 $\pm$ 0.0003 & 0.7361 --- 0.7502 & 0.2867 $\pm$ 0.0003 & 0.2862 --- 0.2867 \\[0.9mm]
56242 & 0.570 & 0.162 & 0.1682 $\pm$ 0.0009 & -4.92$^{+0.00}_{-0.00}$ & 4.178$^{+0.000}_{-0.000}$ & 14.5$^{+0.0}_{-0.0}$ & -4.92 & 0.7703 $\pm$ 0.0004 & 0.7703 & 0.2826 $\pm$ 0.0003 & 0.2826 \\[0.9mm]
56948 & 0.647 & - & 0.1866 $\pm$ 0.0215 & -4.86$^{+0.09}_{-0.11}$ & 3.347$^{+1.669}_{-0.937}$ & 21.4$^{+2.8}_{-2.7}$ & -4.88 --- -4.86 & 0.7520 $\pm$ 0.0014 & 0.7520 --- 0.7548 & 0.2883 $\pm$ 0.0003 & 0.2880 --- 0.2883 \\[0.9mm]
56997 & 0.723 & 0.309 & 0.3677 $\pm$ 0.0013 & -4.45$^{+0.00}_{-0.00}$ & 0.635$^{+0.000}_{-0.000}$ & 10.1$^{+0.0}_{-0.0}$ & -4.57 --- -4.42 & 0.8621 $\pm$ 0.0009 & 0.8275 --- 0.8945 & 0.3185 $\pm$ 0.0004 & 0.3077 --- 0.3247 \\[0.9mm]
57939 & 0.754 & 0.200 & 0.1964 $\pm$ 0.0041 & -4.87$^{+0.01}_{-0.02}$ & 3.473$^{+0.267}_{-0.126}$ & 32.2$^{+0.8}_{-0.4}$ & -4.87 & 0.8310 $\pm$ 0.0007 & 0.8310 & 0.3108 $\pm$ 0.0004 & 0.3108 \\[0.9mm]
58708\sups{SG} & 0.860 & 0.126 & 0.1241 $\pm$ 0.0016 & -5.25$^{+0.01}_{-0.01}$ & 12.438$^{+0.345}_{-0.340}$ & 56.4$^{+0.7}_{-0.6}$ & -5.25 & 0.7023 $\pm$ 0.0010 & 0.7023 & 0.3022 $\pm$ 0.0001 & 0.3022 \\[0.9mm]
60098 & 0.523 & 0.138 & 0.1336 $\pm$ 0.0004 & -5.19$^{+0.01}_{-0.00}$ & 10.470$^{+0.000}_{-0.310}$ & 12.9$^{+0.0}_{-0.1}$ & -5.19 & 0.7065 $\pm$ 0.0003 & 0.7065 & 0.2750 $\pm$ 0.0003 & 0.2750 \\[0.9mm]
60353 & 0.567 & 0.167 & 0.1625 $\pm$ 0.0003 & -4.95$^{+0.00}_{-0.00}$ & 4.664$^{+0.000}_{-0.000}$ & 14.6$^{+0.0}_{-0.0}$ & -4.95 & 0.7467 $\pm$ 0.0007 & 0.7467 & 0.2789 $\pm$ 0.0003 & 0.2789 \\[0.9mm]
61901 & 1.109 & 0.347 & 0.3422 $\pm$ 0.0058 & -4.94$^{+0.01}_{-0.01}$ & 4.496$^{+0.168}_{-0.162}$ & 47.4$^{+0.5}_{-0.5}$ & -4.94 & 0.6985 $\pm$ 0.0021 & 0.6985 & 0.3559 $\pm$ 0.0001 & 0.3559 \\[0.9mm]
62523 & 0.703 & 0.290 & 0.2864 $\pm$ 0.0009 & -4.58$^{+0.00}_{-0.01}$ & 1.237$^{+0.047}_{-0.000}$ & 15.3$^{+0.5}_{-0.0}$ & -4.58 & 0.8162 $\pm$ 0.0004 & 0.8162 & 0.3031 $\pm$ 0.0002 & 0.3031 \\[0.9mm]
64797 & 0.926 & 0.446 & 0.5418 $\pm$ 0.0007 & -4.47$^{+0.00}_{-0.00}$ & 0.731$^{+0.000}_{-0.000}$ & 15.2$^{+0.0}_{-0.0}$ & -4.47 & 0.8620 $\pm$ 0.0038 & 0.8620 & 0.3528 $\pm$ 0.0007 & 0.3528 \\[0.9mm]
65347 & 0.694 & - & 0.3723 $\pm$ 0.0106 & -4.42$^{+0.01}_{-0.02}$ & 0.489$^{+0.097}_{-0.048}$ & 7.9$^{+0.8}_{-0.4}$ & -4.42 & 0.8825 $\pm$ 0.0017 & 0.8825 & 0.3113 $\pm$ 0.0003 & 0.3113 \\[0.9mm]
66147 & 1.026 & 0.565 & 0.5811 $\pm$ 0.0052 & -4.58$^{+0.01}_{-0.00}$ & 1.237$^{+0.000}_{-0.046}$ & 23.6$^{+0.0}_{-0.7}$ & -4.58 & 0.7982 $\pm$ 0.0012 & 0.7982 & 0.3461 $\pm$ 0.0006 & 0.3461 \\[0.9mm]
66774\sups{SG} & 0.847 & - & 0.1153 $\pm$ 0.0058 & -5.30$^{+0.04}_{-0.05}$ & 14.206$^{+1.863}_{-1.423}$ & 59.2$^{+4.5}_{-2.9}$ & -5.30 & 0.7048 $\pm$ 0.0020 & 0.7048 & 0.3049 $\pm$ 0.0002 & 0.3049 \\
67275 & 0.508 & 0.202 & 0.1760 $\pm$ 0.0001 & -4.86$^{+0.00}_{-0.00}$ & 3.347$^{+0.000}_{-0.000}$ & 8.2$^{+0.0}_{-0.0}$ & -4.86 --- -4.84 & 0.7431 $\pm$ 0.0003 & 0.7431 --- 0.7520 & 0.2780 $\pm$ 0.0001 & 0.2780 --- 0.2795 \\[0.9mm]
67422 & 1.110 & 0.736 & 0.7449 $\pm$ 0.1317 & -4.59$^{+0.07}_{-0.09}$ & 1.284$^{+0.474}_{-0.320}$ & 25.0$^{+6.7}_{-5.2}$ & -4.59 & 0.8244 $\pm$ 0.0013 & 0.8244 & 0.3666 $\pm$ 0.0014 & 0.3666 \\[0.9mm]
68184 & 1.040 & 0.235 & 0.2349 $\pm$ 0.0014 & -5.02$^{+0.01}_{-0.00}$ & 6.000$^{+0.000}_{-0.209}$ & 50.1$^{+0.0}_{-0.5}$ & -5.02 & 0.6875 $\pm$ 0.0009 & 0.6875 & 0.3365 $\pm$ 0.0001 & 0.3365 \\[0.9mm]
71181 & 0.997 & 0.446 & 0.3807 $\pm$ 0.0021 & -4.73$^{+0.00}_{-0.00}$ & 2.091$^{+0.000}_{-0.000}$ & 33.9$^{+0.0}_{-0.0}$ & -4.73 & 0.7569 $\pm$ 0.0013 & 0.7569 & 0.3408 $\pm$ 0.0002 & 0.3408 \\[0.9mm]
71631 & 0.626 & 0.530 & 0.5838 $\pm$ 0.0018 & -4.13$^{+0.00}_{-0.00}$ & 0.000$^{+0.000}_{-0.000}$ & 0.8$^{+0.0}_{-0.0}$ & -4.25 --- -4.07 & 1.0759 $\pm$ 0.0024 & 1.0319 --- 1.0976 & 0.3881 $\pm$ 0.0006 & 0.3777 --- 0.4049 \\[0.9mm]
\hline
\end{tabular}
\end{center}
\end{table}
\end{landscape}

\addtocounter {table} {-1}

\begin{landscape}
\begin{table}
\scriptsize
\begin{center}
\caption{continued.} 
\begin{tabular}{lccccccccccc}
\hline
HIP & (B-V) & S-index & S-index & Log(R$^{\prime}_{\rm{HK}}$) & Chromospheric & Chromospheric & Log(R$^{\prime}_{\rm{HK}}$) & Ca\subs{IRT}-index & Ca\subs{IRT}-index & H\subs{\alpha}-index & H\subs{\alpha}-index\\
no. & (HIPP.) & (Wright) & (This work) & & age (Gyr) & period (d) & range & & range & & range\\
\hline
72848 & 0.841 & - & 0.5039 $\pm$ 0.0028 & -4.40$^{+0.00}_{-0.00}$ & 0.393$^{+0.000}_{-0.000}$ & 10.3$^{+0.0}_{-0.0}$ & -4.40 & 0.8848 $\pm$ 0.0022 & 0.8848 & 0.3347 $\pm$ 0.0006 & 0.3347 \\[0.9mm]
74432 & 0.680 & 0.145 & 0.1466 $\pm$ 0.0027 & -5.10$^{+0.02}_{-0.01}$ & 7.888$^{+0.263}_{-0.509}$ & 31.6$^{+0.3}_{-0.6}$ & -5.10 & 0.7200 $\pm$ 0.0008 & 0.7200 & 0.2908 $\pm$ 0.0003 & 0.2908 \\[0.9mm]
76114 & 0.656 & 0.160 & 0.1614 $\pm$ 0.0044 & -4.99$^{+0.03}_{-0.03}$ & 5.392$^{+0.608}_{-0.555}$ & 25.8$^{+0.7}_{-0.7}$ & -4.99 --- -4.97 & 0.7537 $\pm$ 0.0010 & 0.7537 --- 0.7550 & 0.2897 $\pm$ 0.0002 & 0.2894 --- 0.2897 \\[0.9mm]
79578 & 0.646 & - & 0.2114 $\pm$ 0.0020 & -4.75$^{+0.01}_{-0.01}$ & 2.244$^{+0.081}_{-0.078}$ & 18.0$^{+0.3}_{-0.3}$ & -4.76 --- -4.73 & 0.7801 $\pm$ 0.0016 & 0.7594 --- 0.7801 & 0.2932 $\pm$ 0.0002 & 0.2913 --- 0.2932 \\[0.9mm]
79672 & 0.652 & 0.169 & 0.1834 $\pm$ 0.0028 & -4.87$^{+0.01}_{-0.02}$ & 3.473$^{+0.267}_{-0.126}$ & 22.2$^{+0.5}_{-0.3}$ & -4.99 --- -4.84 & 0.7483 $\pm$ 0.0016 & 0.7324 --- 0.7656 & 0.2870 $\pm$ 0.0012 & 0.2835 --- 0.2890 \\[0.9mm]
81300 & 0.827 & 0.355 & 0.3577 $\pm$ 0.0035 & -4.56$^{+0.00}_{-0.01}$ & 1.146$^{+0.046}_{-0.000}$ & 19.4$^{+0.6}_{-0.0}$ & -4.56 & 0.8246 $\pm$ 0.0011 & 0.8246 & 0.3203 $\pm$ 0.0003 & 0.3203 \\[0.9mm]
82588 & 0.749 & 0.392 & 0.3859 $\pm$ 0.0380 & -4.45$^{+0.05}_{-0.06}$ & 0.635$^{+0.284}_{-0.242}$ & 10.9$^{+3.0}_{-2.3}$ & -4.45 & 0.8753 $\pm$ 0.0029 & 0.8753 & 0.3232 $\pm$ 0.0003 & 0.3232 \\[0.9mm]
86193 & 0.693 & 0.180 & 0.1781 $\pm$ 0.0055 & -4.91$^{+0.02}_{-0.03}$ & 4.026$^{+0.470}_{-0.287}$ & 27.8$^{+0.9}_{-0.6}$ & -4.91 & 0.7229 $\pm$ 0.0002 & 0.7229 & 0.2894 $\pm$ 0.0001 & 0.2894 \\[0.9mm]
86400 & 0.959 & 0.284 & 0.3082 $\pm$ 0.0016 & -4.78$^{+0.00}_{-0.01}$ & 2.498$^{+0.092}_{-0.000}$ & 36.6$^{+0.6}_{-0.0}$ & -4.78 & 0.7594 $\pm$ 0.0007 & 0.7594 & 0.3333 $\pm$ 0.0004 & 0.3333 \\[0.9mm]
86974 & 0.750 & 0.145 & 0.1323 $\pm$ 0.0035 & -5.20$^{+0.03}_{-0.03}$ & 10.785$^{+0.977}_{-0.929}$ & 43.9$^{+1.4}_{-1.3}$ & -5.20 & 0.6637 $\pm$ 0.0004 & 0.6637 & 0.2909 $\pm$ 0.0001 & 0.2909 \\[0.9mm]
88194 & 0.635 & 0.159 & 0.1599 $\pm$ 0.0055 & -4.99$^{+0.03}_{-0.04}$ & 5.392$^{+0.822}_{-0.555}$ & 23.2$^{+0.9}_{-0.7}$ & -5.00 --- -4.98 & 0.7483 $\pm$ 0.0008 & 0.7483 --- 0.7523 & 0.2870 $\pm$ 0.0003 & 0.2870 --- 0.2885 \\[0.9mm]
88945 & 0.633 & 0.450 & 0.3907 $\pm$ 0.0009 & -4.35$^{+0.00}_{-0.00}$ & 0.184$^{+0.000}_{-0.000}$ & 4.2$^{+0.0}_{-0.0}$ & -4.38 --- -4.30 & 0.9455 $\pm$ 0.0006 & 0.9188 --- 0.9602 & 0.3270 $\pm$ 0.0003 & 0.3205 --- 0.3303 \\[0.9mm]
88972 & 0.876 & 0.186 & 0.1788 $\pm$ 0.0024 & -5.00$^{+0.01}_{-0.00}$ & 5.589$^{+0.000}_{-0.197}$ & 45.3$^{+0.0}_{-0.4}$ & -5.00 & 0.7316 $\pm$ 0.0006 & 0.7316 & 0.3208 $\pm$ 0.0004 & 0.3208 \\[0.9mm]
90729\sups{SG} & 0.795 & 0.136 & 0.1387 $\pm$ 0.0028 & -5.16$^{+0.02}_{-0.02}$ & 9.557$^{+0.000}_{-0.197}$ & 46.7$^{+0.9}_{-0.9}$ & -5.16 & 0.6823 $\pm$ 0.0004 & 0.6823 & 0.2969 $\pm$ 0.0001 & 0.2969 \\[0.9mm]
91043 & 0.649 & - & 0.5489 $\pm$ 0.0027 & -4.17$^{+0.00}_{-0.01}$ & 0.000$^{+0.000}_{-0.000}$ & 1.3$^{+0.1}_{-0.0}$ & -4.18 --- -4.07 & 1.1101 $\pm$ 0.0019 & 1.0930 --- 1.1714 & 0.3773 $\pm$ 0.0003 & 0.3751 --- 0.4087 \\[0.9mm]
92984 & 0.583 & 0.331 & 0.3162 $\pm$ 0.0000 & -4.45$^{+0.00}_{-0.00}$ & 0.635$^{+0.000}_{-0.000}$ & 5.1$^{+0.0}_{-0.0}$ & -4.46 --- -4.44 & 0.9018 $\pm$ 0.0000 & 0.8915 --- 0.9076 & 0.3094 $\pm$ 0.0000 & 0.3064 --- 0.3138 \\[0.9mm]
95253 & 0.458 & - & 0.2147 $\pm$ 0.0008 & -4.68$^{+0.00}_{-0.01}$ & 1.758$^{+0.062}_{-0.000}$ & 3.6$^{+0.1}_{-0.0}$ & -4.68 --- -4.67 & 0.8376 $\pm$ 0.0008 & 0.8376 --- 0.8433 & 0.2914 $\pm$ 0.0002 & 0.2914 --- 0.2929 \\[0.9mm]
95962 & 0.640 & 0.160 & 0.1544 $\pm$ 0.0040 & -5.03$^{+0.03}_{-0.03}$ & 6.214$^{+0.680}_{-0.626}$ & 24.7$^{+0.7}_{-0.7}$ & -5.03 --- -4.97 & 0.7325 $\pm$ 0.0011 & 0.7325 --- 0.7378 & 0.2876 $\pm$ 0.0006 & 0.2867 --- 0.2876 \\[0.9mm]
96085 & 0.922 & 0.546 & 0.5328 $\pm$ 0.0063 & -4.47$^{+0.00}_{-0.01}$ & 0.731$^{+0.048}_{-0.000}$ & 15.2$^{+0.6}_{-0.0}$ & -4.47 & 0.8507 $\pm$ 0.0014 & 0.8507 & 0.3442 $\pm$ 0.0004 & 0.3442 \\[0.9mm]
96100 & 0.786 & 0.206 & 0.2050 $\pm$ 0.0017 & -4.85$^{+0.00}_{-0.01}$ & 3.226$^{+0.121}_{-0.000}$ & 33.8$^{+0.4}_{-0.0}$ & -4.85 --- -4.81 & 0.7529 $\pm$ 0.0007 & 0.7509 --- 0.7760 & 0.3027 $\pm$ 0.0002 & 0.3027 --- 0.3087 \\[0.9mm]
96895 & 0.643 & 0.145 & 0.1556 $\pm$ 0.0011 & -5.02$^{+0.01}_{-0.00}$ & 6.000$^{+0.000}_{-0.209}$ & 24.9$^{+0.0}_{-0.2}$ & -5.06 --- -5.02 & 0.7489 $\pm$ 0.0003 & 0.7382 --- 0.7489 & 0.2882 $\pm$ 0.0002 & 0.2869 --- 0.2882 \\[0.9mm]
96901 & 0.661 & 0.148 & 0.1537 $\pm$ 0.0005 & -5.04$^{+0.00}_{-0.00}$ & 6.435$^{+0.000}_{-0.000}$ & 27.6$^{+0.0}_{-0.0}$ & -5.06 --- -5.03 & 0.7476 $\pm$ 0.0009 & 0.7350 --- 0.7512 & 0.2897 $\pm$ 0.0001 & 0.2874 --- 0.2897 \\[0.9mm]
98921 & 0.654 & 0.350 & 0.3246 $\pm$ 0.0006 & -4.47$^{+0.00}_{-0.00}$ & 0.731$^{+0.000}_{-0.000}$ & 8.4$^{+0.0}_{-0.0}$ & -4.52 --- -4.38 & 0.8526 $\pm$ 0.0013 & 0.8391 --- 0.8850 & 0.3048 $\pm$ 0.0001 & 0.2996 --- 0.3112 \\[0.9mm]
100511 & 0.498 & - & 0.2008 $\pm$ 0.0008 & -4.74$^{+0.01}_{-0.00}$ & 2.166$^{+0.000}_{-0.075}$ & 6.2$^{+0.0}_{-0.1}$ & -4.74 & 0.8020 $\pm$ 0.0010 & 0.8020 & 0.2865 $\pm$ 0.0002 & 0.2865 \\[0.9mm]
100970 & 0.662 & 0.147 & 0.1521 $\pm$ 0.0041 & -5.05$^{+0.03}_{-0.03}$ & 6.662$^{+0.717}_{-0.662}$ & 28.0$^{+0.7}_{-0.7}$ & -5.05 & 0.7154 $\pm$ 0.0007 & 0.7154 & 0.2870 $\pm$ 0.0003 & 0.2870 \\[0.9mm]
101875 & 0.610 & 0.161 & 0.1618 $\pm$ 0.0018 & -4.97$^{+0.01}_{-0.01}$ & 5.016$^{+0.185}_{-0.179}$ & 19.8$^{+0.2}_{-0.2}$ & -4.97 & 0.7773 $\pm$ 0.0028 & 0.7773 & 0.2858 $\pm$ 0.0007 & 0.2858 \\[0.9mm]
104214 & 1.069 & - & 0.6420 $\pm$ 0.0030 & -4.60$^{+0.00}_{-0.00}$ & 1.331$^{+0.000}_{-0.000}$ & 25.4$^{+0.0}_{-0.0}$ & -4.69 --- -4.39 & 0.7607 $\pm$ 0.0002 & 0.7324 --- 0.8085 & 0.3669 $\pm$ 0.0001 & 0.3599 --- 0.3920 \\[0.9mm]
107350 & 0.587 & - & 0.3330 $\pm$ 0.0014 & -4.42$^{+0.00}_{-0.00}$ & 0.489$^{+0.000}_{-0.000}$ & 4.6$^{+0.0}_{-0.0}$ & -4.50 --- -4.38 & 0.9147 $\pm$ 0.0005 & 0.8819 --- 0.9310 & 0.3153 $\pm$ 0.0003 & 0.3050 --- 0.3203 \\[0.9mm]
108473 & 0.577 & 0.147 & 0.1400 $\pm$ 0.0022 & -5.13$^{+0.02}_{-0.02}$ & 8.696$^{+0.568}_{-0.545}$ & 18.5$^{+0.3}_{-0.3}$ & -5.13 & 0.7530 $\pm$ 0.0022 & 0.7530 & 0.2806 $\pm$ 0.0005 & 0.2806 \\[0.9mm]
108506\sups{SG} & 0.971 & 0.125 & 0.1284 $\pm$ 0.0010 & -5.25$^{+0.00}_{-0.01}$ & 12.438$^{+0.345}_{-0.000}$ & 60.8$^{+0.7}_{-0.0}$ & -5.25 & 0.6225 $\pm$ 0.0035 & 0.6225 & 0.3153 $\pm$ 0.0006 & 0.3153 \\[0.9mm]
109378 & 0.773 & 0.155 & 0.1534 $\pm$ 0.0007 & -5.07$^{+0.01}_{-0.00}$ & 7.134$^{+0.000}_{-0.239}$ & 41.2$^{+0.0}_{-0.4}$ & -5.07 & 0.7024 $\pm$ 0.0012 & 0.7024 & 0.3000 $\pm$ 0.0002 & 0.3000 \\[0.9mm]
109439\sups{SG} & 0.688 & 0.215 & 0.2279 $\pm$ 0.0010 & -4.72$^{+0.01}_{-0.00}$ & 2.019$^{+0.000}_{-0.069}$ & 20.5$^{+0.0}_{-0.4}$ & -4.72 & 0.7695 $\pm$ 0.0004 & 0.7695 & 0.2949 $\pm$ 0.0001 & 0.2949 \\[0.9mm]
109572 & 0.508 & - & 0.1479 $\pm$ 0.0003 & -5.04$^{+0.00}_{-0.01}$ & 6.435$^{+0.227}_{-0.000}$ & 9.8$^{+0.1}_{-0.0}$ & -5.04 & 0.7186 $\pm$ 0.0012 & 0.7186 & 0.2749 $\pm$ 0.0002 & 0.2749 \\[0.9mm]
109674\sups{SG,B} & 0.498 & - & 0.1786 $\pm$ 0.0006 & -4.84$^{+0.00}_{-0.00}$ & 3.109$^{+0.000}_{-0.000}$ & 7.2$^{+0.0}_{-0.0}$ & -4.84 & 0.8917 $\pm$ 0.0010 & 0.8917 & 0.3256 $\pm$ 0.0003 & 0.3256 \\[0.9mm]
111274 & 0.668 & 0.147 & 0.1486 $\pm$ 0.0023 & -5.08$^{+0.02}_{-0.01}$ & 7.379$^{+0.251}_{-0.484}$ & 29.5$^{+0.3}_{-0.5}$ & -5.08 & 0.7361 $\pm$ 0.0004 & 0.7361 & 0.2894 $\pm$ 0.0003 & 0.2894 \\[0.9mm]
113357 & 0.666 & 0.148 & 0.1528 $\pm$ 0.0021 & -5.06$^{+0.01}_{-0.02}$ & 6.895$^{+0.484}_{-0.233}$ & 28.7$^{+0.5}_{-0.3}$ & -5.06 --- -5.00 & 0.7154 $\pm$ 0.0014 & 0.7154 --- 0.7283 & 0.2900 $\pm$ 0.0002 & 0.2900 --- 0.2904 \\[0.9mm]
113421 & 0.744 & 0.150 & 0.1494 $\pm$ 0.0029 & -5.09$^{+0.02}_{-0.02}$ & 7.630$^{+0.521}_{-0.496}$ & 39.0$^{+0.7}_{-0.7}$ & -5.09 & 0.6861 $\pm$ 0.0006 & 0.6861 & 0.2979 $\pm$ 0.0003 & 0.2979 \\[0.9mm]
113829 & 0.620 & 0.316 & 0.3272 $\pm$ 0.0015 & -4.44$^{+0.00}_{-0.01}$ & 0.586$^{+0.049}_{-0.000}$ & 6.2$^{+0.3}_{-0.0}$ & -4.44 & 0.8745 $\pm$ 0.0013 & 0.8745 & 0.3102 $\pm$ 0.0004 & 0.3102 \\[0.9mm]
113896 & 0.581 & 0.152 & 0.1561 $\pm$ 0.0010 & -5.00$^{+0.00}_{-0.01}$ & 5.589$^{+0.203}_{-0.000}$ & 17.0$^{+0.2}_{-0.0}$ & -5.00 & 0.7662 $\pm$ 0.0014 & 0.7662 & 0.2829 $\pm$ 0.0003 & 0.2829 \\[0.9mm]
113994\sups{SG} & 0.886 & 0.164 & 0.1677 $\pm$ 0.0004 & -5.04$^{+0.00}_{-0.00}$ & 6.435$^{+0.000}_{-0.000}$ & 47.3$^{+0.0}_{-0.0}$ & -5.04 & 0.6675 $\pm$ 0.0012 & 0.6675 & 0.3004 $\pm$ 0.0001 & 0.3004 \\[0.9mm]
\hline
\end{tabular}
\end{center}
\end{table}
\end{landscape}

\addtocounter {table} {-1}

\begin{landscape}
\begin{table}
\scriptsize
\begin{center}
\caption{continued.} 
\begin{tabular}{lccccccccccc}
\hline
HIP & (B-V) & S-index & S-index & Log(R$^{\prime}_{\rm{HK}}$) & Chromospheric & Chromospheric & Log(R$^{\prime}_{\rm{HK}}$) & Ca\subs{IRT}-index & Ca\subs{IRT}-index & H\subs{\alpha}-index & H\subs{\alpha}-index\\
no. & (HIPP.) & (Wright) & (This work) & & age (Gyr) & period (d) & range & & range & & range\\
\hline
114378 & 0.607 & 0.343 & 0.3541 $\pm$ 0.0007 & -4.39$^{+0.00}_{-0.00}$ & 0.347$^{+0.000}_{-0.000}$ & 4.5$^{+0.0}_{-0.0}$ & -4.39 --- -4.36 & 0.9351 $\pm$ 0.0011 & 0.9351 --- 0.9432 & 0.3224 $\pm$ 0.0002 & 0.3224 --- 0.3255 \\[0.9mm]
114456 & 0.750 & 0.231 & 0.1747 $\pm$ 0.0009 & -4.95$^{+0.01}_{-0.00}$ & 4.664$^{+0.000}_{-0.168}$ & 34.8$^{+0.0}_{-0.4}$ & -4.96 --- -4.95 & 0.7157 $\pm$ 0.0007 & 0.7157 & 0.2995 $\pm$ 0.0002 & 0.2995 \\[0.9mm]
114622 & 1.000 & - & 0.2590 $\pm$ 0.0012 & -4.91$^{+0.00}_{-0.01}$ & 4.026$^{+0.151}_{-0.000}$ & 44.3$^{+0.5}_{-0.0}$ & -4.91 & 0.6887 $\pm$ 0.0009 & 0.6887 & 0.3340 $\pm$ 0.0000 & 0.3340 \\[0.9mm]
115951 & 0.620 & - & 0.1477 $\pm$ 0.0012 & -5.07$^{+0.00}_{-0.01}$ & 7.134$^{+0.245}_{-0.000}$ & 23.0$^{+0.2}_{-0.0}$ & -5.07 & 0.7138 $\pm$ 0.0004 & 0.7138 & 0.2835 $\pm$ 0.0002 & 0.2835 \\[0.9mm]
116085 & 0.839 & - & 0.1666 $\pm$ 0.0008 & -5.02$^{+0.00}_{-0.00}$ & 6.000$^{+0.000}_{-0.000}$ & 44.2$^{+0.0}_{-0.0}$ & -5.02 & 0.7197 $\pm$ 0.0012 & 0.7197 --- 0.7321 & 0.3128 $\pm$ 0.0001 & 0.3126 --- 0.3128 \\[0.9mm]
116106 & 0.531 & - & 0.1567 $\pm$ 0.0010 & -4.98$^{+0.01}_{-0.01}$ & 5.201$^{+0.191}_{-0.185}$ & 11.4$^{+0.1}_{-0.1}$ & -4.98 & 0.7823 $\pm$ 0.0016 & 0.7823 & 0.2832 $\pm$ 0.0002 & 0.2832 \\[0.9mm]
116421 & 0.604 & - & 0.1512 $\pm$ 0.0006 & -5.04$^{+0.00}_{-0.01}$ & 6.435$^{+0.227}_{-0.000}$ & 20.4$^{+0.2}_{-0.0}$ & -5.04 & 0.7804 $\pm$ 0.0007 & 0.7804 & 0.2883 $\pm$ 0.0007 & 0.2883 \\[0.9mm]
116613 & 0.665 & - & 0.3149 $\pm$ 0.0024 & -4.50$^{+0.01}_{-0.00}$ & 0.873$^{+0.000}_{-0.047}$ & 9.9$^{+0.0}_{-0.4}$ & -4.50 & 0.8414 $\pm$ 0.0007 & 0.8414 & 0.3038 $\pm$ 0.0004 & 0.3038 \\[0.9mm]
hd131156a & 0.72\sups{W} & 0.437 & 0.4746 $\pm$ 0.0013 & -4.31$^{+0.00}_{-0.00}$ & 0.072$^{+0.000}_{-0.000}$ & 4.8$^{+0.0}_{-0.0}$ & -4.39 --- -3.79 & 0.9369 $\pm$ 0.0007 & 0.8425 --- 0.9559 & 0.3426 $\pm$ 0.0001 & 0.3270 --- 0.3979 \\[0.9mm]
hd131156b & 1.16\sups{W} & 1.280 & 1.2803 $\pm$ 0.0173 & -4.44$^{+0.01}_{-0.01}$ & 0.586$^{+0.049}_{-0.049}$ & 14.6$^{+0.7}_{-0.7}$ & -4.90 --- -4.35 & 0.8473 $\pm$ 0.0046 & 0.8380 --- 0.9355 & 0.3912 $\pm$ 0.0007 & 0.3367 --- 0.4096 \\[0.9mm]
hd179958 & 0.65\sups{W} & 0.148 & 0.1602 $\pm$ 0.0067 & -5.00$^{+0.04}_{-0.04}$ & 5.589$^{+0.847}_{-0.751}$ & 25.3$^{+0.9}_{-0.9}$ & -5.06 --- -5.00 & 0.7395 $\pm$ 0.0006 & 0.7390 --- 0.7395 & 0.2890 $\pm$ 0.0005 & 0.2877 --- 0.2890 \\[0.9mm]
moon & 0.656\sups{W} & 0.166 & 0.1662 $\pm$ 0.0061 & -4.96$^{+0.03}_{-0.04}$ & 4.837$^{+0.751}_{-0.503}$ & 25.0$^{+1.0}_{-0.8}$ & -4.96 & 0.7698 $\pm$ 0.0038 & 0.7698 & 0.2877 $\pm$ 0.0002 & 0.2877 \\
\hline
\end{tabular}
\end{center}
\end{table}
\end{landscape}

\begin{table}
\begin{center}
\caption{Table of the coefficients for Equation~\ref{Sindex_eqn} we have determined for our NARVAL and ESPaDOnS observations.}
\label{Tab_Sindex_params}
\begin{tabular}{lcc}
\hline
Coefficient & NARVAL & ESPaDOnS\\
\hline
$a$ & 12.873 & 7.999 \\
$b$ & 2.502 & -3.904 \\
$c$ & 8.877 & 1.150 \\
$d$ & 4.271 & 1.289 \\
$e$ & 1.183 $\times$ 10\sups{-3} & -0.069 \\
\hline
\end{tabular}
\end{center}
\end{table}

\subsection{The Calcium infrared triplet and H$\alpha$ emission}\label{Sec_irtha}

In addition to the \Ca H \& K index the activity of the stars was also analysed using the emission from the \Ca Infrared Triplet (IRT) lines at 8498.02, 8542.09 and 8662.14 \AA, along with the emission in the H$\alpha$ line (6562.85 \AA). The Ca\subs{IRT}-index and the H$\alpha$-index were calculated using Equations~\ref{IRT-index_eqn} and~\ref{Ha-index_eqn} respectively.
\BE
{\rm Ca_{IRT}-index} = \frac{F_{8498} + F_{8542} + F_{8662}}{F_{V_{IRT}} + F_{R_{IRT}}}, \label{IRT-index_eqn}
\EE
where $F_{8498}, F_{8542}$ and $F_{8662}$ are the fluxes measured in 2 \AA\/ rectangular bandpasses centered on the respective \Ca IRT lines, while $F_{V_{IRT}}$ and $F_{R_{IRT}}$ are the fluxes in 5 \AA\/ rectangular bandpasses centered on two continuum points, 8475.8 \AA\/ and 8704.9 \AA, either side of the IRT lines.

\BE
{\rm H\alpha -index} = \frac{F_{H\alpha}}{F_{V_{H\alpha}} + F_{R_{H\alpha}}}, \label{Ha-index_eqn}
\EE
where $F_{H\alpha}$ is the flux in a 3.6 \AA\/ rectangular bandpass centered on the H$\alpha$ line and $F_{V_{H\alpha}}$ and $F_{R_{H\alpha}}$ are the fluxes in two 2.2 \AA\/ rectangular bandpasses centered on the continuum points 6558.85 \AA\/ and 6567.30 \AA. Equation~\ref{IRT-index_eqn} is taken from \citet[][Equation 1]{PetitP:2013} while Equation~\ref{Ha-index_eqn} is taken from \citet*[][Table 3]{GizisJE:2002}.

The results of the Ca\subs{IRT}-index and H$\alpha$-index are given in Table~\ref{Bcool_Activity}, with once again the errors in the indices being calculated from the standard deviation of the measurements from the four exposures and the range of the indices being the range seen over all of the multiple observations of a star (again ignoring the small error bars). 

\section{Results and Discussion}\label{Sec_res}

\subsection{Longitudinal field measurements}\label{Sec_ResBl}

One of the unique results that can be obtained from spectropolarimetric observations is a measure of the mean longitudinal magnetic field ($B_{l}$) on the visible stellar surface. One goal of this paper is to investigate how \Bl varies with basic stellar parameters, such as T\subs{eff}, age and rotation rate, as well as how it relates to other more traditional activity proxies. 

Unlike the traditional activity indicators (such as the S-index) which are sensitive to the magnetic field strength and therefore retain the contribution of most visible magnetic features, the measured longitudinal field is also sensitive to the field polarities. It is therefore very dependent upon the distribution and polarity mix of the magnetic regions across the stellar surface, and its value relates to the largest-scale component of the magnetic field. Even in the most favourable case of a purely dipolar magnetic field, the value of \Bl underestimates the polar field strength by at least a factor of three \citep{AuriereM:2007}. If the magnetic field distribution is more complex with different polarities across the visible hemisphere of the star (which is a likely situation with Sun-like stars), the measurement of \Bl will even further underestimate the local stellar magnetic field strength. 

As most of the stars in our sample have very low \vsinis values, which makes the detection of complex magnetic fields especially difficult due to magnetic flux cancellation, this means that our measurements of the magnetic field are restricted to that of the large scale field component on the stellar surface. However, a number of significant dynamo characteristics (e.g. magnetic cycles) are primarily expected to be seen on the largest spatial scales of the surface magnetic field. Thus spectropolarimetric measurements are a useful probe of the underlying stellar dynamo, in spite of its intrinsically limited spatial resolution.  

With this in mind, we have compared \Bl to other activity indicators computed from the same spectra. As explained in Section~\ref{Sec_Bl}, for stars with multiple observations we have chosen to analyse the observation that has given the largest absolute value of $B_{l}$, as this should correspond to a rotational phase that best represents the global magnetic field.  It should be noted that this maximal value is still an underestimate of the local field strength in individual magnetic regions.

In the discussion of the results in following sections, we have broadly classified those stars with T\subs{eff} $<$ 5000 K to be K-stars, those with 5000 K $\le$ T\subs{eff} $\le$ 6000 K to be G-stars and those with T\subs{eff} $>$ 6000 K to be F-stars. 

\subsection{Magnetic field detection rates}\label{Sec_Resdet}

\begin{figure}
  \centering
  \includegraphics[trim =0 15 0 30, clip, angle=-90, width=\columnwidth]{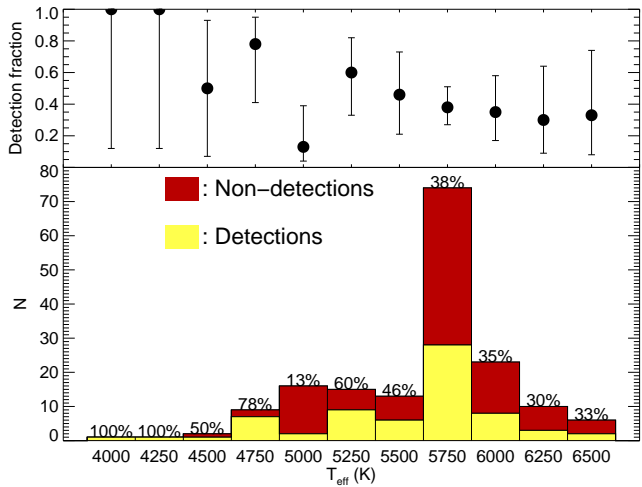}
  \caption{Histogram of detections/non-detections of surface magnetic fields against T\subs{eff} for our Bcool solar-type stars sample. The figure is as described in Figure~\ref{Fig_Hist_Err}.} 
  \label{Fig_Hist_Temp}
\end{figure}

The overall detection rate of magnetic fields on the surface of our stars (including both marginal and definite detections) is 39\% (67 out of 170 stars) meaning that we have detected a magnetic field on 39\% of our stars, excluding multiple observations. There appears to be a slight increase in the detection rate for K-stars (12/21 = 57\%) over G-stars (49/130 = 38\%) and F-stars (6/19 = 32\%), but the numbers of K- and F-stars are quite small (compared to G-stars) and these percentages may be affected by statistical fluctuations or a selection bias (in particular their rotation or age, see discussions below). In Figure~\ref{Fig_Hist_Temp} we have plotted the temperature distribution of our sample, showing the detection rate for each 250 K bin. This shows the strong bias towards stars around the solar temperature and the relative dearth of F- and K-stars in our sample.

\subsubsection{Correlation with age}

\begin{figure}
  \centering
  \includegraphics[trim =0 15 0 25, clip, angle=-90, width=\columnwidth]{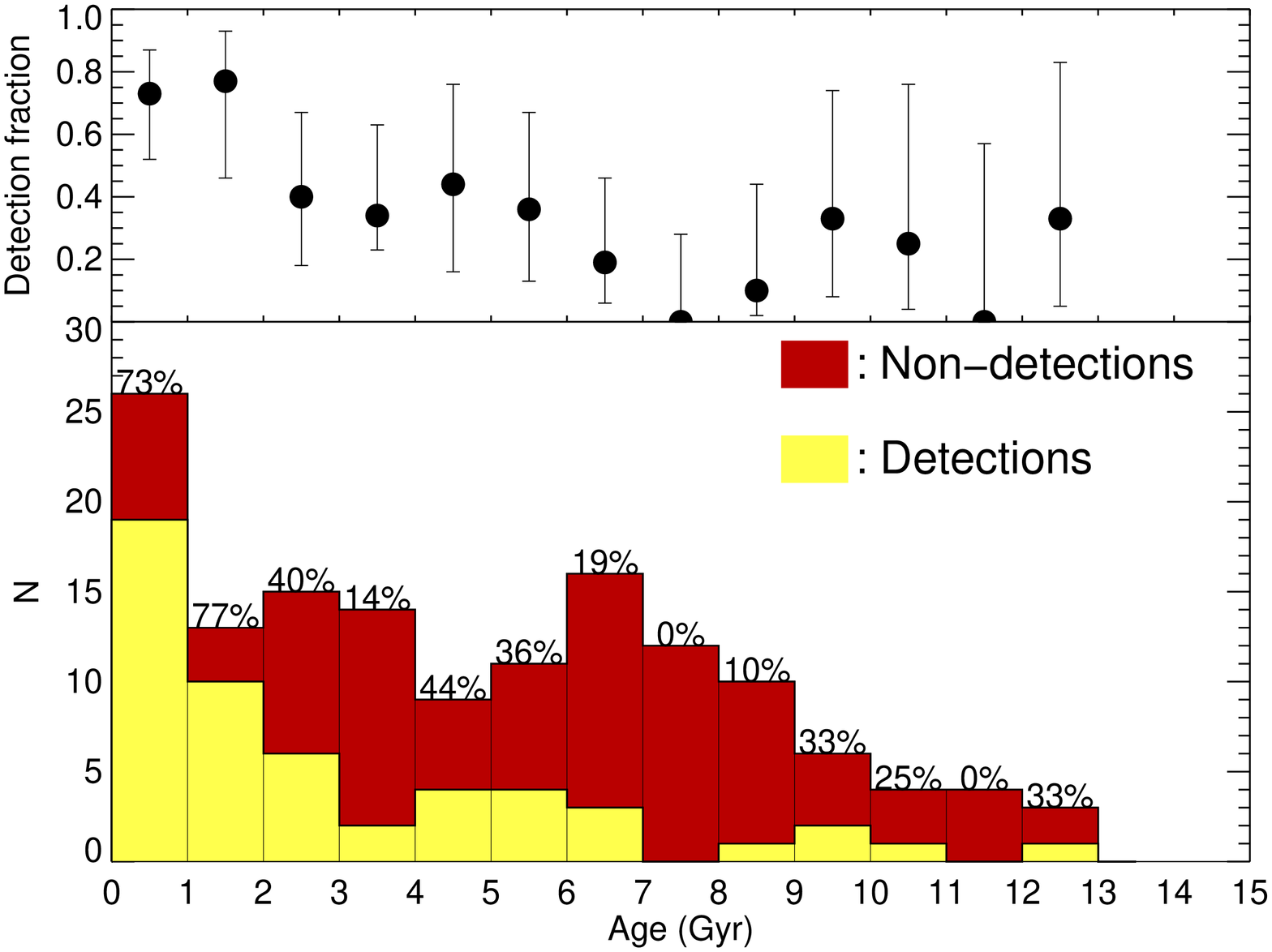}
  \includegraphics[trim =0 15 0 25, clip, angle=-90, width=\columnwidth]{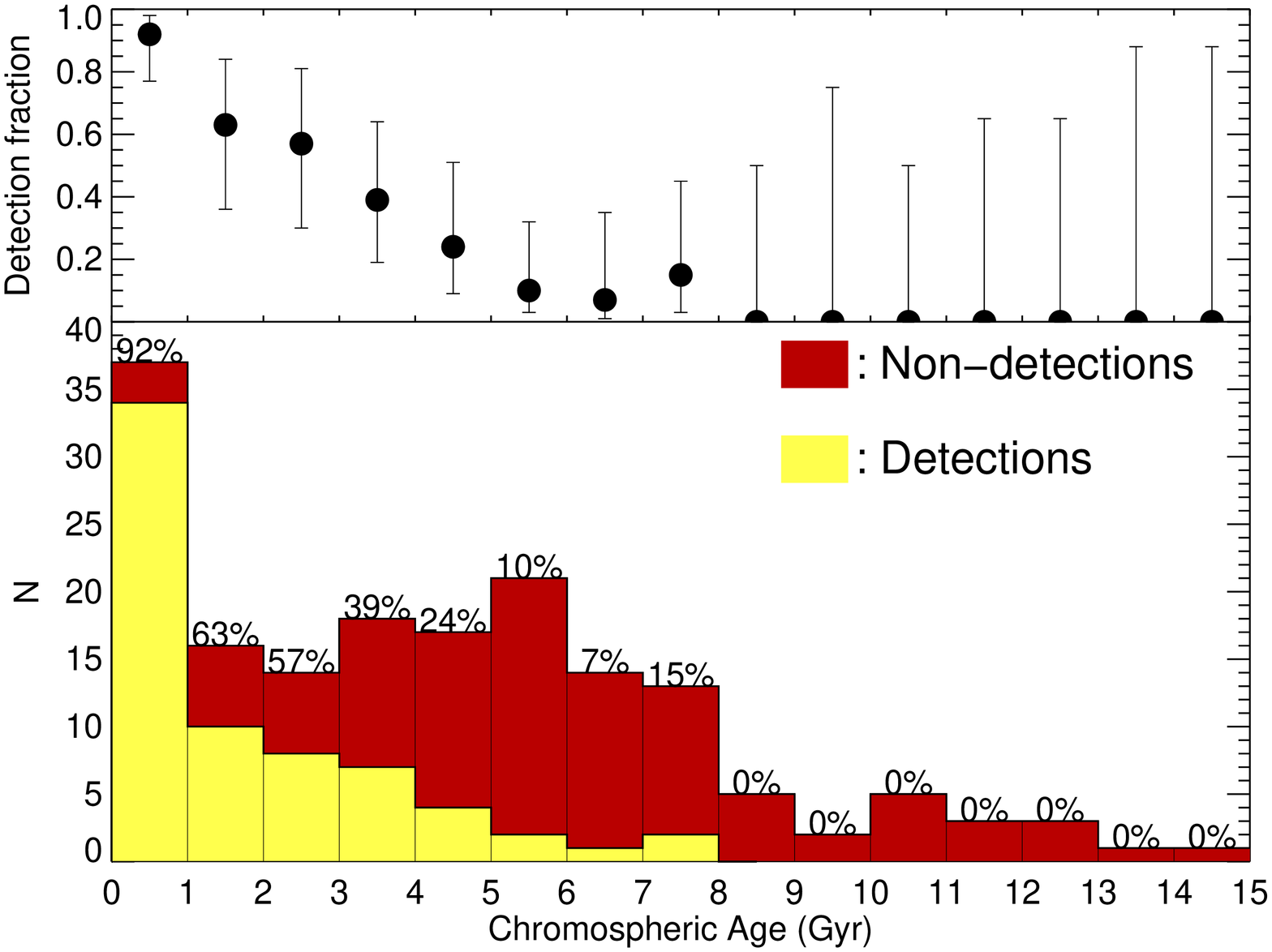}
  \caption{Histogram of detections/non-detections of surface magnetic fields against age (upper) and chromospheric age (lower), with the figure as described in Figure~\ref{Fig_Hist_Err}. The chromospheric age has been determined from the Calcium II HK emission from Table~\ref{Bcool_Activity} using \citet[][Equation 15]{WrightJT:2004}}
  \label{Fig_Hist_Age}
\end{figure}

In Figure~\ref{Fig_Hist_Age} we have plotted the age distribution of our sample, showing the detection rate for each 1 Gyr bin. We consider two independent age estimates by taking ages computed by \citet{TakedaG:2007} from evolutionary tracks (top panel) and chromospheric ages (bottom panel) calculated from our data, using the relation of \citet[][Equation 15]{WrightJT:2004} based on the individual measurements of Log(R$^{\prime}_{\rm{HK}}$) given in Table~\ref{Bcool_Activity}. A number of stars in \citet{TakedaG:2007} have ages greater than recent estimates of the age of the universe \citep{PlanckCollaboration:2013}, and we chose to ignore these stars in the plot, since some of the oldest objects are obviously much younger  (\citet{TakedaG:2007} actually acknowledge that their method produces an accumulation of stars with extreme ages). A clear example is given by the visual binary system HD 131156, for which the primary component is estimated to be very young (less than 1 Gyr), while the secondary is listed with a much older age of 13 Gyr. In this case, the chromospheric ages come up with a more consistent overlapping estimate of 0.18 to 0.35 Gyrs for the two companions (see Appendix~\ref{Appendix_LSD}). Similarly, a number of stars in \citet{TakedaG:2007} have estimated ages less than 0 Gyrs and for these stars we have assigned an age of 0 Gyrs. 
 
As can be seen our sample has a broad distribution of ages for stars up to at least 9 Gyrs and the majority of our sample are mature aged solar-type stars. Figure~\ref{Fig_Hist_Age} shows that there is an unsurprisingly higher detection rate for younger stars, with stars around an age of 0 - 2 Gyrs showing a detection rate of over 65\%. For stars older than $\sim$1 - 2 Gyrs, the detection rate appears to drop, fluctuating between 0\% and 57\%, with a smoother decrease in detection rate when the chromospheric age is considered.

The average age of all the stars in the sample, for which we have an age estimate from \citet{TakedaG:2007} (excluding those stars older than the age of the universe), is 4.8 Gyrs (not taking into account the large error bars for each measurement). This values drops to 4.3 Gyrs if chromospheric ages are used. For F-stars the average age is 2.4 and 3.0 Gyrs (respectively for evolutionary tracks and chromospheric ages), while for G-stars it is 5.3 and 4.5 Gyrs, and 2.7 and 4.4 Gyrs for K-stars. Thus the higher detection rates and activity seen on the K-stars (over that of the G-stars) in Figures~\ref{Fig_Hist_Temp} and~\ref{Fig_Blong_Temp} may be related to the youthfulness of the K-stars. However, a lower detection rate is obtained for F-stars (which are similarly young), suggesting that, at a given age, stellar large-scale magnetic fields are inversely correlated to the stellar mass.

Previously there have been very few magnetic field detections on mature solar-type stars, as stellar activity (and therefore magnetic field strength) declines significantly as a star ages. If we define mature as an age greater than 2 Gyrs (ignoring the error bars in the age estimates) then one mature age solar-type star with a magnetic field detection was presented by \citet{PetitP:2008}, HIP 79672 (18 Sco), which is part of this data set, while \citet{FaresR:2012} presented magnetic field detections and maps of the mature age planet-hosting star HD 179949. A further three mature age solar-type stars (HD 70642, HD 117207 and HD 154088) with magnetic detections were presented by \citet{FossatiL:2013}. If we discard magnetic field detections on subgiant members of our sample we have 21 mature age solar-stars with magnetic field detections, 20 of which have not previously been reported.

\subsubsection{Correlation with field strength}

\begin{figure}
  \centering
  \includegraphics[trim =0 15 0 15, clip, angle=-90, width=\columnwidth]{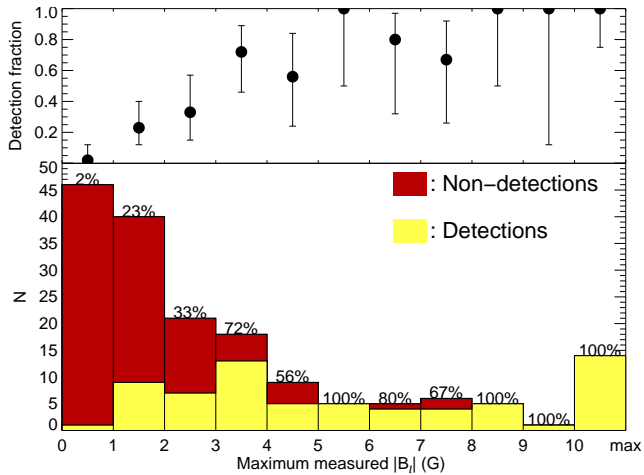}
  \caption{Histogram of detections/non-detections of surface magnetic fields against the maximum measured $|B_{l}|$, with the figure as described in Figure~\ref{Fig_Hist_Err}. We emphasize that yellow bins (detections) correspond to actual field values, while red bins (non-detections) correspond to upper limits.} 
  \label{Fig_Hist_Beff}
\end{figure}

We plot in Figure~\ref{Fig_Hist_Beff} field detections as a function of \Bla (yellow bins), and over plot non-detection (red bins) as a function of their upper limit on \Bla. From this histogram, it is clear that it is virtually impossible to detect mean surface magnetic fields of less than $\sim$1 G on our stars even with the excellent signal-to-noise ratio of most of our observations, see Figure~\ref{Fig_Hist_SNR} and Table~\ref{Bcool_LSD} (the one detection for 0 - 1 G shown in Figure~\ref{Fig_Hist_Beff} is for HIP 8362 with \Bl = -0.9 $\pm$ 0.2 G). Detection of surface magnetic fields of less than 1 G has been achieved previously for some cool stars \citep[i.e.][]{AuriereM:2009} but this has only been possible with the addition of a large number of spectropolarimetric observations. The detection rate correlates directly with the magnitude of the mean magnetic field.  For field strengths of 5 G or greater we obtain an almost 100\% detection rate. The few stars with fields greater than 5 G that did not give detections are due to the low signal-to-noise ratio of the observations (SNR\subs{LSD} $<$ 20,000). 

\subsubsection{Correlation with signal to noise}

\begin{figure}
  \centering
  \includegraphics[trim =0 15 0 30, clip, angle=-90, width=\columnwidth]{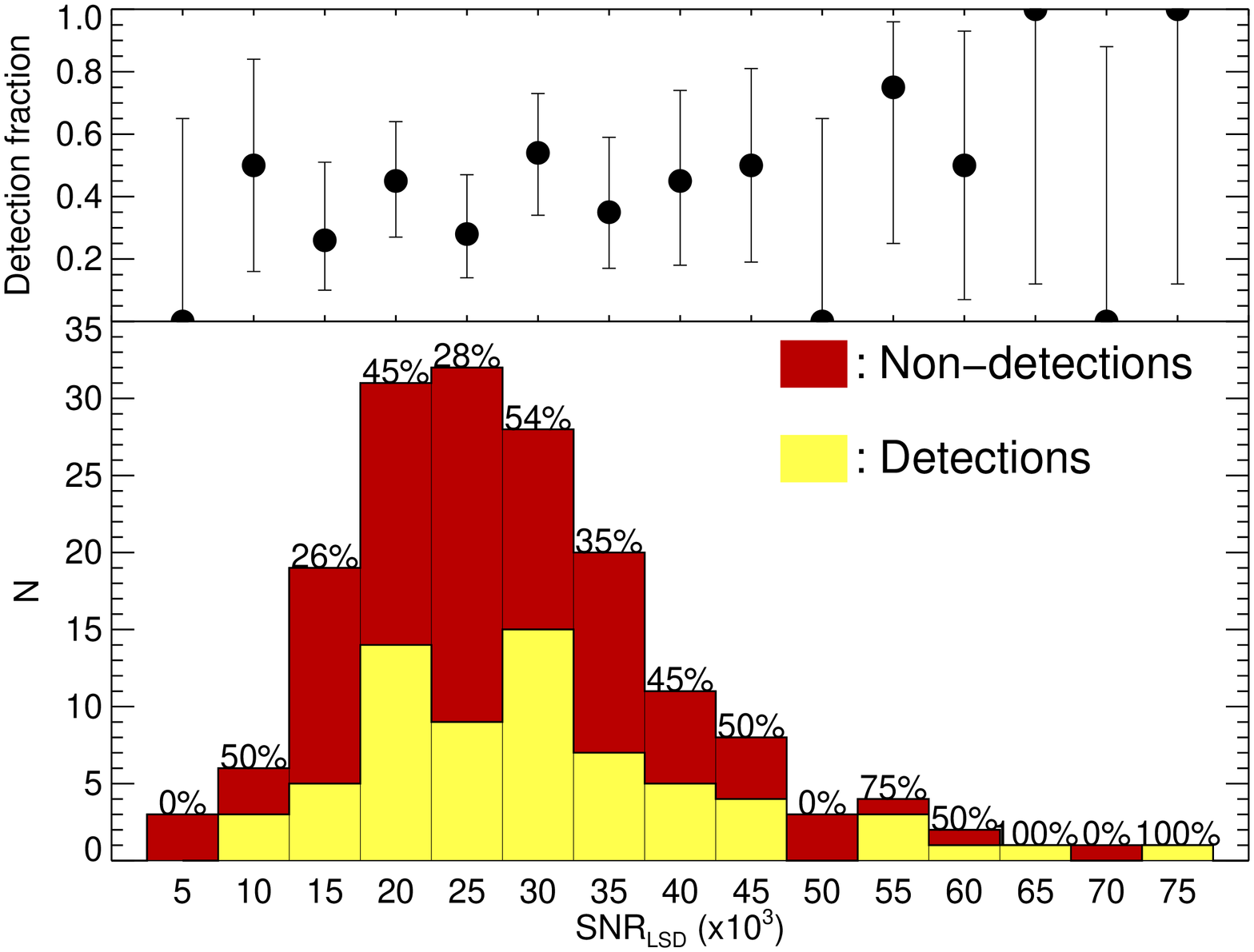}
  \caption{Histogram of detections/non-detections of surface magnetic fields against SNR\subs{LSD}, with the figure as described in Figure~\ref{Fig_Hist_Err}.} 
  \label{Fig_Hist_SNR}
\end{figure}

In Figure~\ref{Fig_Hist_SNR} the sample magnetic field detection rate is plotted as a function of the signal-to-noise ratio of the Stokes V LSD profile. A significant majority (125/170 = 74\%) of our sample have observations with SNR\subs{LSD} $>$ 20,000. For those bins with 5 or more observations in them, Figure~\ref{Fig_Hist_SNR} shows that the detection rate varies between 26\% and 54\% with little evidence of an increase in the detection rate with an increase in SNR\subs{LSD}.  This is mostly because longer exposure times were adopted for a number of low-activity targets with observations occasionally reaching SNR\subs{LSD} $>$ 50,000, but even this was not enough to detect the magnetic field on the surface of many mature low-activity Suns. Consequently it is difficult to determine the lowest signal-to-noise ratio required to detect a surface magnetic field in any cool star, but given that the three stars with \Bla over 5 G that showed no detection have in common SNR\subs{LSD} $<$ 20,000, we adopt this value as the minimum signal-to-noise ratio to be aimed for in future observations.

\subsubsection{Correlation with Ca H\&K S-index}

\begin{figure}
  \centering
  \includegraphics[trim =0 10 0 20, clip, angle=-90, width=\columnwidth]{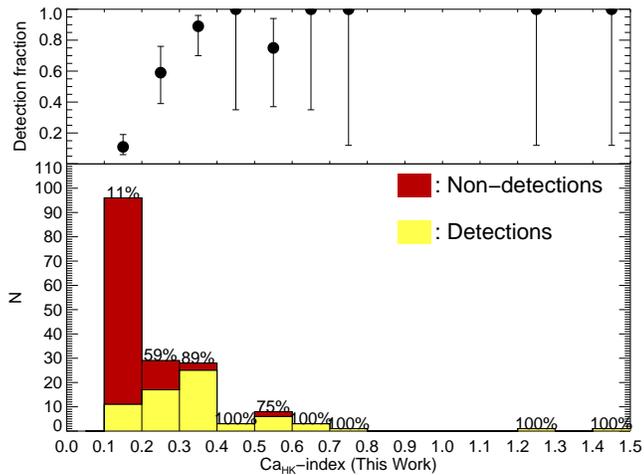}
  \caption{Histogram of detections/non-detections of surface magnetic fields against the \Ca H \& K index, with the figure as described in Figure~\ref{Fig_Hist_Err}.} 
  \label{Fig_Hist_Sindex}
\end{figure}

Figure~\ref{Fig_Hist_Sindex} shows the detection rate as a function of the \Ca H \& K-index (S-index).  Clearly from this plot, stars with an S-index $\ga$ 0.2 have a very high (77\%) rate of magnetic detection, while those stars with an S-index $\ga$ 0.3 are almost certain (89\%) to have a detectable magnetic field on the stellar surface, while for those stars with an S-index $\la$ 0.2 the probability of detecting a magnetic field is very low (11\%). The star with the smallest S-index which shows a definite detection of a magnetic field is the subgiant HIP 113944 (S-index = 0.1677).

\subsubsection{Correlation with vsini}

\begin{figure}
  \centering
  \includegraphics[trim =0 15 0 15, clip, angle=-90, width=\columnwidth]{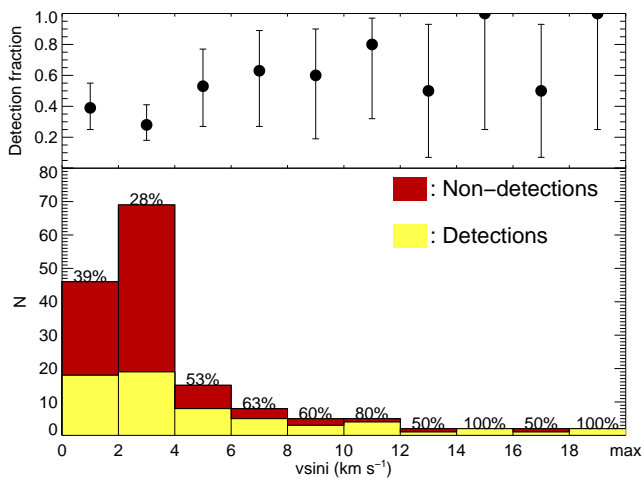}
  \caption{Histogram of detections/non-detections of surface magnetic fields against the stellar \vsini, with the figure as described in Figure~\ref{Fig_Hist_Err}.} 
  \label{Fig_Hist_vsini}
\end{figure}

Figure~\ref{Fig_Hist_vsini} shows the detection rate as a function of the \vsinis of the star (with those stars with \vsinis greater than 20 km/s being counted in the 18+ box). The detection rate for stars with \vsinis values less than 4 \kmss is 32\% (37/155) while for those stars with \vsinis values between 4 and 10 \kmss the detection rate is 57\% (16/28). For stars with \vsinis values greater than 10 \kmss the detection rate increases to 77\% (10/13) with only three stars with \vsinis values this high not giving magnetic detections. 

\subsection{The variation of \Bla with basic stellar parameters}\label{Sec_Resvar}

Because the measurement of \Bl is based on a averaged measure of the magnetic fields on the visible surface of a star, and because it is a measure of the amount of magnetic flux on the stellar surface, it is expected that \Bl should show a correlation with other activity indicators (such as the \Ca H \& K S-index).  The shear scale of 
the large sample of stars with measured stellar parameters that we have in the Bcool spectropolarimetric survey (from \citealt{ValentiJA:2005} and \citealt{TakedaG:2007}) means that we can also investigate the relationship between \Bla and basic stellar parameters in a statistically significant manner.

In the following figures we have only plotted those stars that have the relevant stellar parameters available (see Table~\ref{Bcool_params}). For stars with multiple observations (i.e. those that show a range in their measurements of activity in columns 6, 8 and 10 in Table~\ref{Bcool_Activity}) we have plotted the single measurement (i.e. the value from columns 4, 5, 7 and 9 in Table~\ref{Bcool_Activity}) that can be directly compared to the \Bl measurement taken from the same observation (see Table~\ref{Bcool_LSD}). 

\subsubsection{Correlation with spectral type}

\begin{figure}
  \centering
  \includegraphics[trim =30 25 10 5, clip, angle=-90, width=\columnwidth]{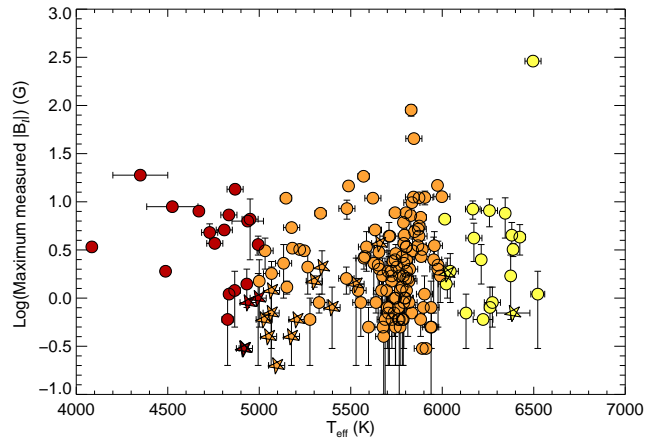}
  \caption{Plot of Log(maximum measured $|B_{l}|$) against T\subs{eff}. The symbols are the same as in Figure~\ref{Fig_HR}.} 
  \label{Fig_Blong_Temp}
\end{figure}

A correlation of magnetic activity with stellar mass is commonly seen in activity indices, such as the S-index where for stars with similar ages, K-stars are generally more active than G-stars and G-stars are more active than F-stars \citep[i.e.][]{NoyesRW:1984, BaliunasSL:1995, BasriG:2010}. It is therefore of particular interest to determine if K-stars also have stronger large-scale magnetic fields compared to G- and F-stars. In Figure~\ref{Fig_Blong_Temp} we have plotted the maximum measured \Bla against the stellar effective temperature. This plot shows little evidence of a significant increase in the maximum field strength measured on K-stars compared to that seen on G-stars (ignoring the three most active stars in our sample, HIP 25486 [HR 1817], HIP 71631 [EK Dra] and HIP 91043 [V889 Her, HD171488]) although there may be an indication of a weaker maximum field strength for F-stars. The mean \Bla measured on F-, G- and K-stars (excluding the subgiants and the 3 very active stars) is 3.3 G, 3.2 G and 5.7 G respectively, with the mean \Bla being 3.5 G. Thus the mean measured field strength is somewhat higher for K-stars, compared to F- and G-stars. However, the number of main-sequence K-stars in our sample is comparitively small (17) and some of the K dwarfs were specifically selected for our observations because they were known to be active stars.  This may result in a bias towards stronger magnetic fields and bearing in mind that this bias is mostly equivalent to the bias towards low ages mentioned in Section~\ref{Sec_Resdet}.

\subsubsection{Correlation with stellar age}

\begin{figure}
  \centering
  \includegraphics[trim =30 25 0 15, clip, angle=-90, width=\columnwidth]{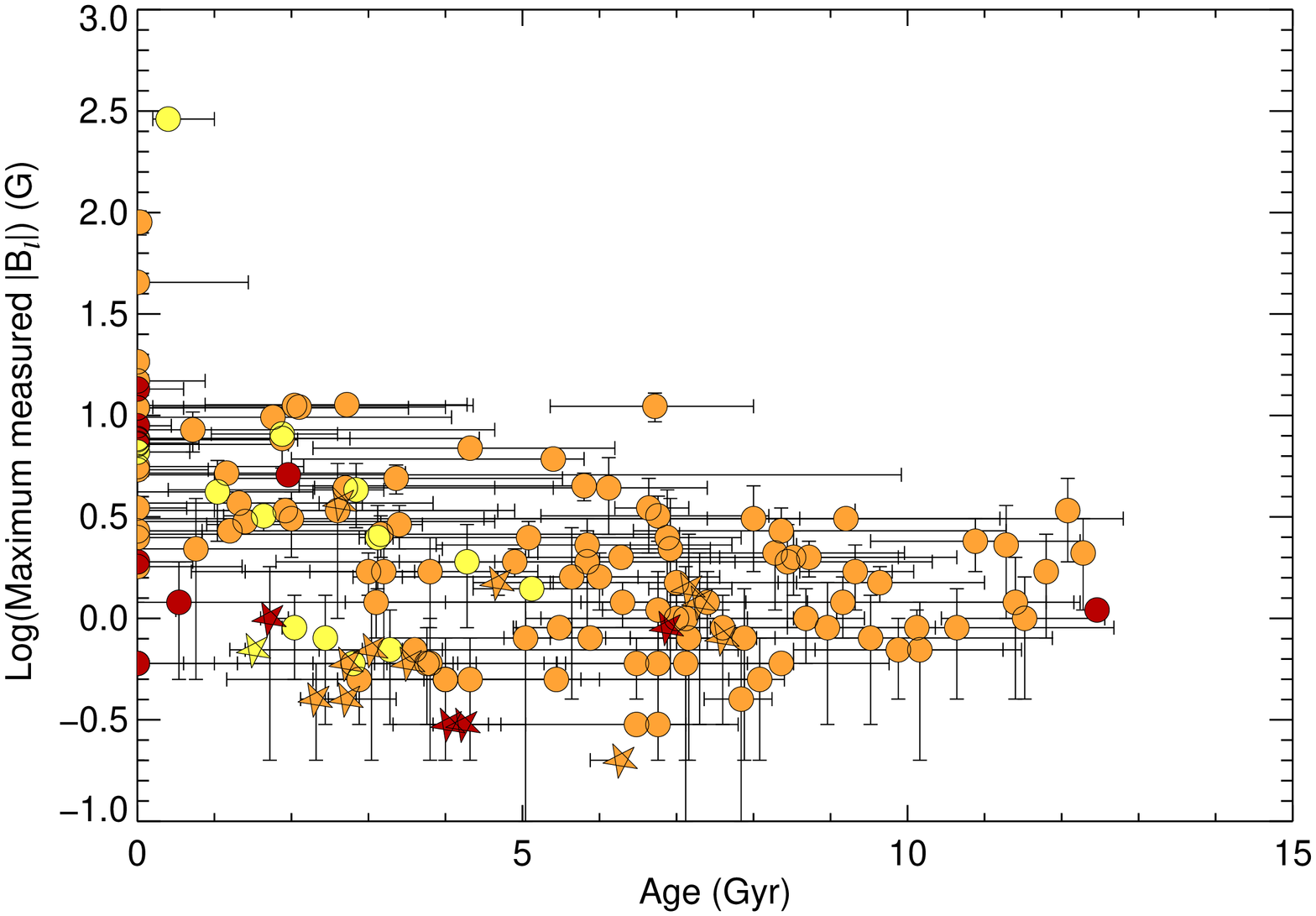}
  \includegraphics[trim =30 25 10 15, clip, angle=-90, width=\columnwidth]{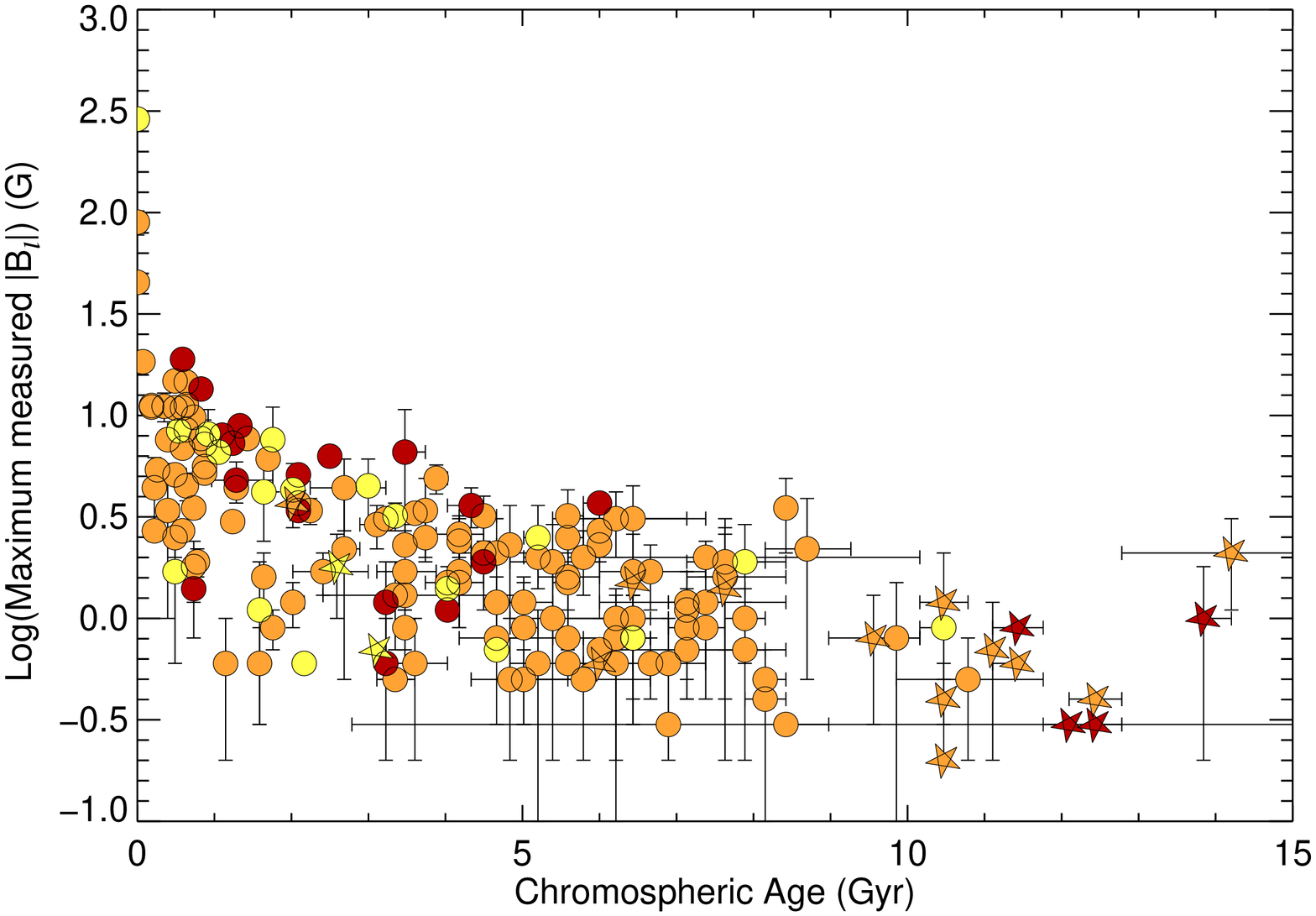}
  \caption{Plot of Log(maximum measured $|B_{l}|$) against age (upper) and chromospheric age (lower), with the chromospheric age being determined as described in Figure~\ref{Fig_Hist_Age}. The symbols are the same as in Figure~\ref{Fig_HR}.} 
  \label{Fig_Blong_Age}
\end{figure}

When plotted against the stellar age, the maximum measured \Bla shows a general decrease in the upper envelope of the measured magnetic field strength as a star ages (see Figure~\ref{Fig_Blong_Age}), similar to that observed in other activity indicators \citep[i.e.][]{PaceG:2013}. This trend is more pronounced when the chromospheric age is used and can be linked to the evolution with stellar age of the detection rate (see Figure~\ref{Fig_Hist_Age}). The oldest stars of our list systematically display \Bla values of $\la$ 5 G, with an age limit that depends on the age determination method (around 7 Gyrs from evolutionary models, versus 4 Gyrs from chromospheric emission). Using age values from \citealt{TakedaG:2007}, HIP 114378 is a notable exception to the general trend with \Bla around 12 G (see Appendix~\ref{Appendix_LSD}). We also note that most subgiants of our sample show up in the high-age tail of chromospheric ages, while their position is scattered at all ages using evolutionary tracks. 

\subsubsection{Correlation with stellar rotation}\label{Sec_rot}

\begin{figure}
  \centering
  \includegraphics[trim =30 25 10 15, clip, angle=-90, width=\columnwidth]{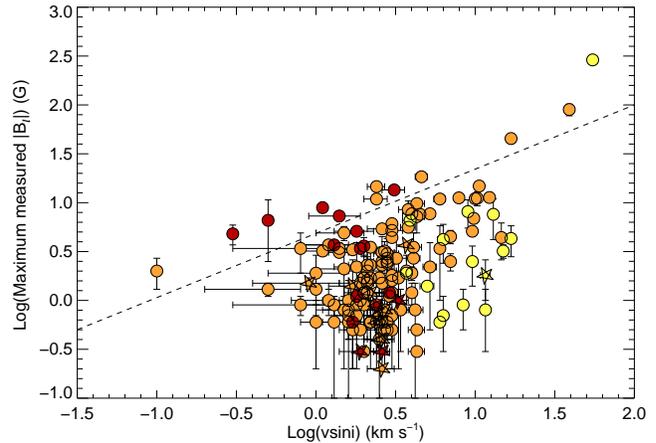}
  \caption{Plot of Log(maximum measured $|B_{l}|$) against Log(\vsini). The symbols are the same as in Figure~\ref{Fig_HR}, with the stars above the dashed line being defined as having high \Bla values (see Section~\ref{Sec_rot}).} 
  \label{Fig_Blong_Vsini}
\end{figure}

It is widely documented that the rotation rate of a star is tightly correlated to its stellar activity \citep[e.g.][]{NoyesRW:1984}. To understand if there is the same correlation with the large-scale magnetic field strength, we used the projected rotational velocity as a proxy of stellar rotation (Figure~\ref{Fig_Blong_Vsini}). The upper envelope of the \Bla values do appear to increase with rotation rate. We emphasize that, as a star's \vsinis increases, the spatial resolution of the stellar surface also increases. This effect is however not expected to strongly bias the \Bla estimates, as the integrals in Equation \ref{eq:bl} cover the width of the Stokes profile and thus only measure the largest-scale magnetic component, independently of the star's \vsinis value.  

There are a few stars which appear to have significantly larger \Bla values than other stars with similar \vsinis values (shown above the dashed line in Figure~\ref{Fig_Blong_Vsini}). Some of these deviant points are likely to be a result of an inclination effect where for a given rotation period, a low stellar inclination angle will result in a low \vsinis value. A good example is HD 131156 A, for which an inclination angle of 28$^\circ$ was proposed by Petit et al. (2005), in spite of a relatively rapid rotation of about 6~d. Several other deviant stars have been further studied using Zeeman Doppler Imaging \citep*[ZDI,][]{SemelM:1989} to reconstruct their large-scale surface magnetic topologies (Petit et al., in prep.).  This allows an approximate determination of the stellar inclination angle by using the combined knowledge of the stellar radius, rotation period (estimated through the ZDI method) and \vsinis value. For at least two of these stars (HIP 104214 and HIP 56997) the estimated stellar inclination angle is above 60\degrees although it should be noted that they suffer from large error bars in the estimated inclination angle. We finally note that the twelve stars with higher \vsinis values are mostly G and K dwarfs. 

\subsection{The correlation of \Bla with other stellar activity indicators}

\begin{figure}
  \centering
  \includegraphics[trim =30 25 0 10, clip, angle=-90, width=\columnwidth]{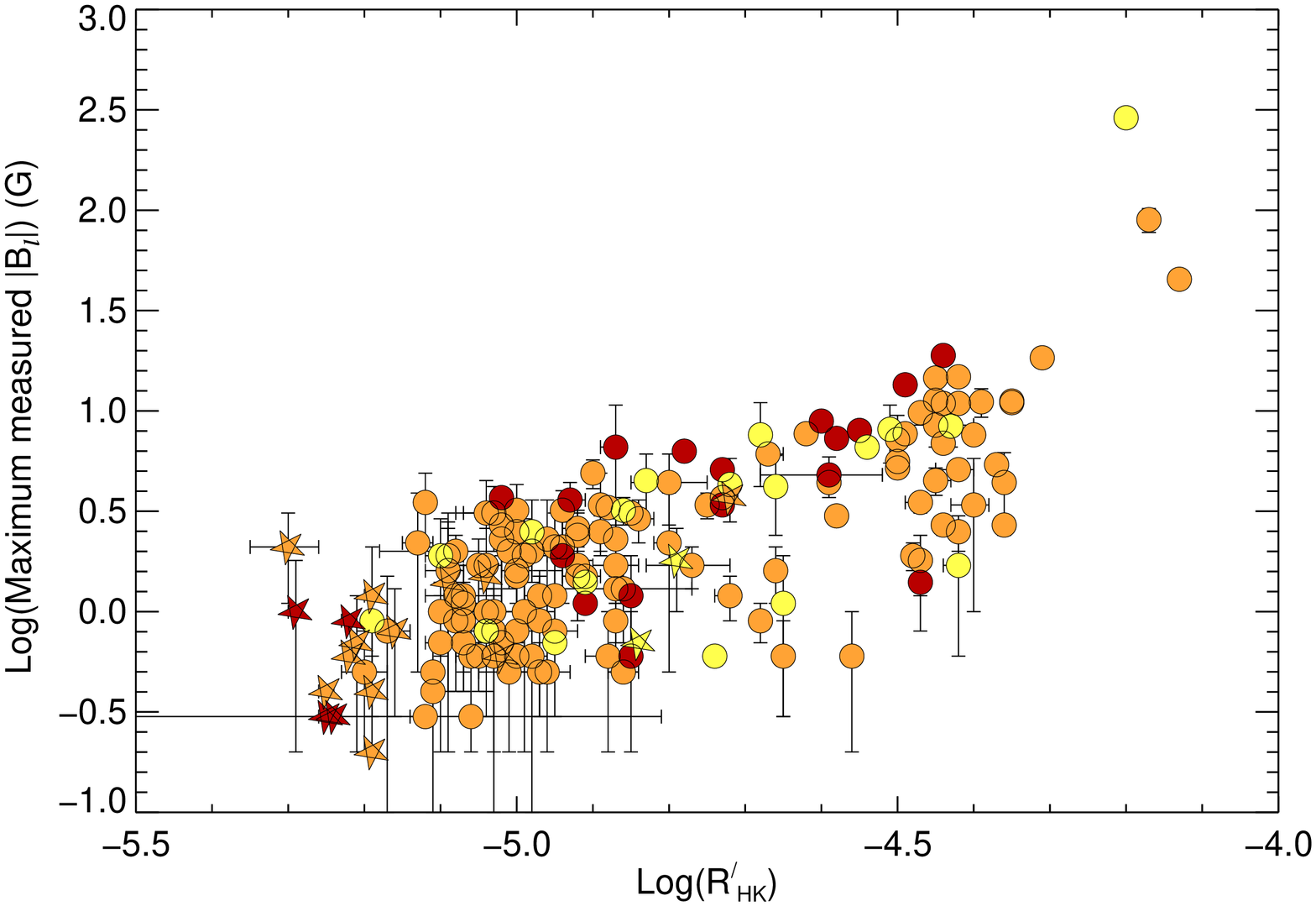}
  \includegraphics[trim =30 25 10 10, clip, angle=-90, width=\columnwidth]{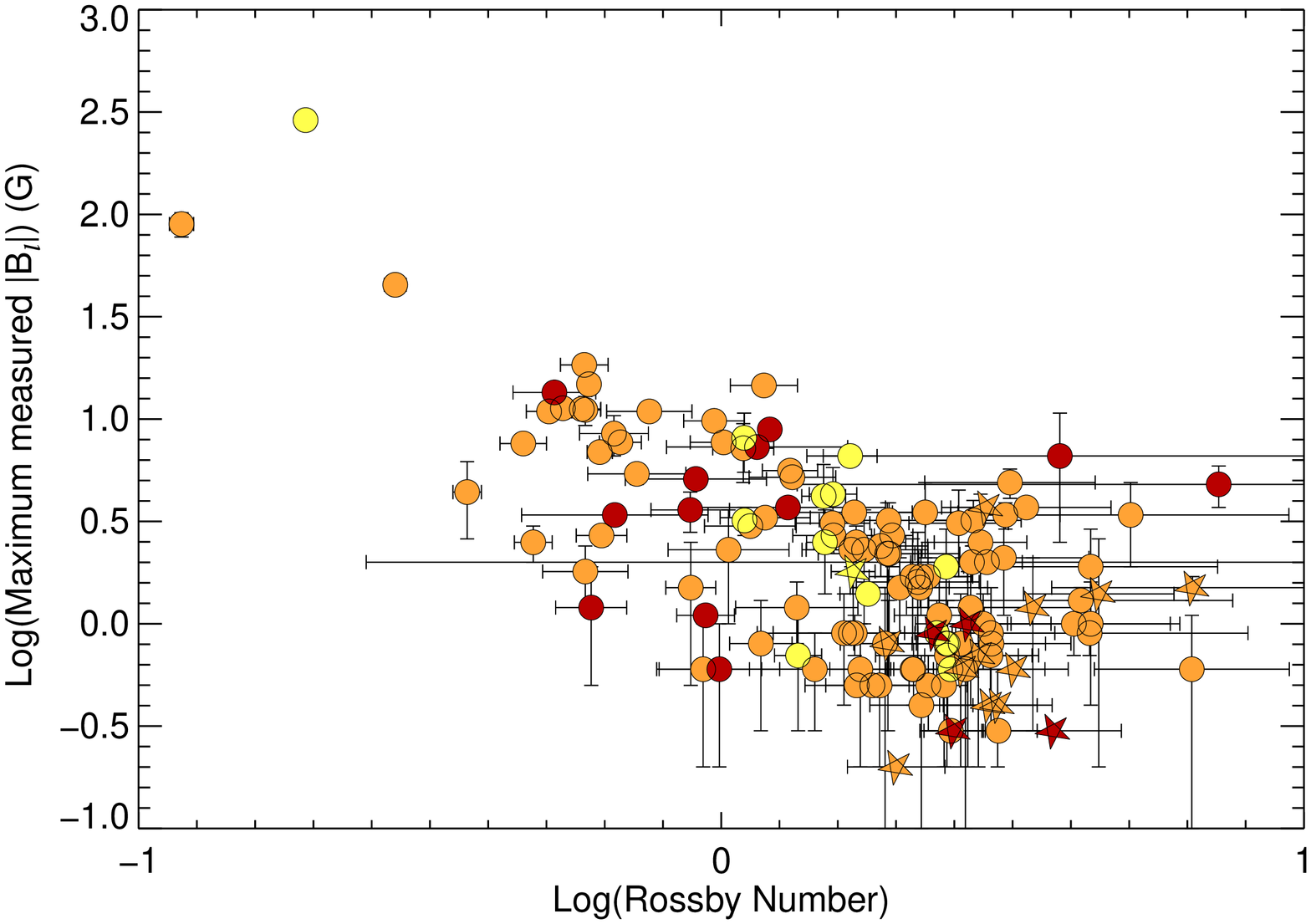}
  \caption{Plot of Log(maximum measured $|B_{l}|$) against Log(R\sups{\prime}\subs{HK}) (top) and Log(Rossby Number) (bottom). The symbols are the same as in Figure~\ref{Fig_HR}.}   
  \label{Fig_Blong_HK}
\end{figure}

\begin{figure}
  \centering
  \includegraphics[trim =30 25 0 10, clip, angle=-90, width=\columnwidth]{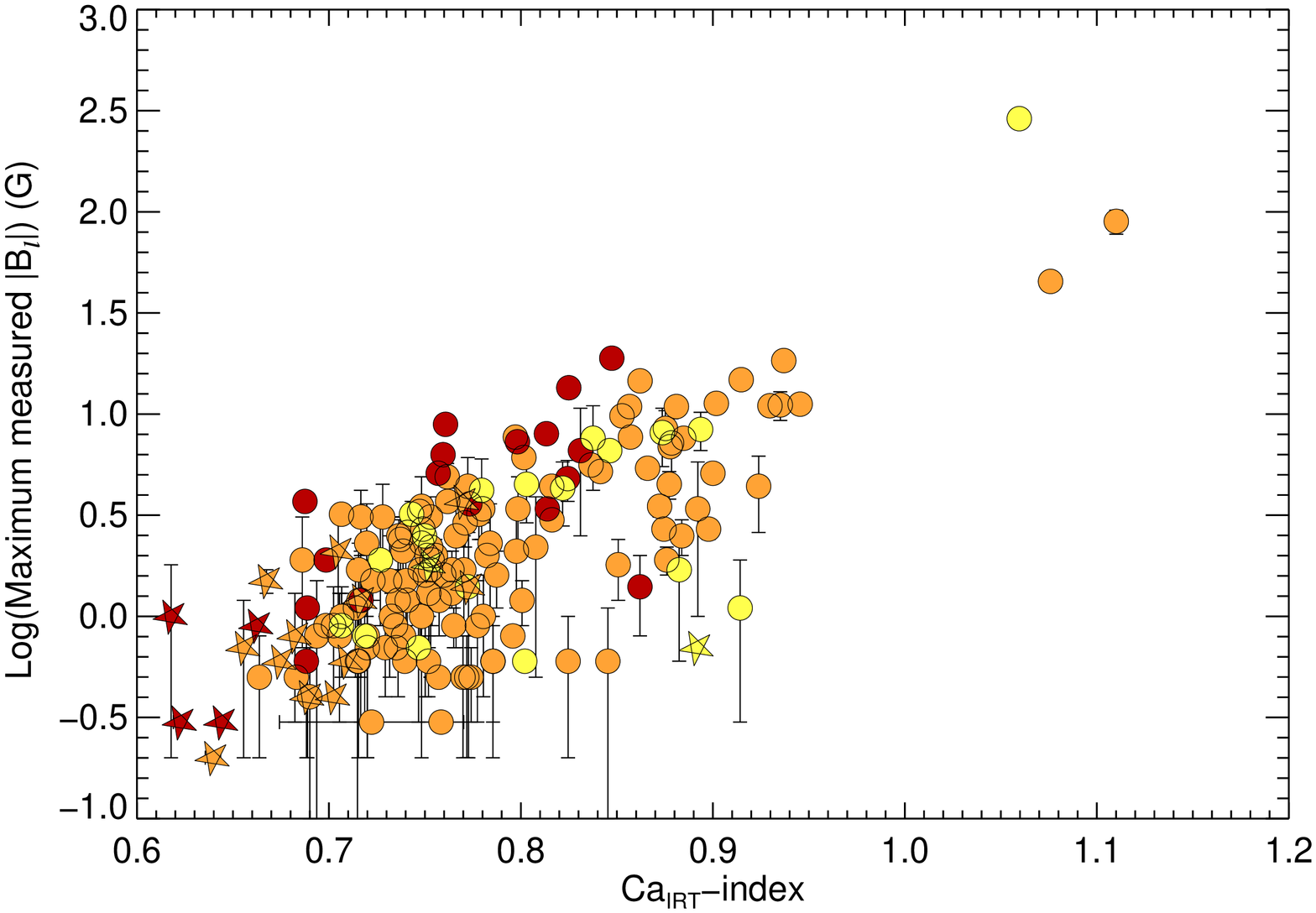}
  \includegraphics[trim =30 25 10 10, clip, angle=-90, width=\columnwidth]{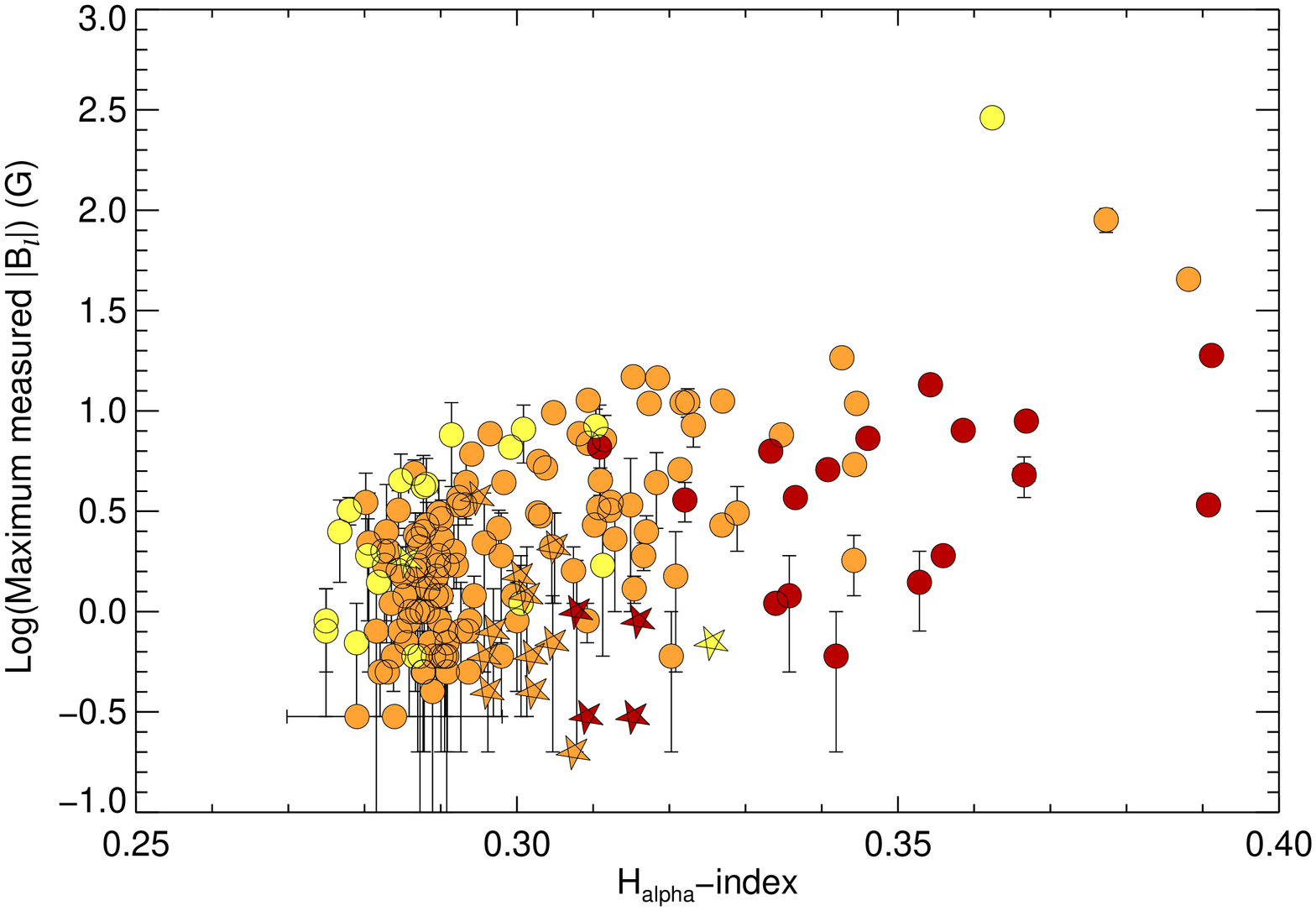}
  \caption{Plot of Log(maximum measured $|B_{l}|$) against the Ca\subs{IRT} activity index (top) and the H\subs{\alpha} activity index (bottom). The symbols are the same as in Figure~\ref{Fig_HR}.} 
  \label{Fig_Blong_indexes}
\end{figure}

The maximum measured \Bla shows a clear correlation with other activity indicators (see Figures~\ref{Fig_Blong_HK} and~\ref{Fig_Blong_indexes}) with the upper envelope of \Bla increasing with the other activity indices. When plotted against the Rossby number, \Bla shows a similar increase to that of Log(R\sups{\prime}\subs{HK}) taking note that the Rossby number has been determined using the formula from \citet[][Equations 13 and 14]{WrightJT:2004} which is based on the Log(R\sups{\prime}\subs{HK}) value of the star. The outlying stars shown at low \vsinis (see above) follow the general trend in this plot, confirming the hypothesis of an inclination effect to explain their relatively high surface magnetic fields.

\begin{figure}
  \centering
  \includegraphics[trim =25 25 5 15, clip, angle=-90, width=\columnwidth]{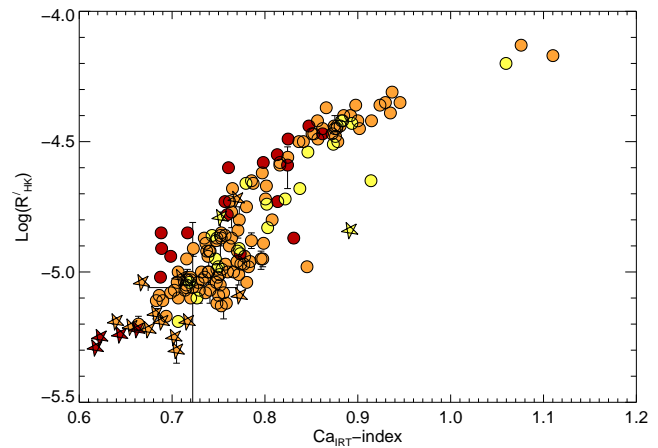}
  \caption{Plot of Log(R\sups{\prime}\subs{HK}) against the Ca\subs{IRT} activity index. The symbols are the same as in Figure~\ref{Fig_HR}.} 
  \label{Fig_HK_Cairt}
\end{figure}

We have extracted both the \Ca H \& K and \Ca IRT emission from our observations, as described in Section~\ref{Sec_irtha}.  These two activity indicators have been observed simultaneously and are compared in Figure~\ref{Fig_HK_Cairt}, where they are seen to show a strong correlation.  In contrast, the comparison of maximum measured \Bla with the H$\alpha$-index (Figure~\ref{Fig_Blong_indexes}) indicates a different trend.  This is due to a strong temperature dependence for the H$\alpha$-index, where F-stars show a lower core emission than K-stars.  This temperature dependence for H$\alpha$ is consistent with previous studies by \citep{BoppBW:1988, StrassmeierKG:1990} and is not relevant for the other activity proxies (Log(R\sups{\prime}\subs{HK}) and Ca\subs{IRT}-index).

\subsection{The comparison with magnetic field measurements from Stokes I observations} 
 
As previously discussed, the measurements of \Bla reflect only a fraction of the full magnetic flux on the star. Observations of Stokes I and V have shown that typical values for the ratio between the full average field (from Stokes I) and the large scale field (Stokes V) are on the order of 10\% or lower \citep{ReinersA:2012,ReinersA:2009}. From our sample, measurements of average field strengths from Stokes I are available for five stars: HIP 3765 \citep{MarcyGW:1989}, HIP 15457 \citep{SaarSH:1992}, HIP 16537, HIP 96100 \citep{ValentiJA:1995}, and HIP 88972 \citep{BasriG:1988}, see \cite{ReinersA:2012} for an overview and a discussion of different methods. In four of the five stars, measured average fields from Stokes I are around 200-300 G, only for HIP 96100 the field is lower with an upper detection limit of 100 G. The ratios between field strengths in Stokes I and V are below 7\% in all five stars. Measurements in Stokes I and V were taken at different times in all cases, hence the systematic error in this fraction probably depends mostly on the field geometry visible at the time of observation (for both Stokes I and V).

\section{Conclusions}

One of the strongest threads of the Bcool project is the study of main-sequence solar-type stars, aimed at understanding of how magnetic fields are generated in stars with physical parameters close to solar values. We have observed a large sample (170) of solar-type stars (on the main sequence or at the subgiant stage) using spectropolarimetric observations, in order to detect and characterise their surface magnetic fields. This dataset will continue to expand and forms the basis of the Bcool solar-type stars sample, on a sub-sample of which further analysis will be undertaken, such as the mapping of the magnetic topologies (Petit et al., in prep.) or long-term magnetic monitoring of specific targets in a search for magnetic cycles.

Prior to this paper, the majority of the magnetic field detections for solar-type stars have been achieved on young stars \citep[e.g.][]{DonatiJF:1997, DonatiJF:2003, JeffersSV:2011, MarsdenSC:2011a}, with magnetic field detections for mature solar-type stars limited to only five such targets \citep{PetitP:2008, FaresR:2012, FossatiL:2013}. This project has detected magnetic fields on 67 stars from our stellar sample, with 21 of our stars with detections (and age estimates) being classified as mature age solar-type stars, 20 of which have not been reported previously, and so constituting a significant addition to previous studies.

Of the 18 subgiant stars that have been observed as part of the survey, we report a marginal detection of a surface magnetic field on one of the stars (HIP 5493) and definite detections of magnetic fields on two of them (HIP 109439 and HIP 113994). Further investigation is required to interpret these isolated magnetic detections on evolved objects for which slow rotation is expected and should lead to a low surface activity (as observed for the rest of the subgiants in our sample).  

We have shown that for stars with a \Ca H \& K S-index $\ga$ 0.3, we are almost certain to detect the presence of a magnetic field while for those stars with an S-index $\la$ 0.2 the detection rate falls to around 10\%, while the minimum signal-to-noise ratio required in the Stokes V LSD profile should be $>$ 20,000 to ensure a good detection rate. For stars with \Bla $\ga$ 5 G we are very likely to detect the surface magnetic field (given a reasonable signal-to-noise ratio) while for stars with \Bla $\la$ 1 G it is almost impossible to detect the surface magnetic field from individual measurements. 

The results presented here indicate that the upper envelope of the absolute value of the mean longitudinal surface magnetic field ($|B_{l}|$) appears to be directly related to chromospheric activity indicators. The strength of \Bl drops off with age and increases with rotation rate. The mean value of \Bla found from our measurements is 3.5 G, with K-stars having a higher mean \Bla (5.7 G) than F-stars (3.3 G) and G-stars (3.2 G), although the number of F- and K-stars in our sample is currently limited.   

This paper presents an extensive and unique dataset of the magnetic fields of solar-type stars, and a demonstration of the feasibility of more detailed follow-up studies. Subsequent mapping of the surface magnetic fields of a large number and wide range of these solar-type stars (including many sun-like stars) would be of great scientific value, by providing an empirical basis for understanding how dynamos vary according to fundamental stellar properties, and may vary over time to produce magnetic cycles or other variability. In addition, magnetic maps would provide an empirical basis for the numerical modelling of stellar coronae and winds, and possible wind impacts on any orbiting planets \citep[i.e.][]{VidottoAA:2012}. In overall terms the magnetic detections presented here are thus likely to provide the basis for an extensive set of detailed magnetic studies that can provide fresh insights into the nature of stellar magnetism.

\section*{Acknowledgments}

The authors would like to thank the staff at the TBL and CFHT for their excellent support in helping with the observations for this paper. S.V.J. acknowledges research funding by the Deutsche Forschungsgemeinschaft (DFG) under grant SFB 963/1, project A16. J.M. was funded by a fellowship of the Alexander von Humboldt Foundation in G\"{o}ttingen. R.F. acknowledges funding from the STFC. This research has made use of the SIMBAD database, operated at CDS, Strasbourg, France. Partly based on data from the Brazilian CFHT time allocation under the proposals 09.BB03, 11.AB05, Pi: J.D. do Nascimento.

\appendix

\section{Notes on individual stars}\label{Appendix_LSD}

\subsection{HIP 16537 ($\epsilon$ Eri)} \label{App_epseri}

\begin{figure}
  \centering
  \includegraphics[angle=-90, width=\columnwidth]{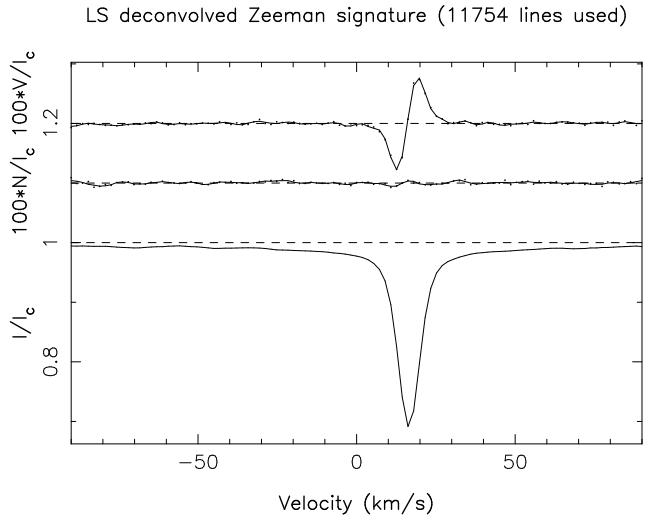}
  \caption{Plot of the LSD profile for HIP 16537 (Epsilon Eri), with the upper plot showing the Stokes V LSD profile (expanded by 100 times and shifted up by 0.2), the middle plot showing the Null LSD profile (again expanded by 100 times and shifted up by 0.1) and the lower plot showing the Stokes I LSD profile.} 
  \label{Fig_epsiloneri}
\end{figure}

Epsilon Eridani is an active K-star (K2V according to the SIMBAD data base) and a possible planet hosting star \citep{HatzesAP:2000, ButlerRP:2006, BenedictGF:2006, AngladaEscudeG:2012} and as such its magnetic field properties are of significant interest in the study of the impact of magnetic activity on the formation and evolution of orbiting planets. We have fifty-eight observations of eps Eri with a vast majority (fifty-two) showing definite detections of a magnetic field. The star is very young, with an estimated age of 0.0 to 0.6 Gyrs \citep[according to][]{TakedaG:2007}.  The star is a slow rotator with a \vsinis of 2.4 $\pm$ 0.4 \kmss and shows a very strong \Bl (-10.9 $\pm$ 0.2 G) for such low rotational broadening (it is one of the stars above the line in Figure~\ref{Fig_Blong_Vsini}). An individual map of eps Eridani is presented in Petit et al. (in prep.), while an analysis of the full data set (spanning 5 epochs of observations) is presented in Jeffers et al. (in prep.). The LSD profile of the star is shown in Figure~\ref{Fig_epsiloneri}.  

\subsection{HIP 71631 (EK Dra)}

\begin{figure}
  \centering
  \includegraphics[angle=-90, width=\columnwidth]{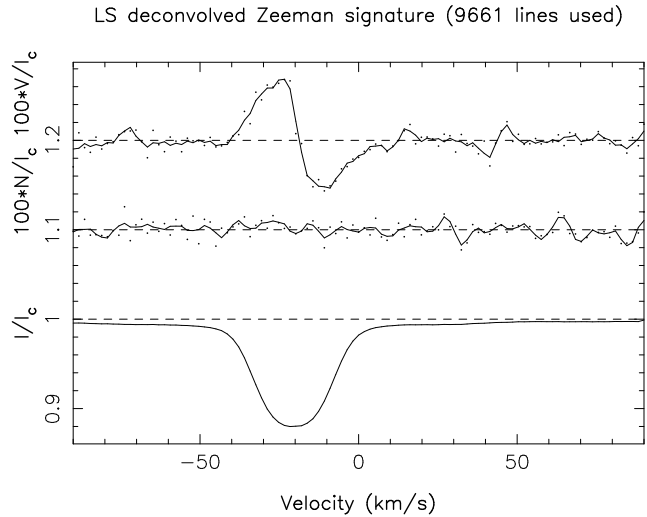}
  \caption{Plot of the LSD profile for HIP 71631 (EK Dra), with the plots as described in Figure~\ref{Fig_epsiloneri}.} 
  \label{Fig_ekdra}
\end{figure}

EK Dra is often described as an infant Sun (SIMBAD spectral type of G1.5V) with T\subs{eff} = 5845 $\pm$ 44 K, a mass of 1.044$^{+0.014}_{-0.020}$ \Msolar and an age of 0.00 to 1.44 Gyrs \citep{TakedaG:2007}. It is one of the most rapidly rotating, and hence active, stars in our sample with \vsinis = 16.8 $\pm$ 0.5 \kmss and \Bl = +45.3 $\pm$ 2.8 G. The LSD profile, showing its large Stokes V signature, is given in Figure~\ref{Fig_ekdra} and the magnetic maps of this star (using these observations) are presented in Waite et al. (in prep.), showing the evolution of the magnetic topology over several epochs of observations.

\subsection{HIP 79672 (18 Sco)}

\begin{figure}
  \centering
  \includegraphics[angle=-90, width=\columnwidth]{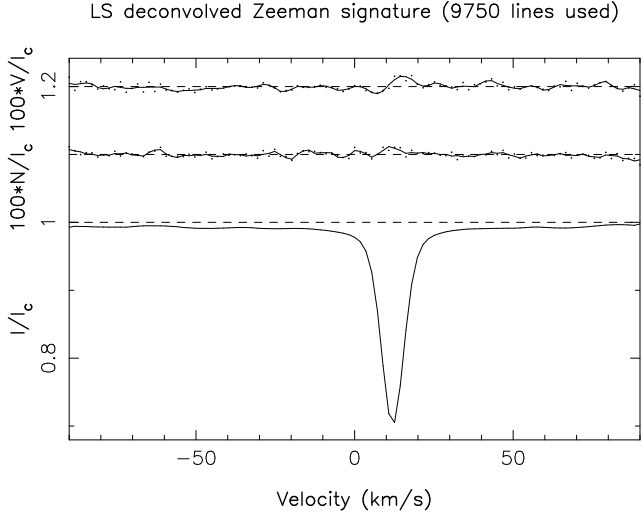}
  \caption{Plot of the LSD profile for HIP 79672 (18 Sco), with the plots as described in Figure~\ref{Fig_epsiloneri}.} 
  \label{Fig_18sco}
\end{figure}

18 Scorpius is an interesting target as it is often proposed as one of the best solar twins among bright stars (SIMBAD spectral type of G2Va) with T\subs{eff} = 5791 $\pm$ 44 K, Log(g) = 4.42 $\pm$ 0.03 cm s\sups{-2}, Log(M/H) = +0.03 $\pm$ 0.03, age = 5.84$^{+1.88}_{-1.96}$ Gyrs, mass = 1.005$^{+0.028}_{-0.024}$ and \vsinis = 2.6 $\pm$ 0.5 \kmsn. We have fifty-seven observations of 18 Sco, but only twelve have resulted in marginal or definite detections of a polarised signature. The maximum measured value of \Bl is only a few gauss (-2.3 $\pm$ 0.4 G) and as can be seen in Figure~\ref{Fig_18sco}, the strength of the magnetic signal in the Stokes V LSD profile is very weak. 18 Sco is very close to the limit of our observational capabilities with an S-index of 0.1834 at the epoch of the observation illustrated here. The magnetic topology of 18 Sco has previously been published (using the observations presented here) by \citet{PetitP:2008}. 

\subsection{HIP 91043 (V889 Her)}

\begin{figure}
  \centering
  \includegraphics[angle=-90, width=\columnwidth]{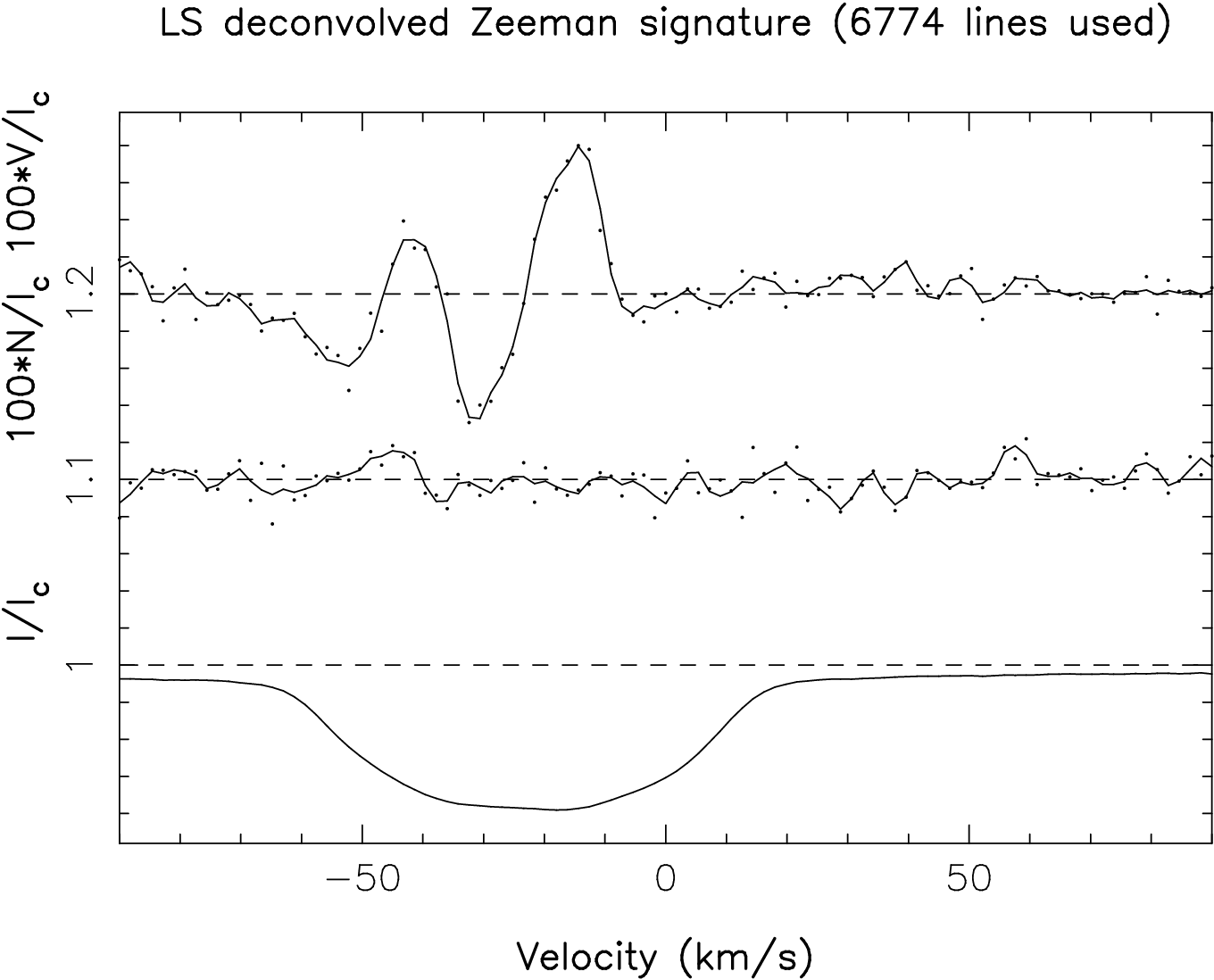}
  \caption{Plot of the LSD profile for HIP 91043 (V889 Her), with the plots as described in Figure~\ref{Fig_epsiloneri}.} 
  \label{Fig_hd171488}
\end{figure}

V889 Her is a young (age 30 - 50 Myrs, \citealt{StrassmeierKG:2003}) and active G-star (SIMBAD spectral type G2V). With the exception of its low metallicity (Log(M/H) = -0.5) V889 Her could well be considered an active young Sun and is typical of the active stars more frequently observed using spectropolarimetry. Given its youthfulness and rapid rotation (\vsinis = 39.0 $\pm$ 0.5 \kmsn) it is unsurprising that we have a large number of magnetic field detections for the star (48 out of 53 observations are definite detections). The maximum measured mean longitudinal field is significantly higher than that seen on the more mature (and less rapidly rotating) solar-type stars with a value of +89.7 $\pm$ 12.2 G. The LSD profile of the star is given in Figure~\ref{Fig_hd171488} while numerous magnetic field maps of the star have been previously published \citep{MarsdenSC:2006, JeffersSV:2008, JeffersSV:2011}. Additionally, because of its large \vsini, the spot features of V889 Her can also be mapped using Doppler Imaging and have also been published \citep{MarsdenSC:2006, JeffersSV:2008, JeffersSV:2011}.

\subsection{HIP 104214 (61 Cyg A)}

\begin{figure}
  \centering
  \includegraphics[angle=-90, width=\columnwidth]{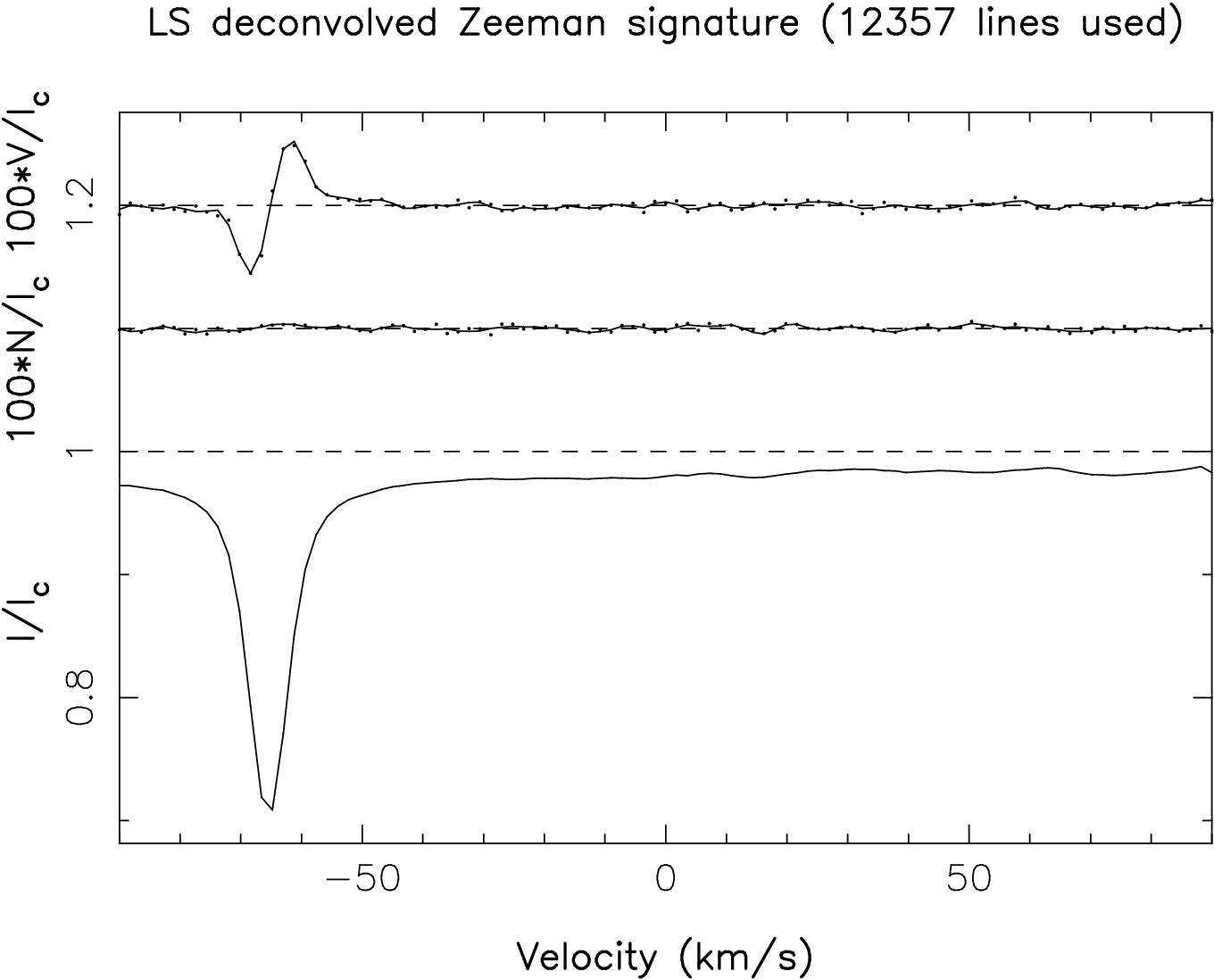}
  \caption{Plot of the LSD profile for HIP 104214 (61 Cyg A), with the plots as described in Figure~\ref{Fig_epsiloneri}.} 
  \label{Fig_61cyga}
\end{figure}

61 Cyg A is one of the coolest stars in our sample (along with $\xi$ Boo B) with T\subs{eff} = 4525 $\pm$ 140 K, is very young (an age of 0.00 to 0.44 Gyrs) and has a small \vsinis value ($\sim$1.1 \kmsn). The star's SIMBAD spectral type is given as K5V. Sixty-eight observations of 61 Cyg A have been obtained with over two-thirds (forty-nine) showing marginal or definite detection of a surface magnetic field. The maximum measured value of \Bl is quite high (-8.9 $\pm$ 0.3 G) given its low \vsinis. The LSD profile of the star is given in Figure~\ref{Fig_61cyga} and a map of the magnetic topology of the star is given in Petit et al. (in prep.).

\subsection{HIP 109439} \label{App_hip109439}

\begin{figure}
  \centering
  \includegraphics[angle=-90, width=\columnwidth]{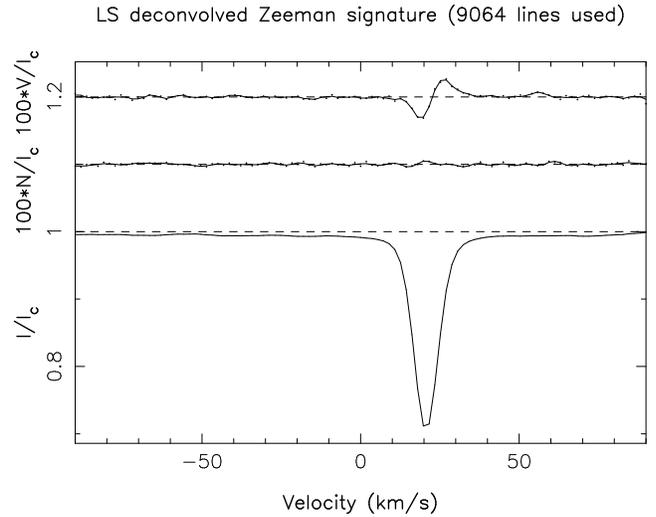}
  \caption{Plot of the LSD profile for HIP 109439, with the plots as described in Figure~\ref{Fig_epsiloneri}.} 
  \label{Fig_hip109439}
\end{figure}

HIP 109439 is an interesting target as it is a subgiant member of our sample, with T\subs{eff} = 5658 $\pm$ 44 K, Log(g) = 3.79$^{+0.04}_{-0.02}$ cm s\sups{-2}, a mass of 1.49$^{+0.23}_{-0.20}$ \Msolar and a radius of 2.95$^{+0.11}_{-0.02}$ \Rsolar. Most observed subgiant stars are low-activity objects, as can be seen in their \Ca H \& K measurements (see Table~\ref{Bcool_Activity}), although HIP 109439 is moderately active with Log(R$^{\prime}_{\rm{HK}}$) = -4.72. The Stokes V LSD profile of HIP 109439 is given in Figure~\ref{Fig_hip109439}. From our single observation we obtain a definite detection of a surface magnetic field. The estimated mass suggests that HIP 109439 may be the descendant of a main sequence star with a very shallow convective envelope, so that the generation of a detectable large-scale magnetic field may be relatively recent and linked to the growth of its post-main-sequence convective envelope. 

\subsection{HIP 113357 (51 Pegase)}

\begin{figure}
  \centering
  \includegraphics[angle=-90, width=\columnwidth]{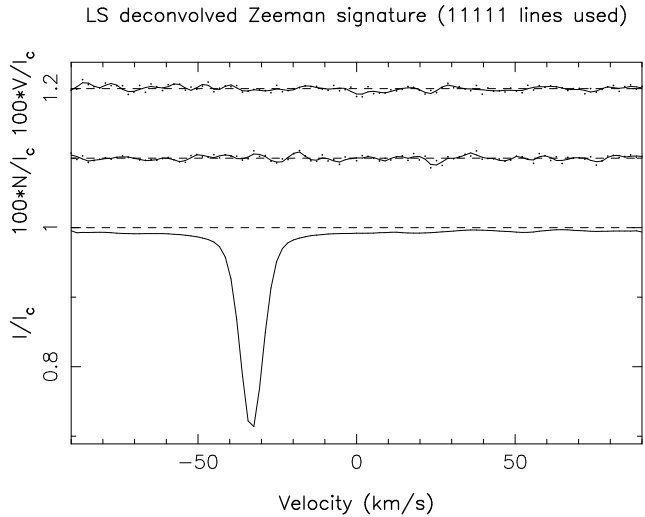}
  \caption{Plot of the LSD profile for HIP 113357 (51 Pegase), with the plots as described in Figure~\ref{Fig_epsiloneri}.} 
  \label{Fig_51pegase}
\end{figure}

51 Pegasi was the first Sun-like star (the SIMBAD spectral type is given as G2.5IVa) to be found to have a planetary companion. 51 Peg has similar properties to that of our own Sun with T\subs{eff} = 5787 $\pm$ 25 K, Log(M/H) = +0.15 $\pm$ 0.02,  \vsinis = 2.6 $\pm$ 0.3 \kmss and  an age of $\sim$7 Gyrs. The age and slow rotation rate of the star make it a challenging target for spectropolarimetry and unfortunately we have no detections of a magnetic field in the three observations we have for the star. The largest magnetic field we have been able to estimate on the star is \Bl = +0.6 $\pm$ 0.4 G. The LSD profile of the star is given in Figure~\ref{Fig_51pegase}.

\subsection{HIP 113994}

\begin{figure}
  \centering
  \includegraphics[angle=-90, width=\columnwidth]{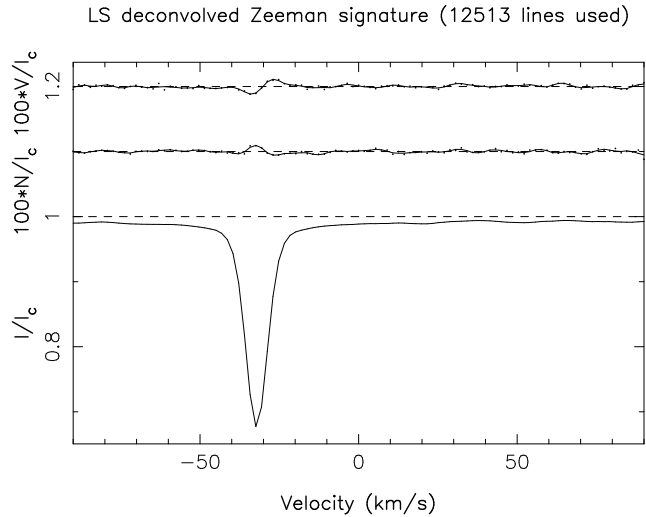}
  \caption{Plot of the LSD profile for HIP 113994, with the plots as described in Figure~\ref{Fig_epsiloneri}.} 
  \label{Fig_hip113994}
\end{figure}

Like HIP 109439 (see Section~\ref{App_hip109439}) HIP 113994 is classified by us as a subgiant star with T\subs{eff} = 5301 $\pm$ 44 K, Log(g) = 3.96$^{+0.02}_{-0.01}$ cm s\sups{-2}, a mass of 2.02 $\pm$ 0.31 \Msolar and a radius of 2.48$^{+0.02}_{-0.09}$ \Rsolar. The SIMBAD spectral type for the star is given as G8IV. We have only one observation of this star, corresponding to a definite detection (false alarm probability = 2.243 $\times$ 10\sups{-14}) of a surface magnetic field, in spite of its rather low Log(R$^{\prime}_{\rm{HK}}$) value of -5.04. The star only just has a Log(g) below 4.0 cm s\sups{-2} but the mass and radius, compared to its effective temperature, would indicate that this is a subgiant star worthy of further study, since it may be the descendant of a main-sequence star too massive to have deep convective outer layers, and therefore not suited to generate an efficient solar-type dynamo. The Stokes V profile of the star is given in Figure~\ref{Fig_hip113994}. We also record a bump in the Null profile (with an amplitude smaller than that of the Stokes V profile), so that the star should be re-examined to determine if this detection can be confirmed.

\subsection{HIP 114378}

\begin{figure}
  \centering
  \includegraphics[angle=-90, width=\columnwidth]{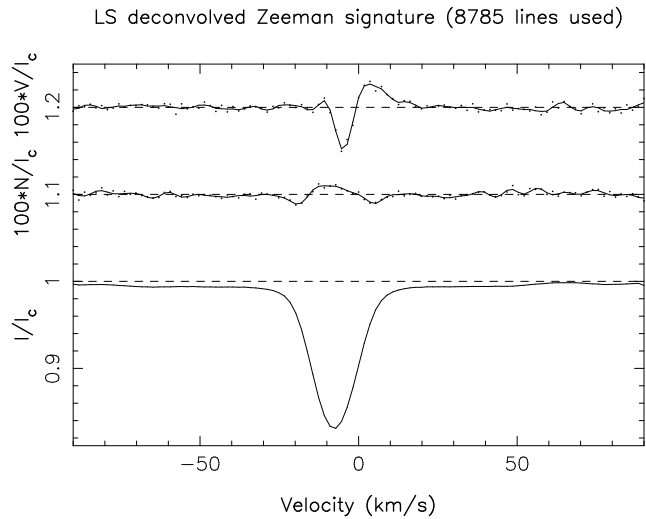}
  \caption{Plot of the LSD profile for HIP 114378, with the plots as described in Figure~\ref{Fig_epsiloneri}.} 
  \label{Fig_hip114378}
\end{figure}

HIP 114378 is listed in Table~\ref{Bcool_params} as a G-star (SIMBAD spectral type of G0V) with a mass near solar, an age of $\sim$7 Gyrs and a \vsinis of $\sim$10 \kmsn. The star has been observed twice, with both observations yielding definite detections. The star is slightly unusual in that the magnetic field measured ($B_{l}$) is rather strong (-11.1 $\pm$ 1.8 G) for its age, as derived from evolutionary tracks. Using equation 13 from \citet{WrightJT:2004} gives an age of 0.22 to 0.35 Gyr, based on the range of its Log(R$^{\prime}_{\rm{HK}}$) values (Table~\ref{Bcool_Activity}). If the evolutionary age of \citet{TakedaG:2007} is correct, the high activity of this star may indicate that the star is a member of a close binary system, although no evidence of a secondary companion is seen in its Stokes I LSD profile (Figure~\ref{Fig_hip114378}). The Null profile of this star shows a statistically significant signature in the line profile, but the variations in the Stokes V LSD profile are well above those seen in the diagnostic profile. We therefore consider the magnetic field detection to be robust, although error bars on \Bl may be underestimated. 

\begin{figure}
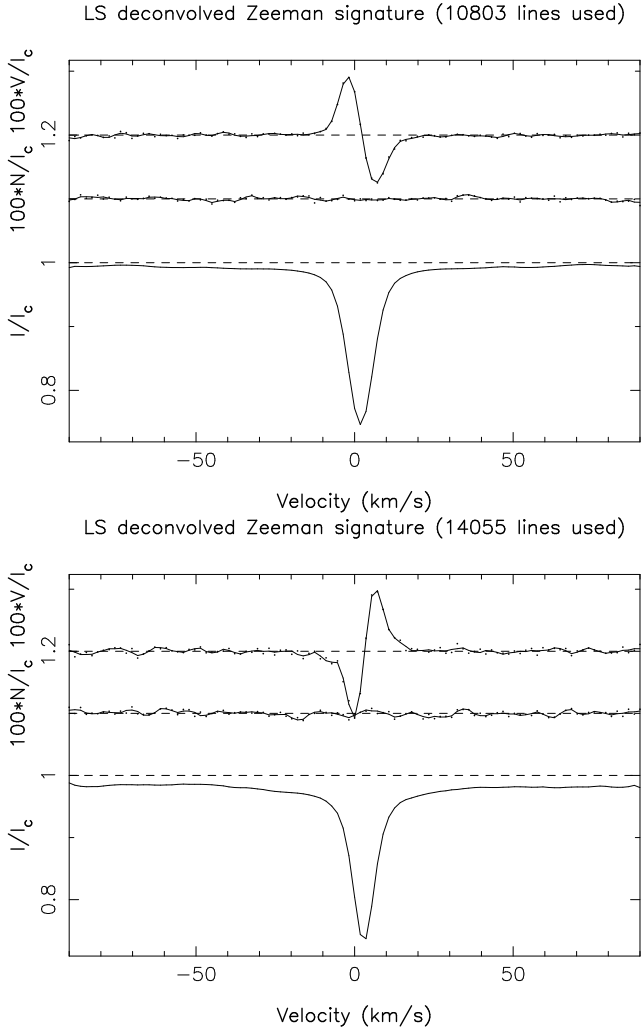

  \centering
  \includegraphics[angle=-90, width=\columnwidth]{FigA10a.ps}
  \includegraphics[angle=-90, width=\columnwidth]{FigA10b.ps}
  \caption{Plot of the LSD profile for $\xi$ Boo system, HD 131156A (top) and HD 131156B (bottom), with the plots as described in Figure~\ref{Fig_epsiloneri}.} 
  \label{Fig_xiboo}
\end{figure}

\begin{figure}
  \centering
  \includegraphics[angle=-90, width=\columnwidth]{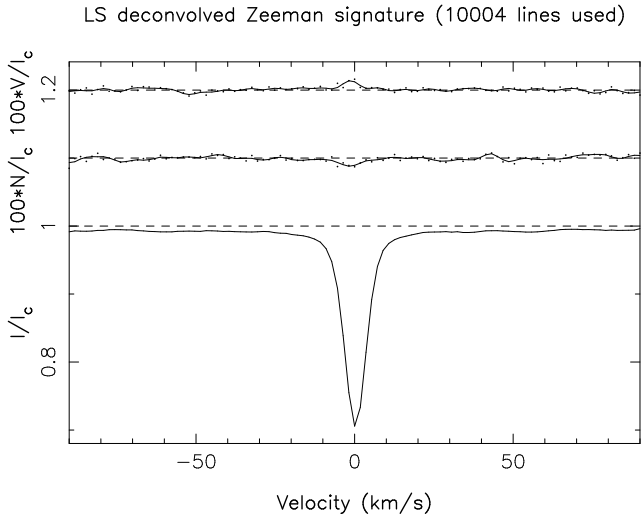}
  \caption{Plot of the LSD profile for the Moon, with the plots as described in Figure~\ref{Fig_epsiloneri}.} 
  \label{Fig_moon}
\end{figure}

\subsection{HD 131156 A \& B ($\xi$ Boo A \& B)}

The ksi Boo system is a very long period binary (orbital period $\sim$151 years, \citealt{WielenR:1962}) with the A component being an active G-star (SIMBAD spectral type of G8V) while the B component is an active K-star (SIMBAD spectral type of K4V). Ksi Boo A has been studied previously as part of the Bcool solar-type star project \citep{MorgenthalerA:2012} with magnetic topologies reconstructed for seven different epochs from 2007 to 2011, showing magnetic field evolution but with no cycle yet determined. The age of the two components of the system (according to \citealt{TakedaG:2007}) is markedly different (0.00 to 0.76 Gyrs for A and 12.60 to 13.80 Gyrs for B), but given the activity of the two components, the system is most likely young, with the activity-based ages from \citet{WrightJT:2004} being 0.00 to 0.35 Gyrs for A and 0.18 to 3.89 Gyrs for B, based on the range of their Log(R$^{\prime}_{\rm{HK}}$) values (Table~\ref{Bcool_Activity}). The radii and masses of the system (again determined by \citealt{TakedaG:2007}) are very similar (with B being slightly higher than A) despite their significant temperature differences, as thus the derived parameters for these two stars should be treated with caution. Both stars gave definite detections for surface magnetic fields and the LSD profiles of these stars are given in Figure~\ref{Fig_xiboo}. A magnetic map of ksi Boo B is given in Petit et al. (in prep.).

\subsection{The Moon}

We use here an observation of solar light reflected on the moon as a proxy of a Sun-as-a-star spectropolarimetric observation. The low activity of the Sun makes it a difficult target for spectropolarimetry and although the LSD profile (Figure~\ref{Fig_moon}) shows as small bump in Stokes V (along with a small dip in the Null parameter) we did not get a magnetic detection for our only spectrum.

\section{Summary of Spectropolarimetric Observations}\label{Appendix_OBS}

As mentioned previously (Section~\ref{Sec_obs}) the data for this project have been obtained during 25 observing runs at the TBL and CFHT using the NARVAL and ESPaDOnS spectropolarimeters, respectively. The observations were obtained from late-2006 through to mid-2013 with the majority of the observing runs (19/25 = 76\%) being carried out at the TBL. The majority of the targets have only been observed once (117/170 = 69\%) or at only one epoch of observations (135/170 = 79\%), however, a number of stars have been observed at multiple epochs to look for temporal evolution in their magnetic field properties. These stars will be the subjects of forthcoming Bcool papers. A summary of the number of observations for each star at each observing epoch is given in Table~\ref{Bcool_obs}.  

\begin{landscape}
\begin{table}
\scriptsize
\begin{center}
\caption{Summary of the number of observations of each star at each observing epoch in the Bcool solar-type stars sample. The top section of the table lists the telescope used (TBL: Telescope Bernard Lyot, CFH: Canada-France-Hawaii Telescope) and the year, start month and end month of the observations. \sups{SG}: identifies the star as a subgiant, see Figure~\ref{Fig_HR}.} \label{Bcool_obs} 
\begin{tabular}{lccccccccccccccccccccccccc}
\hline
Telescope: & TBL & CFH & TBL & TBL & TBL & TBL & CFH & TBL & TBL & CFH & TBL & CFH & TBL & TBL & TBL & CFH & TBL & TBL & TBL & CFH & TBL & TBL & TBL & TBL & TBL\\
Year (2000+): & 06 & 06 & 07 & 07 & 07 & 07 & 07 & 07/08 & 08 & 08 & 08 & 08 & 08 & 09 & 09 & 09/10 & 09/10 & 10 & 10/11 & 11 & 11 & 11/12 & 12 & 12/13 & 13 \\
Start: & nov & nov & jun & may & jul & nov & dec & dec & jan & mar & may & jun & jul & jan & may & sep & dec & mar & sep & feb & mar & oct & mar & oct & mar \\
End: & nov & dec & feb & may & aug & nov & dec & jan & feb & mar & may & jul & aug & jan & jul & jan & feb & aug & feb & jul & aug & feb & aug & jan & aug \\
HIP \# &  &  &  &  &  &  &  &  &  &  &  &  &  &  &  &  &  &  &  &  &  &  &  & \\
\hline
400 & - & - & - & - & - & - & - & - & - & - & - & - & - & - & - & - & - & - & 1 & - & - & - & - & - & -\\
544 & - & - & - & - & - & - & - & - & - & - & - & - & - & - & - & - & - & - & 1 & - & - & - & - & - & -\\
682 & - & - & - & - & - & - & - & - & - & - & - & - & - & - & - & - & - & - & 1 & - & - & - & - & - & -\\
1499 & - & - & - & - & - & - & - & - & - & - & - & - & - & - & - & - & - & - & 1 & - & - & - & - & - & -\\
1813 & - & - & - & - & - & - & - & - & - & - & - & - & - & - & - & - & - & - & 1 & - & - & - & - & - & -\\
3093 & - & - & - & - & 2 & - & - & - & - & - & - & - & - & - & - & - & - & 13 & - & - & 3 & - & 10 & - & -\\
3203 & - & - & - & - & - & - & - & - & - & - & - & - & - & - & - & - & - & - & 1 & - & - & - & - & - & -\\
3206 & - & - & - & - & - & - & - & - & - & - & - & - & - & - & - & - & - & - & 1 & - & - & - & - & - & -\\
3765 & - & - & - & - & 1 & - & - & - & - & - & - & - & - & - & - & - & - & - & 1 & - & - & - & - & - & -\\
3821 & - & - & - & - & - & - & - & - & - & - & - & - & - & - & - & - & - & - & 1 & - & - & - & - & - & -\\
3979 & - & - & - & - & - & - & - & - & - & - & - & - & - & - & - & - & - & - & 1 & - & - & - & - & - & -\\
4127 & - & - & - & - & - & - & - & - & - & - & - & - & - & - & - & - & - & - & 1 & - & - & - & - & - & -\\
5315\sups{SG} & - & - & - & - & - & - & - & - & - & - & - & - & - & - & - & - & - & - & 1 & - & - & - & - & - & -\\
5493\sups{SG} & - & - & - & - & 4 & - & - & - & - & - & - & - & - & - & - & - & - & - & - & - & - & - & - & - & -\\
5985 & - & - & - & - & - & - & - & - & - & - & - & - & - & - & - & - & - & - & 1 & - & - & - & - & - & -\\
6405 & - & - & - & - & - & - & - & - & - & - & - & - & - & - & - & - & - & - & 1 & - & - & - & - & - & -\\
7244 & - & - & - & - & - & - & - & - & - & - & - & - & - & - & - & 1 & - & - & - & - & - & - & - & - & -\\
7276 & - & - & - & - & - & - & - & - & - & - & - & - & - & - & - & - & - & - & 1 & - & - & - & - & - & -\\
7339 & - & - & - & - & - & - & - & - & - & - & - & - & - & - & - & - & - & - & 1 & - & - & - & - & - & -\\
7513 & 3 & - & - & - & - & - & - & - & - & - & - & - & - & - & - & - & - & - & - & - & - & - & - & - & -\\
7585 & - & - & - & - & - & - & - & - & 20 & - & - & - & - & - & - & - & - & - & - & - & - & 11 & - & 13 & -\\
7734 & - & - & - & - & - & - & - & - & - & - & - & - & - & - & - & - & - & - & 1 & - & - & - & - & - & -\\
7918 & 1 & - & - & - & - & - & - & - & - & - & - & - & - & - & - & 1 & - & - & - & - & - & - & - & - & -\\
7981 & - & - & 7 & - & 13 & - & - & - & 19 & - & - & - & - & - & - & - & - & - & - & - & - & - & - & - & -\\
8159 & - & - & - & - & - & - & - & - & - & - & - & - & - & - & - & - & - & - & 1 & - & - & - & - & - & -\\
8362 & - & - & - & - & - & - & - & - & - & - & - & - & - & - & - & - & - & - & 1 & - & - & - & - & - & -\\
8486 & - & - & - & - & - & - & - & - & - & - & - & - & - & - & - & 1 & - & - & - & - & - & - & - & - & -\\
9349 & - & - & - & - & - & - & - & - & - & - & - & - & - & - & - & 1 & - & - & - & - & - & - & - & - & -\\
9406\sups{SG} & - & - & - & - & - & - & - & - & - & - & - & - & - & - & - & - & - & - & 1 & - & - & - & - & - & -\\
9829 & - & - & - & - & - & - & - & - & - & - & - & - & - & - & - & - & - & - & 1 & - & - & - & - & - & -\\
9911 & - & - & - & - & - & - & - & - & - & - & - & - & - & - & - & - & - & - & 1 & - & - & - & - & - & -\\
10339 & - & - & - & - & - & - & - & - & - & - & - & - & - & - & - & 1 & - & - & - & - & - & - & - & - & -\\
10505 & - & - & - & - & - & - & - & - & - & - & - & - & - & - & - & - & - & - & 1 & - & - & - & - & - & -\\
11548 & - & - & - & - & - & - & - & - & - & - & - & - & - & - & - & - & - & - & 1 & - & - & - & - & - & -\\
12048 & - & - & - & - & - & - & - & - & - & - & - & - & - & - & - & - & - & - & 1 & - & - & - & - & - & -\\
12114 & - & - & 1 & - & - & - & - & - & - & - & - & - & - & - & - & - & - & - & 1 & - & - & - & - & - & -\\
13702 & - & - & 1 & - & - & - & - & - & - & - & - & - & - & - & - & - & - & - & - & - & - & - & - & - & -\\
14150 & - & - & - & - & - & - & - & - & - & - & - & - & - & - & - & - & - & - & 1 & - & - & - & - & - & -\\
15457 & - & - & 1 & - & - & - & - & - & - & - & - & - & - & - & - & - & - & - & 1 & - & - & - & - & 14 & -\\
15776\sups{SG} & - & - & - & - & - & - & - & - & - & - & - & - & - & - & - & - & - & - & 1 & - & - & - & - & - & -\\
16537 & - & - & 8 & - & - & - & - & - & 21 & - & - & - & - & - & - & - & 4 & - & - & - & - & 12 & - & 13 & -\\
16641\sups{SG} & - & - & - & - & - & - & - & - & - & - & - & - & - & - & - & - & - & - & 1 & - & - & - & - & - & -\\
17027\sups{SG} & - & - & - & - & - & - & - & - & - & - & - & - & - & - & - & - & - & - & 1 & - & - & - & - & - & -\\
17147 & - & - & - & - & - & - & - & - & - & - & - & - & - & - & - & - & - & - & 1 & - & - & - & - & - & -\\
17183\sups{SG} & - & - & - & - & - & - & - & - & - & - & - & - & - & - & - & - & - & - & 1 & - & - & - & - & - & -\\
17378\sups{SG} & - & - & - & - & - & - & - & - & - & - & - & - & - & - & - & - & - & - & 1 & - & - & - & - & - & -\\
18106 & - & - & - & - & - & - & - & - & - & - & - & - & 7 & - & - & - & - & - & - & - & - & - & - & - & -\\
18267 & - & - & - & - & - & - & - & - & - & - & - & - & - & - & - & - & - & - & 1 & - & - & - & - & - & -\\
18606\sups{SG} & - & - & - & - & - & - & - & - & - & - & - & - & - & - & - & - & - & - & 1 & - & - & - & - & - & -\\
19076 & - & - & - & - & - & - & - & - & - & - & - & - & - & - & - & - & - & - & 1 & - & - & - & - & - & -\\
19849 & - & - & - & - & - & - & - & - & - & - & - & - & - & - & - & - & - & - & 1 & - & - & - & - & - & -\\
\hline
\end{tabular}
\end{center}
\end{table} 
\end{landscape}

\addtocounter {table} {-1}

\begin{landscape}
\begin{table}
\scriptsize
\begin{center}
\caption{continued.} 
\begin{tabular}{lccccccccccccccccccccccccc}
\hline
Telescope: & TBL & CFH & TBL & TBL & TBL & TBL & CFH & TBL & TBL & CFH & TBL & CFH & TBL & TBL & TBL & CFH & TBL & TBL & TBL & CFH & TBL & TBL & TBL & TBL & TBL\\
Year (2000+): & 06 & 06 & 07 & 07 & 07 & 07 & 07 & 07/08 & 08 & 08 & 08 & 08 & 08 & 09 & 09 & 09/10 & 09/10 & 10 & 10/11 & 11 & 11 & 11/12 & 12 & 12/13 & 13 \\
Start: & nov & nov & jun & may & jul & nov & dec & dec & jan & mar & may & jun & jul & jan & may & sep & dec & mar & sep & feb & mar & oct & mar & oct & mar \\
End: & nov & dec & feb & may & aug & nov & dec & jan & feb & mar & may & jul & aug & jan & jul & jan & feb & aug & feb & jul & aug & feb & aug & jan & aug \\
HIP \# &  &  &  &  &  &  &  &  &  &  &  &  &  &  &  &  &  &  &  &  &  &  &  & \\
\hline
19925 & - & - & - & - & - & - & - & - & - & - & - & - & - & - & - & 1 & - & - & - & - & - & - & - & - & -\\
20800 & - & - & - & - & - & - & - & - & - & - & - & - & - & - & - & - & - & - & 1 & - & - & - & - & - & -\\
22319\sups{SG} & - & - & - & - & - & - & - & - & - & - & - & - & - & - & - & - & - & - & 1 & - & - & - & - & - & -\\
22336 & - & - & - & - & - & - & - & - & - & - & - & - & - & - & - & - & - & - & 1 & - & - & - & - & - & -\\
22449 & - & - & - & - & - & - & - & - & - & - & - & - & - & - & - & - & - & - & - & - & - & - & - & 1 & -\\
22633\sups{SG} & - & - & - & - & - & - & - & - & - & - & - & - & - & - & - & - & - & - & 1 & - & - & - & - & - & -\\
23311 & - & - & - & - & - & - & - & - & - & - & - & - & - & - & - & - & - & - & - & - & - & - & - & 1 & -\\
24813 & - & - & - & - & - & - & - & - & 1 & - & - & - & - & - & - & - & - & - & - & - & - & - & - & - & -\\
25278 & - & - & 7 & - & - & - & - & - & 23 & - & - & - & - & 1 & - & - & - & - & - & - & - & - & - & - & -\\
25486 & - & - & - & - & - & - & 22 & 38 & - & - & - & - & - & - & - & - & - & - & - & - & - & - & - & - & -\\
27913 & - & - & 9 & - & - & - & - & - & 21 & - & - & - & - & 1 & - & - & - & - & - & - & - & 13 & - & - & -\\
29568 & - & - & - & - & - & - & - & - & - & - & - & - & - & - & - & 1 & - & - & - & - & - & - & - & - & -\\
30476 & - & - & - & - & - & - & - & - & - & 1 & - & - & - & - & - & - & - & - & - & - & - & - & - & - & -\\
31965 & - & - & - & - & - & - & - & - & - & - & - & - & - & - & - & 1 & - & - & - & - & - & - & - & - & -\\
32673 & - & - & - & - & - & - & - & - & - & - & - & - & - & - & - & 1 & - & - & - & - & - & - & - & - & -\\
32851 & - & - & - & - & - & - & - & - & - & - & - & - & - & - & - & - & - & - & - & - & - & 1 & - & - & -\\
33277 & - & - & 3 & - & - & - & - & - & - & - & - & - & - & - & - & - & - & - & - & - & - & - & - & - & -\\
35185 & - & - & - & - & - & - & - & - & - & - & - & - & - & - & - & 1 & - & - & - & - & - & - & - & - & -\\
35265 & - & - & - & - & - & - & - & - & 21 & - & - & - & - & 1 & - & - & - & - & - & - & - & 10 & - & 4 & -\\
36704 & 1 & - & - & - & - & - & - & - & - & - & - & - & - & - & - & - & - & - & - & - & - & - & - & - & -\\
38018 & - & - & - & - & - & - & - & - & - & - & - & - & - & 1 & - & - & - & - & - & - & - & - & - & - & -\\
38228 & - & - & - & - & - & - & - & - & - & - & - & - & - & - & - & 1 & - & - & - & - & - & - & - & - & -\\
38647 & - & - & - & - & - & - & - & - & - & - & - & - & - & - & - & 1 & - & - & - & - & - & - & - & - & -\\
38747 & - & - & - & - & - & - & - & - & - & - & - & - & - & - & - & 1 & - & - & - & - & - & - & - & - & -\\
41484 & - & - & - & - & - & - & - & - & 1 & 2 & - & - & - & - & - & 1 & - & - & - & - & - & - & - & - & -\\
41526 & - & - & - & - & - & - & - & - & - & - & - & - & - & - & - & 1 & - & - & - & - & - & - & - & - & -\\
41844 & - & - & - & - & - & - & - & - & - & - & - & - & - & - & - & 1 & - & - & - & - & - & - & - & - & -\\
42333 & - & - & 9 & - & - & - & - & - & - & - & - & - & - & 1 & - & - & - & - & - & - & - & 10 & - & - & -\\
42403 & - & - & - & - & - & - & - & - & 1 & - & - & - & - & - & - & - & - & - & - & - & - & - & - & - & -\\
42438 & - & - & 12 & - & - & - & - & - & - & - & - & - & - & - & - & - & - & - & - & - & - & - & - & - & -\\
43410 & 2 & - & 8 & - & - & - & - & - & - & - & - & - & - & 1 & - & - & 10 & - & - & - & - & 10 & - & - & -\\
43557 & - & - & - & - & - & - & - & - & - & - & - & - & - & - & - & 1 & - & - & - & - & - & - & - & - & -\\
43726 & - & - & 9 & - & - & - & - & - & - & - & - & - & - & - & - & - & 11 & - & - & - & - & 10 & - & - & -\\
44897 & - & - & - & - & - & - & - & - & 22 & - & - & - & - & - & - & - & 5 & - & - & - & - & 11 & - & - & -\\
44997 & - & - & - & - & - & - & - & - & - & - & - & - & - & - & - & 1 & - & - & - & - & - & - & - & - & -\\
46066 & - & - & - & - & - & - & - & - & - & - & - & - & - & - & - & 1 & - & - & - & - & - & - & - & - & -\\
46580 & - & - & - & - & - & - & - & - & - & - & - & - & - & - & - & - & - & - & - & - & - & - & - & - & 1\\
46903 & - & - & - & - & - & - & - & - & - & - & - & - & - & - & - & 1 & - & - & - & - & - & - & - & - & -\\
49081 & - & - & - & - & - & - & - & - & 1 & - & - & - & - & - & - & - & - & - & - & - & - & - & - & - & -\\
49350 & - & - & - & - & - & - & - & - & - & - & - & - & - & - & - & 1 & - & - & - & - & - & - & - & - & -\\
49580 & - & - & - & - & - & - & - & - & - & - & - & - & - & - & - & 1 & - & - & - & - & - & - & - & - & -\\
49728 & - & - & - & - & - & - & - & - & - & - & - & - & - & - & - & 1 & - & - & - & - & - & - & - & - & -\\
49756 & - & - & - & - & - & - & - & - & - & 2 & - & - & - & - & - & - & - & - & - & - & - & - & - & - & -\\
49908 & - & - & - & - & - & - & - & - & - & - & - & - & - & - & - & - & - & - & - & - & - & - & - & 1 & -\\
50316 & - & - & 1 & - & - & - & - & - & - & - & - & - & - & - & - & - & - & - & - & - & - & - & - & - & -\\
50505 & 2 & - & - & - & - & - & - & - & - & - & - & - & - & - & - & - & - & - & - & - & - & - & - & - & -\\
53721 & - & - & - & - & - & - & - & - & - & - & - & - & - & - & - & 1 & - & - & - & - & - & - & - & - & -\\
54952 & - & - & - & - & - & - & - & - & - & - & - & - & - & - & - & - & - & - & - & - & - & - & - & - & 1\\
55459 & - & - & - & - & - & - & - & - & - & 2 & - & - & - & - & - & 3 & - & - & - & - & - & - & - & - & -\\
56242 & - & - & 1 & - & - & - & - & - & - & - & - & - & - & - & - & - & - & - & - & - & - & - & - & - & -\\
56948 & - & - & - & - & - & - & - & - & - & - & - & - & - & - & - & 2 & - & - & - & - & - & - & - & - & -\\
56997 & - & - & - & - & - & - & - & - & 21 & - & - & - & - & 1 & - & - & 9 & - & - & - & - & 10 & - & - & -\\
57939 & - & - & 1 & - & - & - & - & - & - & - & - & - & - & - & - & - & - & - & - & - & - & - & - & - & -\\
\hline
\end{tabular}
\end{center}
\end{table} 
\end{landscape}

\addtocounter {table} {-1}

\begin{landscape}
\begin{table}
\scriptsize
\begin{center}
\caption{continued.} 
\begin{tabular}{lccccccccccccccccccccccccc}
\hline
Telescope: & TBL & CFH & TBL & TBL & TBL & TBL & CFH & TBL & TBL & CFH & TBL & CFH & TBL & TBL & TBL & CFH & TBL & TBL & TBL & CFH & TBL & TBL & TBL & TBL & TBL\\
Year (2000+): & 06 & 06 & 07 & 07 & 07 & 07 & 07 & 07/08 & 08 & 08 & 08 & 08 & 08 & 09 & 09 & 09/10 & 09/10 & 10 & 10/11 & 11 & 11 & 11/12 & 12 & 12/13 & 13 \\
Start: & nov & nov & jun & may & jul & nov & dec & dec & jan & mar & may & jun & jul & jan & may & sep & dec & mar & sep & feb & mar & oct & mar & oct & mar \\
End: & nov & dec & feb & may & aug & nov & dec & jan & feb & mar & may & jul & aug & jan & jul & jan & feb & aug & feb & jul & aug & feb & aug & jan & aug \\
HIP \# &  &  &  &  &  &  &  &  &  &  &  &  &  &  &  &  &  &  &  &  &  &  &  & \\
\hline
58708\sups{SG} & - & - & - & - & - & - & - & - & - & - & - & - & - & - & - & - & - & - & - & - & - & 1 & - & - & -\\
60098 & - & - & - & - & - & - & - & - & - & - & - & - & - & - & - & - & - & - & - & - & - & 1 & - & - & -\\
60353 & - & - & - & - & - & - & - & - & - & - & - & - & - & - & - & - & - & - & - & - & - & 1 & - & - & -\\
61901 & - & - & - & - & - & - & - & - & - & - & - & - & - & - & - & - & - & - & - & - & - & - & - & - & 1\\
62523 & - & - & 1 & - & - & - & - & - & - & - & - & - & - & - & - & - & - & - & - & - & - & - & - & - & -\\
64797 & - & - & - & - & - & - & - & - & - & - & - & - & - & - & - & - & - & - & - & - & - & - & - & - & 1\\
65347 & - & - & - & - & - & - & - & - & - & - & - & - & - & - & - & - & - & - & - & 1 & - & - & - & - & -\\
66147 & - & - & - & - & - & - & - & - & - & - & - & - & - & - & - & - & - & - & - & - & - & - & - & - & 1\\
66774\sups{SG} & - & - & - & - & - & - & - & - & - & - & - & - & - & - & - & - & - & - & - & 1 & - & - & - & - & -\\
67275 & - & - & - & - & - & - & - & - & - & - & - & - & - & - & - & - & - & - & - & - & - & - & - & - & 8\\
67422 & - & - & - & - & - & - & - & - & - & - & - & - & - & - & - & - & - & - & - & - & - & - & - & - & 1\\
68184 & - & - & - & - & - & - & - & - & - & - & - & - & - & - & - & - & - & - & - & - & - & - & - & - & 1\\
71181 & - & - & - & - & - & - & - & - & - & - & - & - & - & - & - & - & - & - & - & - & - & - & - & - & 1\\
71631 & - & 9 & 15 & - & - & - & - & - & 5 & - & - & - & - & - & - & - & 1 & - & - & - & - & 10 & - & - & -\\
72848 & - & - & - & - & - & - & - & - & - & - & - & - & - & - & - & - & - & - & - & - & - & - & - & - & 1\\
74432 & - & - & 1 & - & - & - & - & - & - & - & - & - & - & - & - & - & - & - & - & - & - & - & - & - & -\\
76114 & - & - & - & - & - & - & - & - & - & 2 & - & - & - & - & - & - & - & - & - & - & - & - & - & - & -\\
79578 & - & - & - & - & - & - & - & - & - & 3 & - & 1 & - & - & - & - & - & - & - & - & - & - & - & - & -\\
79672 & - & - & 6 & - & 11 & - & - & - & - & - & - & - & 4 & - & 15 & - & 2 & 10 & - & - & - & - & 2 & - & 7\\
81300 & - & - & 1 & - & - & - & - & - & - & - & - & - & - & - & - & - & - & - & - & - & - & - & - & - & -\\
82588 & - & - & 1 & - & - & - & - & - & - & - & - & - & - & - & - & - & - & - & - & - & - & - & - & - & -\\
86193 & - & - & - & - & - & - & - & - & - & - & - & 3 & - & - & - & - & - & - & - & - & - & - & - & - & -\\
86400 & - & - & - & - & 1 & - & - & - & - & - & - & - & - & - & - & - & - & - & - & - & - & - & - & - & -\\
86974 & - & - & - & - & - & - & - & - & - & - & - & - & - & - & - & - & - & - & - & 1 & - & - & - & - & -\\
88194 & - & - & - & - & - & - & - & - & - & - & - & 3 & - & - & - & - & - & - & - & - & - & - & - & - & -\\
88945 & - & - & - & - & 1 & - & - & - & - & - & - & - & - & - & - & - & - & 20 & - & - & 6 & - & 10 & - & -\\
88972 & - & - & - & - & 1 & - & - & - & - & - & - & - & - & - & - & - & - & - & - & - & - & - & - & - & -\\
90729\sups{SG} & - & - & - & - & - & - & - & - & - & - & - & - & - & - & - & - & - & - & - & 1 & - & - & - & - & -\\
91043 & - & - & - & 30 & - & 12 & - & - & - & - & 11 & - & - & - & - & - & - & - & - & - & - & - & - & - & -\\
92984 & - & - & - & - & - & - & - & - & - & - & - & - & 40 & - & - & - & - & - & - & - & - & - & 6 & - & -\\
95253 & - & - & - & - & 2 & - & - & - & - & - & - & - & - & - & - & - & - & - & - & - & - & - & - & - & -\\
95962 & - & - & - & - & - & - & - & - & - & - & - & 4 & - & - & - & - & - & - & - & - & - & - & - & - & -\\
96085 & - & - & - & - & - & - & - & - & - & - & - & - & - & - & - & - & - & - & - & - & - & - & - & - & 1\\
96100 & - & - & - & - & 4 & - & - & - & - & - & - & - & - & - & 7 & - & - & - & - & - & - & - & - & - & -\\
96895 & - & - & - & - & - & - & - & - & - & - & - & - & - & - & - & - & - & - & - & - & 3 & - & - & - & 1\\
96901 & - & - & - & - & - & - & - & - & - & - & - & - & - & - & - & - & - & - & - & - & 3 & - & - & - & 1\\
98921 & - & - & - & - & 16 & - & - & - & - & - & - & - & 10 & - & 14 & - & - & 10 & - & - & 10 & - & 37 & - & -\\
100511 & - & - & - & - & 1 & - & - & - & - & - & - & - & - & - & - & - & - & - & - & - & - & - & - & - & -\\
100970 & - & - & - & - & - & - & - & - & - & - & - & - & - & - & - & - & - & - & - & 1 & - & - & - & - & -\\
101875 & - & - & - & - & 2 & - & - & - & - & - & - & - & - & - & - & - & - & - & - & - & - & - & - & - & -\\
104214 & - & - & - & - & 14 & - & - & - & - & - & - & - & 10 & - & - & - & - & 19 & - & - & - & - & 10 & - & 15\\
107350 & - & - & - & - & 16 & - & - & - & - & - & - & - & 11 & - & 11 & - & - & 13 & - & - & 10 & - & 15 & - & 15\\
108473 & - & - & - & - & - & - & - & - & - & - & - & - & - & - & - & - & - & - & 1 & - & - & - & - & - & -\\
108506\sups{SG} & - & - & - & - & - & - & - & - & - & - & - & - & - & - & - & - & - & - & 1 & - & - & - & - & - & -\\
109378 & - & - & - & - & - & - & - & - & - & - & - & - & - & - & - & - & - & - & 1 & - & - & - & - & - & -\\
109439\sups{SG} & - & - & - & - & - & - & - & - & - & - & - & - & - & - & - & - & - & - & 1 & - & - & - & - & - & -\\
109572 & - & - & - & - & - & - & - & - & - & - & - & - & - & - & - & - & - & - & - & - & - & - & - & - & 1\\
109647\sups{SG,B} & - & - & - & - & - & - & - & - & - & - & - & - & - & - & - & - & - & - & - & - & - & - & - & - & 1\\
111274 & - & - & - & - & - & - & - & - & - & - & - & - & - & - & - & - & - & - & 1 & - & - & - & - & - & -\\
113357 & - & - & - & - & - & - & - & - & - & - & - & - & - & - & 1 & - & - & - & 1 & 1 & - & - & - & - & -\\
113421 & - & - & - & - & - & - & - & - & - & - & - & - & - & - & - & - & - & - & 1 & - & - & - & - & - & -\\
113829 & - & - & - & - & - & - & - & - & - & - & - & - & - & - & - & - & - & - & 1 & - & - & - & - & - & -\\
\hline
\end{tabular}
\end{center}
\end{table} 
\end{landscape}

\addtocounter {table} {-1}

\begin{landscape}
\begin{table}
\scriptsize
\begin{center}
\caption{continued.}
\begin{tabular}{lccccccccccccccccccccccccc}
\hline
Telescope: & TBL & CFH & TBL & TBL & TBL & TBL & CFH & TBL & TBL & CFH & TBL & CFH & TBL & TBL & TBL & CFH & TBL & TBL & TBL & CFH & TBL & TBL & TBL & TBL & TBL\\
Year (2000+): & 06 & 06 & 07 & 07 & 07 & 07 & 07 & 07/08 & 08 & 08 & 08 & 08 & 08 & 09 & 09 & 09/10 & 09/10 & 10 & 10/11 & 11 & 11 & 11/12 & 12 & 12/13 & 13 \\
Start: & nov & nov & jun & may & jul & nov & dec & dec & jan & mar & may & jun & jul & jan & may & sep & dec & mar & sep & feb & mar & oct & mar & oct & mar \\
End: & nov & dec & feb & may & aug & nov & dec & jan & feb & mar & may & jul & aug & jan & jul & jan & feb & aug & feb & jul & aug & feb & aug & jan & aug \\
HIP \# &  &  &  &  &  &  &  &  &  &  &  &  &  &  &  &  &  &  &  &  &  &  &  & \\
\hline
113896 & - & - & - & - & - & - & - & - & - & - & - & - & - & - & - & - & - & - & 1 & - & - & - & - & - & -\\
113994\sups{SG} & - & - & - & - & - & - & - & - & - & - & - & - & - & - & - & - & - & - & 1 & - & - & - & - & - & -\\
114378 & - & - & - & - & - & - & - & - & - & - & - & - & - & - & - & - & - & - & 2 & - & - & - & - & - & -\\
114456 & - & - & - & - & - & - & - & - & - & - & - & - & - & - & - & - & - & - & 2 & - & - & - & - & - & -\\
114622 & - & - & - & - & - & - & - & - & - & - & - & - & - & - & - & - & - & - & 1 & - & - & - & - & - & -\\
115951 & - & - & - & - & - & - & - & - & - & - & - & - & - & - & - & - & - & - & 1 & - & - & - & - & - & -\\
116085 & - & - & - & - & - & - & - & - & - & - & - & - & - & - & - & - & - & - & 1 & - & - & 1 & - & - & -\\
116106 & - & - & - & - & - & - & - & - & - & - & - & - & - & - & - & - & - & - & 1 & - & - & - & - & - & -\\
116421 & - & - & - & - & - & - & - & - & - & - & - & - & - & - & - & - & - & - & 1 & - & - & - & - & - & -\\
116613 & - & - & - & - & - & - & - & - & - & - & - & - & - & - & - & - & - & - & 1 & - & - & - & - & - & -\\
hd131156A & - & - & - & - & 12 & - & - & - & 17 & - & - & - & 2 & - & 13 & - & 11 & 19 & - & - & 11 & 14 & - & - & 13\\
hd131156B & - & - & - & - & - & - & - & - & 20 & - & - & - & - & - & 5 & - & - & 10 & - & - & - & 8 & - & - & -\\
hd179958 & - & - & - & - & - & - & - & - & - & - & - & 3 & - & - & - & - & - & - & - & - & - & - & - & - & -\\
moon & - & - & - & - & - & - & - & - & - & - & - & - & - & - & - & - & - & - & - & 1 & - & - & - & - & -\\
\hline
\end{tabular}
\end{center}
\end{table} 
\end{landscape}



\label{lastpage}

\end{document}